\documentclass[fleqn,usenatbib]{mnras}
\usepackage{newtxtext,newtxmath}
\usepackage{subcaption}
\usepackage[T1]{fontenc}

\DeclareRobustCommand{\VAN}[3]{#2}
\let\VANthebibliography\thebibliography
\def\thebibliography{\DeclareRobustCommand{\VAN}[3]{##3}\VANthebibliography}

\usepackage{graphicx}	
\usepackage{amsmath}	

\title[Evolution of Multi-wavelength AGNs]{A Spatially Resolved Evolutionary Sequence of Multi-wavelength AGN Host Galaxies}

\author[Gaoxiang Jin et al.]{
Gaoxiang Jin,$^{1}$\thanks{E-mail: gxjin@mpa-garching.mpg.de}
Guinevere Kauffmann,$^{1}$
Y. Sophia Dai,$^{2}$
Martin J. Hardcastle,$^{3}$
and Bohan Yue$^{4,5}$
\\
\\
$^{1}$Max Planck Institute for Astrophysics, Karl-Schwarzschild-Str. 1, D-85741 Garching, Germany\\
$^{2}$Chinese Academy of Sciences South America Center for Astronomy, National Astronomical Observatories, CAS, Beijing 100101, People's Republic of China\\
$^{3}$Department of Physics, Astronomy and Mathematics, University of Hertfordshire, College Lane, Hatfield AL10 9AB, UK\\
$^{4}$Institute for Astronomy, University of Edinburgh, Royal Observatory, Blackford Hill, Edinburgh EH9 3HJ, UK\\
$^{5}$Leiden Observatory, Leiden University, PO Box 9513, NL-2300 RA Leiden, The Netherlands
}

\date{Accepted XXX. Received YYY; in original form ZZZ}

\pubyear{\the\year{}}

\begin{document}
\label{firstpage}
\pagerange{\pageref{firstpage}--\pageref{lastpage}}
\maketitle

\begin{abstract}

We study the spatially resolved star formation, gas ionisation, and outflow properties of 1813 active galactic nuclei (AGNs) from the MaNGA survey, which we classify into infrared (IR), broad-line (BL), narrow-line (NL), and radio (RD) AGNs based on their mid-infrared colours, optical spectra, and/or radio photometry. We also provide estimations of AGN power at different wavelengths.
AGN incidence is found to increase with stellar mass following a power-law, with the high-mass end dominated by RDAGNs and the low-mass end dominated by NLAGNs. 
Compared to their mass-matched non-AGN counterparts, we find that IRAGNs, BLAGNs, and NLAGNs on average show enhanced specific star formation rates, younger stellar populations, and harder ionisation towards the centre. RDAGNs, in contrast, show radial profiles similar to quiescent galaxies.
[\ion{O}{III}] outflows are more common and stronger in BL/IRAGNs, while RDAGNs on average show no outflow features. The 
outflow incidence increases with [\ion{O}{III}] luminosity, and the features in BL/IRAGNs on average extend to $\sim$2 kpc from the nuclei.
We further discuss a possible evolutionary sequence of AGNs and their host galaxies, where AGNs with strong emission lines or dust tori are present in
star-forming galaxies. 
Later, young compact radio jets emerge, the
host galaxies gradually quench, and the AGN hosts eventually evolve into globally quiescent systems
with larger radio jets that prevent further gas cooling.
\end{abstract}




\section{Introduction}
\label{sec:intro}
Despite the significant difference in physical scales between galaxies and their central supermassive black holes (SMBHs), 
the masses of SMBHs are found to correlate tightly with the stellar masses of their host galaxies or bulges \citep[e.g., ][and references therein]{1998AJ....115.2285M,2013ApJ...764..184M,2015ApJ...813...82R}. 
This suggests a co-evolution scenario between SMBHs and their host galaxies, 
in which the growth of SMBHs and the mass assembly of galaxies are closely linked through cosmic time, 
with similar merger and accretion histories expected for both \citep[e.g.][]{2013ARA&A..51..511K}. 
Observationally, SMBHs that are actively accreting are known as active galactic nuclei (AGNs). 
On average, the accretion rate density of AGNs and the star formation rate density of galaxies are found to be similar across cosmic time \citep[see][for a review]{2014ARA&A..52..415M},
which strongly supports the idea that there are mechanisms linking the growth of SMBHs and their host galaxies. 
However, the physical processes that drive this co-evolution are still not fully understood.

To reproduce the observed galaxy mass function in cosmological simulations, specifically the upper limit on galaxy masses, 
most galaxy evolution models require AGNs to inject energy into  the gas surrounding their host galaxies and  to suppress star formation \citep[e.g.][and references therein]{2017NatAs...1E.165H}. 
This process, known as AGN feedback, can also explain the co-evolution of SMBHs and galaxies. 
Many simulations have shown that strong AGN activity can simultaneously regulate star formation and black hole growth by fuelling, heating, or expelling gas \citep[e.g.][]{2005MNRAS.361..776S,2005Natur.433..604D,2006ApJS..163....1H}.

In observations, direct evidence of AGN feedback has been found in the form of AGN-driven outflows \citep[e.g.][]{1990ApJS...74..833H,2010MNRAS.402.2211A,2014MNRAS.441.3306H,2025A&A...695A.185Z} and AGN-heated hot bubbles in cool-core galaxy clusters \citep[see][for a review]{2012ARA&A..50..455F,2020NewAR..8801539H}.
However, most of this direct evidence is only found in the most luminous or massive AGNs. 
For the majority of the AGN population with moderate and low luminosities, only indirect evidence is found, 
typically from comparisons with normal galaxies. Examples of such indirect evidence include lower cold gas fractions in AGN hosts \citep[e.g.][]{2024Natur.632.1009W}, disturbed gas kinematics \citep[e.g.][]{2013MNRAS.433..622M}, and  higher  excitation ionized gas \citep[e.g.][]{2018MNRAS.480.3201K}.

AGN samples can be strongly affected by selection biases, especially when selection is carried out at only one wavelength or using only one technique.
A unified model of the internal structure of AGNs and their accretion processes was proposed decades ago to explain the various observed AGN types \citep[e.g.][]{1993ARA&A..31..473A}, 
attributing them to different viewing angles toward the same intrinsic structure. 
However, this model struggles to explain the diversity found in large AGN samples with multi-wavelength observations \citep[e.g.][]{2017A&ARv..25....2P,2019MNRAS.488.3109K}. 
The AGN populations selected at different wavelengths are often found to have different host galaxy properties \citep[see][for a review]{2014ARA&A..52..589H}.

How AGN properties at different wavelengths connect with host galaxy evolution is still an open question.
In recent years, a growing body of  theoretical work  has suggested that the AGN feedback model and  black hole accretion processes can be described by a two-mode scenario: the radiative mode and the kinetic mode \citep[e.g.][]{2007MNRAS.380..877S,2014ARA&A..52..589H}.
Radiative-mode AGNs are characterized by high Eddington ratios, strong radiative winds, and gas-rich environments \citep[e.g.][]{2005Natur.433..604D}, 
and are powered by a radiatively efficient accretion disk \citep[e.g.][]{1973A&A....24..337S}. 
In observations, these AGNs are often identified at X-ray, optical, and infrared wavelengths through their luminous accretion disks, broad and narrow emission line regions, and the dusty torus.
Subsample of them are also found to host radio jet \citep[e.g., radio-loud quasar,][]{2019A&A...631A..46G},
Their host galaxies are often found to have strong star formation activity \citep[e.g.][]{2018MNRAS.478.4238D,2021ApJ...906...38Z}. 
Kinetic-mode AGNs are characterized by low Eddington ratios, powerful radio jets, and gas-poor environments \citep[e.g.][]{2005MNRAS.363L..91C}, 
and are powered by a radiatively inefficient accretion flow \citep[e.g.][]{1995ApJ...452..710N}. 
Observationally, these AGNs are typically identified at radio wavelengths due to their strong and large-scale radio jets (sometimes appearing as giant radio galaxies), 
and their host galaxies are often found to be quiescent, having quenched their star formation long ago \citep[e.g.][]{2005MNRAS.362...25B,2012MNRAS.421.1569B,2025A&A...694A.309J}. 
In simulations, radiative-mode AGNs typically heat the surrounding gas, while kinetic-mode AGNs inject energy through shocks into their host galaxies \citep[e.g.][]{2017MNRAS.465.3291W}, 
though the detailed impact of different AGN feedback modes depends on the accretion model and the gas environment of the host galaxy.

A number of observational studies have gone one step further and suggested that these two modes may be linked by an evolutionary sequence.
Dust-reddened or infrared-bright quasars, typically associated with gas-rich, merger-driven systems, exhibit both active star formation and powerful radiatively driven winds. 
These characteristics suggest they represent a transitional phase bridging heavily obscured growth and unobscured blue quasars \citep[e.g.][]{2012ApJ...757...51G,2012MNRAS.427.2275B,2019MNRAS.488.3109K}.
At later times, as gas and dust are depleted or expelled, the AGN is expected to evolve towards a lower-Eddington, jet-dominated phase in which kinetic feedback from radio jets helps to maintain quiescence in massive galaxies and clusters \citep[e.g.][]{2012MNRAS.421.1569B,2012ARA&A..50..455F}.
Together, these works outline a phenomenological picture in which obscured, radiative-mode AGNs, unobscured quasars, and radio-mode systems represent different stages in the co-evolution of black holes and their hosts.
However, most such studies focus on luminous quasars at intermediate and high redshift, and are based on integrated host properties rather than spatially resolved information in typical, low-redshift galaxies.

Most of the observed scaling relations between SMBHs and their host galaxies mentioned previously, as well as evidence for  AGN feedback, are derived from global galaxy properties.
However, recent large integral field unit (IFU) surveys, 
such as MaNGA \citep[Mapping nearby Galaxies at Apache Point Observatory,][]{2015ApJ...798....7B}, 
CALIFA \citep[Calar Alto Legacy Integral Field Area survey, ][]{2012A&A...538A...8S}, 
and SAMI \citep[Sydney-AAO Multi-object Integral-field spectrograph, ][]{2015MNRAS.447.2857B}, 
have shown that the evolution of galaxies can better be understood if physical properties are resolved spatially \citep[e.g.][]{2018MNRAS.477.3014B}.
Local massive galaxies ($M_{\star}>10^{9} \rm M_{\odot}$) are found to grow their stellar mass inside-out on average \citep[e.g.][]{2018MNRAS.480.2544R}.
This raises several key questions:
How are the  AGN-galaxy scaling relations and the AGN feedback phenomenon manifested at different galactocentric radii?
What is the typical scale over which AGNs affect their host galaxies? Do these scales differ for AGNs selected at different wavelengths or using different techniques?

MaNGA, as the largest completed IFU survey, 
is ideal for investigating these questions.
It provides spatially resolved spectra for a sample of $\sim$10\,000 galaxies \citep{2015ApJ...798....7B} with a wide range of stellar masses \citep[$10^{9}-10^{12}\rm M_{\odot}$, ][]{2017AJ....154...86W}.
The IFU data can resolve star formation and gas ionisation properties on kpc scales and out to at least 1.5 effective radii ($R_{\rm e}$) for most galaxies.
Together with photometric data from large sky surveys at other wavelengths, we can robustly classify AGNs into different populations and study their host galaxy properties in a spatially resolved manner.
This combination is particularly well suited to test, in the local universe, whether radiative-mode and kinetic-mode AGNs can be arranged along a coherent evolutionary sequence that is reflected in the spatially resolved properties of their hosts.
Multi-wavelength selection within the MaNGA sample allows us to identify infrared, broad-line, narrow-line, and radio AGNs, which are commonly associated with the dusty torus, the broad- and narrow-line regions, and radio jets, respectively.
By combining these classifications with MaNGA IFU data, we can trace how star formation, stellar age, ionisation state, and ionised outflows vary as a function of radius for each AGN class, and compare them to carefully mass-matched non-AGN controls.
In this way, we can extend the red-blue quasar evolution scenario proposed in \citet{2019MNRAS.488.3109K} to a larger, lower luminosity population of low-redshift AGNs, and ask whether the transition from radiative to kinetic dominance is accompanied by systematic, kpc-scale changes in their host galaxies.

\citet{2020ApJ...901..159C}  analyzed  a multi-wavelength AGN catalogue (X-ray, infrared, and radio AGNs) for $\sim$60\% of the full MaNGA sample. This work 
demonstrated the potential of combining MaNGA IFU data with multi-wavelength AGN selection. In practice, the IR and X-ray AGN samples were too small
to yield statistically robust results, so the comparisons were restricted to radio (1.4 GHz)  and optical emission line samples.
Since then, the LOFAR Two-metre Sky Survey \citep[LoTSS, ][]{2017A&A...598A.104S,2022A&A...659A...1S} has provided a significant improvement in radio data.
The LoTSS survey is one to two orders of magnitude deeper than previous large sky radio surveys, 
doubling the number of radio detections for the MaNGA galaxies \citep{2025A&A...694A.309J}.
We will use the latest data release of the LoTSS survey (LoTSS DR3, Shimwell et al., submitted), 
which will cover almost the entire MaNGA sample, 
allowing us to complete the radio view of the MaNGA AGNs.

In summary, we have IFU data from MaNGA to classify broad- and narrow-line AGNs based on their emission line widths and ratios \citep[e.g.][]{2020MNRAS.499.5749J}.
We have mid-infrared (mid-IR) photometry from the Wide field Infrared Survey Explorer \citep[WISE, ][]{2010AJ....140.1868W} to select infrared AGNs based on their mid-infrared colours \citep[e.g.][]{2024ApJ...977..102P}.
We also have deep radio photometry at 144 MHz from LoTSS DR3 to identify radio AGNs based on their excess radio emission \citep[e.g.][]{2025A&A...694A.309J}.
In addition to AGN classification, we can also use these high-quality data to estimate the luminosity of (or upper limits on the luminosity)  of different AGN structures such as the jet or dusty torus.
For example, the broad line region (BLR) power can be traced by the broad component of the Balmer emission lines;
the narrow line region (NLR) power can be traced by high-ionisation emission lines such as [\ion{O}{III}]$\lambda 5008$;
the torus power can be estimated from spectral energy distribution (SED) fitting of photometry from the ultraviolet (UV) to the infrared (IR);
and the jet power can be estimated from the excess radio luminosity at 144 MHz compared to the radio-SFR relation \citep[e.g.][]{2025A&A...694A.309J}.
By combining these data with the IFU information of the host galaxies, we can investigate global and spatially resolved star formation, gas ionisation, and outflow properties.
Finally, using the spatially resolved velocity dispersion as a proxy for SMBH mass, we can test the SMBH-galaxy co-evolution scenarios in a spatially resolved manner.

The paper is structured as follows:
In Section~\ref{sec:data}, we introduce the sample selection, observational data, and the calculation of physical parameters.
In Section~\ref{sec:agn}, we describe the classification criteria of different AGN populations and the estimation of AGN luminosities at different wavelengths.
In Section~\ref{sec:results}, we compare the global and radial gradients of star formation, gas ionisation, and outflow properties among different AGN populations and with their mass-matched non-AGN counterparts.
In Section~\ref{sec:discussion}, we discuss how to interpret our classifications and results in the context of an evolutionary sequence of AGNs and their host galaxies.
Finally, we summarize our main conclusions in Section~\ref{sec:conclusion}.

Throughout this paper, we convert all radio luminosities to rest-frame 144 MHz assuming a radio spectral index $\alpha=-0.7$ \citep{2002AJ....124..675C}, defined such that $L_{\nu_{1}}/L_{\nu_{2}}=(\nu_{1}/\nu_{2})^{\alpha}$.
We adopt a flat cosmology with $H_{0}=70\,\rm km\,s^{-1}\, Mpc^{-1}$, $\Omega_{m}=0.3$, and $\Omega_{\Lambda}=0.7$.
All stellar masses and star formation rates are based on the Chabrier initial mass function \citep[IMF,][]{2003PASP..115..763C}.
When comparing literature results based on different IMFs, we follow \citet{2014ARA&A..52..415M} and use the conversion factors (for both stellar masses and star formation rates): Salpeter\,:\,Kroupa\,:\,Chabrier\,=\,1\,:\,0.67\,:\,0.63.
Throughout this paper, optical magnitudes are reported in the AB system, while infrared magnitudes from WISE are given in the Vega system.

\section{Sample and Data}\label{sec:data}
\subsection{Parent sample selection}
\label{sec:sample}
To classify different AGN populations and study their spatially resolved properties, 
we require a large sample of galaxies with both IFU data and photometric data from large sky surveys at optical, mid-infrared, and radio wavelengths.
In addition, we must take into account the depth of each survey to address selection effects and ensure that the distribution of AGN luminosity in host galaxies of different stellar masses is not biased by incompleteness.

MaNGA is the largest IFU survey to date, which provides spatially resolved spectra for more than 10,000 nearby galaxies \citep{2015ApJ...798....7B}.
We start with the full MaNGA sample as our parent sample, 
which is designed to achieve a flat mass distribution in the range of $9.5<\rm log(\it M_{\star}/\rm M_{\odot})<11.5$.
This means that massive galaxies are oversampled.
To enable correction of this sample to a volume-limited one, 
a volume weight is provided for each galaxy \citep{2017AJ....154...86W}.
The resulting sample is divided into the Primary+ and Secondary samples, 
for which the IFU covers at least 1.5 and 2.5 effective radii ($R_{\rm e}$) of the galaxies, respectively.
Elliptical-Petrosian photometric measurements spanning from the ultraviolet (UV) to optical wavelengths are available for each MaNGA galaxy \citep{2011AJ....142...31B}.

For the selection of infrared AGNs (IRAGNs), we require mid-infrared photometry.
The WISE mission provides all-sky imaging in four mid-infrared bands,
among which the W1 and W2 bands are most useful for IRAGN selection at low redshift \citep{2010AJ....140.1868W}.
MaNGA galaxies are selected to have an $r$-band magnitude brighter than 17 \citep{2017AJ....154...86W}.
Assuming a colour index of $r$-W1$>3$ for normal galaxies \citep[e.g.][]{2016ApJ...818...88L},
the majority of the MaNGA sample is expected to be well-detected in the W1 and W2 bands \citep[limiting magnitudes $\sim 19$, ][]{2013wise.rept....1C}.

For the classification of radio AGNs (RDAGNs), we use the latest LoTSS DR3 data, which provides deep radio imaging for most of the northern sky at 144 MHz.
LoTSS DR3 has a median sensitivity of 112 $\rm \mu Jy\,beam^{-1}$ and a resolution of 6 arcsec (Shimwell et al., submitted), enabling the detection of luminous radio AGNs and star-forming galaxies, and also providing meaningful upper limits on radio luminosities for non-detections.

We require the final sample to be detected in the W2 band (S/N > 3) for robust IRAGN selection.
Galaxies are also required to be within the LoTSS DR3 sky coverage for RDAGN classification.
We also exclude galaxies at $z<0.01$ to avoid cases where the MaNGA IFU spatial coverage may be insufficient for our analysis.

\begin{figure}
	\includegraphics[width=\columnwidth]{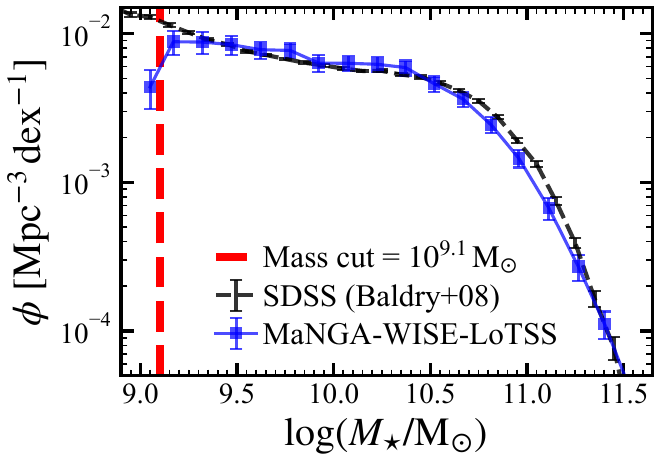}
    \caption{The galaxy stellar mass function of our parent sample (blue), 
    recovered using volume weights. 
    The red line shows the mass function of the full MaNGA Primary+ and Secondary sample. Our sample follows a mass function similar to that derived from the low-redshift SDSS sample \citep[][]{2008MNRAS.388..945B}, 
    and is thus mass-complete at $M_{\star} > \sim 10^{9.1}\rm M_{\odot}$.}
    \label{fig:gsmf}
\end{figure}

After applying the volume weights, our sample recovers the galaxy stellar mass function of the local universe \citep{2008MNRAS.388..945B} for $M_{\star}>10^{9.1}\rm M_{\odot}$, as shown in Figure~\ref{fig:gsmf}.
Therefore, we apply an additional mass cut at $10^{9.1}\rm M_{\odot}$ to ensure mass completeness.
The final parent sample contains 9222 unique galaxies, which have photometric information at UV, optical, mid-infrared, and radio wavelengths, as well as IFU data from MaNGA.

\subsection{Observational data}
\label{sec:obsdata}
The IFU data products used in this paper are provided by the survey-led data analysis pipeline {\tt MaNGA DAP} \citep{2019AJ....158..231W,2019AJ....158..160B}, published as part of the SDSS DR17 \citep{2022ApJS..259...35A}.
We derive the spatially resolved spectra and spectral properties from the {\tt LOGCUBE-SPX} and {\tt MAPS-SPX} files, respectively.
{\tt LOGCUBE-SPX} files provide spectra covering the 360--1030\,nm range for each spaxel \citep{2015AJ....149...77D}.
{\tt MAPS-SPX} files provide the properties of emission lines and spectral indices.
These spaxel-by-spaxel data, without binning, preserve the original high spatial resolution (pixel scale of 0.5\arcsec), and are thus suitable for our stacking analysis.

UV and optical photometry for our sample are taken directly from MaNGA's parent catalogue, the NASA-Sloan-Atlas \citep[NSA; ][]{2011AJ....142...31B,2017AJ....154...86W}.
Each MaNGA galaxy has elliptical Petrosian aperture photometry measurements in the GALEX FUV and NUV bands \citep{2005ApJ...619L...1M}, as well as the SDSS $ugriz$ bands \citep{1996AJ....111.1748F}.

For mid-IR photometry, we match the MaNGA galaxies to the DESI Legacy Surveys DR10 catalogue \citep{2019AJ....157..168D} using a 3 arcsec matching radius.
In the Legacy catalogue, WISE fluxes are measured from all imaging through year 7 of NEOWISE-Reactivation \citep{2014ApJ...792...30M}, using forced photometry on the unWISE images \citep{2014AJ....147..108L} based on the optical locations and shapes.
This approach is necessary because MaNGA galaxies are often more extended than the WISE point spread function (PSF), and their flux would otherwise be underestimated by the ALLWISE catalogue \citep{2015ApJS..219....8C}.

The radio photometry is based on the latest LoTSS DR3 observations (Shimwell et al., submitted).
For galaxies within the LoTSS DR2 coverage, we follow the approach of \citet{2025A&A...694A.309J}, which used the measurements from the LoTSS-optical catalogue \citep{2023A&A...678A.151H}.
The LoTSS-optical catalogue combines the likelihood-ratio cross-match method and visual inspection to identify the optical counterparts of the LoTSS radio sources and provides more accurate photometry for well-extended radio jets.
The visual inspection is still ongoing for the new LoTSS DR3 images,
so we match the rest of the MaNGA galaxies with the catalogue created by the Gaussian-fitting-based radio source extraction code {\tt PyBDSF} \citep{2015ascl.soft02007M},
using a 6 arcsec matching radius (typical LoTSS beam size) to get the radio photometry at 144 MHz.
For our sample, we find 4326 detections with flux and morphological measurements in LoTSS DR3.
According to the source completeness analysis in \citet{2024MNRAS.527.6540H},
we define the flux upper limit for each non-detected MaNGA galaxy as $f_{\rm upper}=f_{\rm 6arcsec} + 7.5\times \rm RMS$,
where $f_{\rm 6arcsec}$ is the flux measured within a 6 arcsec aperture at the optical position on the LoTSS flux map, and RMS is the local noise level estimated from the root-mean-square (RMS) map.

\subsection{Physical parameters}
\label{sec:param}
For global properties, the stellar mass ($M_{\star}$) and effective radius ($R_{\rm e}$) are taken from the NSA catalogue.
Global SFRs are estimated from integrated $\rm H\alpha$ luminosities within the IFU coverage using the conversion in \citet{2012ARA&A..50..531K}:
${\rm SFR\,(M_{\odot}\,yr^{-1})}=L_{\rm H\alpha SF}\, {(\rm erg\,s^{-1})} \times 10^{-41.30}$,
where $L_{\rm H\alpha SF}$ is corrected for both dust attenuation and AGN contamination.
The dust attenuation is estimated from the Balmer decrement assuming an intrinsic $\rm H\alpha/H\beta=2.86$ and the \citet{2000ApJ...533..682C} extinction curve $k(\lambda)$.
Therefore, all the emission line luminosities at wavelength $\lambda$ used in this paper are corrected by Eq.~\ref{eq:dust}:
\begin{equation}
    \frac{L_{\rm Line}}{L_{\rm Obs}} = \left( \frac{F_{\rm H\alpha} / F_{\rm H\beta}}{2.86} \right)^{\frac{k(\lambda)}{1.30}}
\label{eq:dust}
\end{equation}

AGN contributions to the $\rm H\alpha$ luminosity are estimated through different line ratios and are removed from the total $\rm H\alpha$ luminosity.
The method is described in detail in Section~\ref{sec:agn}.
Specifically, for broad-line AGNs, we use the global $M_{\star}$ and SFR derived from the spectral energy distribution (SED) fitting as described in Section~\ref{sec:lagn},
because the strong AGN continuum and broad emission lines can significantly affect the measurements from the optical spectra.

For spatially resolved properties, SFR surface density ($\Sigma_{\rm SFR}$) is calculated in a similar way as that of global SFR,
except the spectra are from MaNGA spaxels.
The star formation history (SFH) of each spaxel is derived from the full-spectrum fitting using the {\tt pyPipe3D} pipeline \citep{2022NewA...9701895L,2022ApJS..262...36S}.
The SFHs provide the stellar masses and their assembly histories of different regions in galaxies, and can be used to calculate the longer-timescale SFR averaged over several hundred Myrs.

The properties of the central SMBHs, such as the black hole mass ($M_{\rm BH}$) and Eddington ratio ($\lambda_{\rm Edd}$), are difficult to estimate directly from MaNGA data due to the kpc-scale spatial resolution.
Here we use empirical scaling relations to estimate these properties.
The $M_{\rm BH}$ is estimated from the stellar velocity dispersion ($\sigma_{\star}$) using the relation in \citet{2013ApJ...764..184M}: $\log M_{\rm BH}=8.32+5.64\, \log(\sigma_{\star}/200\rm km\,s^{-1})$,
where $\sigma_{\star}$ is the stellar velocity dispersion within the effective radius.
The bolometric luminosity ($L_{\rm Bol}$) can be roughly converted from the [\ion{O}{III}]$\lambda 5008$ luminosity ($L_{\rm [OIII]}$) by assuming a bolometric correction factor of $\sim 600 \pm 150$ \citep{2009MNRAS.397..135K},
and serves as a tracer of the radiative power of the AGN.
The Eddington ratio is then calculated as $\lambda_{\rm Edd}=L_{\rm Bol}/L_{\rm Edd}$, where $L_{\rm Edd}=1.26\times10^{38}(M_{\rm BH}/\rm M_{\odot})\rm \, erg\,s^{-1}$.

Similarly, the kinetic power of the radio jets $L_{\rm Kin}$ can be very roughly estimated from the empirical relation between the 1.4 GHz luminosity and the jet power \citep[e.g. Eq. 2 in][]{2014ARA&A..52..589H}: 
\begin{equation}
    L_{\rm Kin} = 2.8\times 10^{37} \left(\frac{L_{\rm 1.4GHz}}{10^{25}\rm \, W\,Hz^{-1}}\right)^{0.68} \rm W
\label{eq:ljet}
\end{equation}
The 1.4 GHz luminosity is converted from the excess 144\,MHz luminosity (see Sec.~\ref{sec:lagn}) assuming a spectral index of $\alpha=-0.7$.
The scaled kinetic power is then calculated as $\lambda_{\rm Kin} = L_{\rm Kin}/L_{\rm Edd}$.
However, this conversion introduces uncertainties due to potential differences in jet morphology observed at 1.4\,GHz and 144\,MHz. 
Furthermore, this scaling relation is calibrated using extended radio jets in galaxy clusters and thus may not be strictly applicable to unresolved sources or field AGNs.

We note that these scaling relations all have large uncertainties; thus, the estimated $M_{\rm BH}$, $\lambda_{\rm Edd}$, and $\lambda_{\rm Kin}$ are only used for statistical comparisons.

\section{The AGN sample}
\label{sec:agn}
\subsection{Classification of emission line, infrared, and radio AGNs}
\label{sec:agnclass}

\begin{figure*}
	\includegraphics[width=0.8\textwidth]{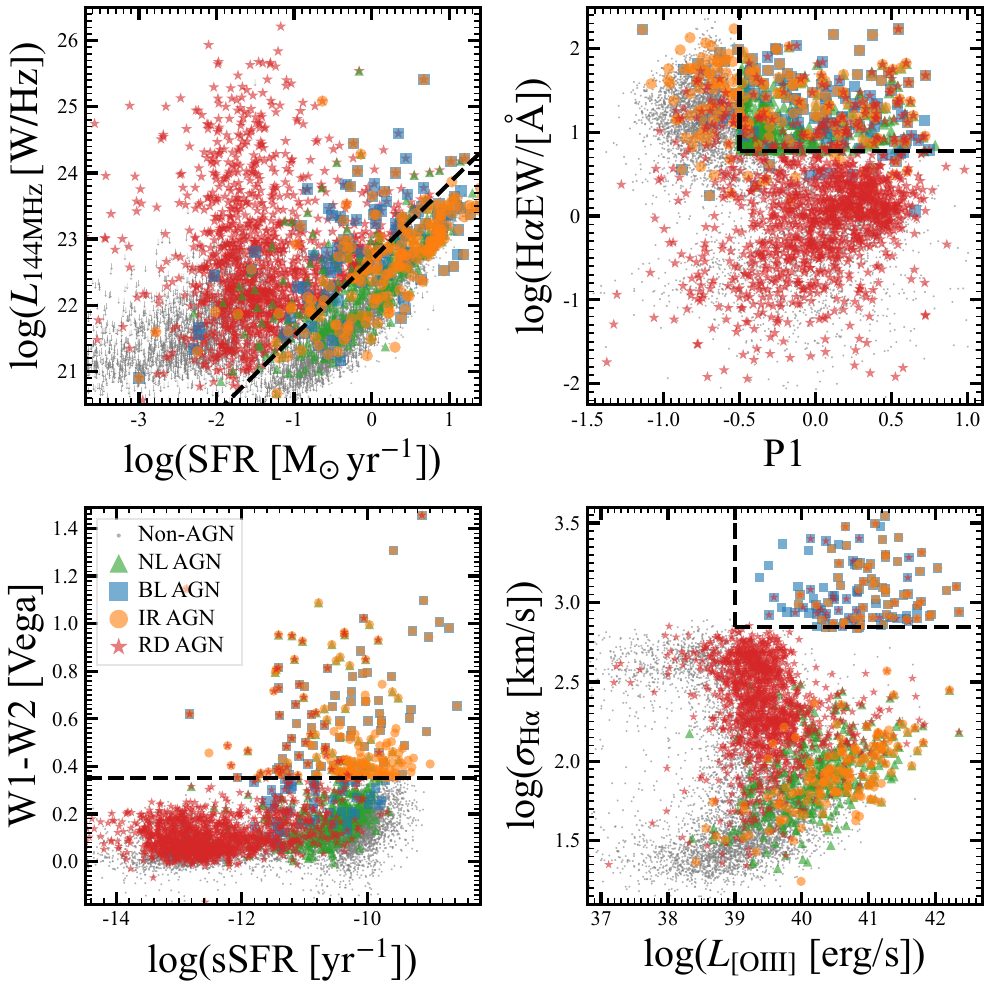}
    \caption{AGN classification diagnostics. The criteria are shown as black dashed lines in each panel. BLAGNs, NLAGNs, IRAGNs, RDAGNs, and non-AGNs are shown in blue squares, green triangles, orange circles, red stars, and grey dots, respectively.
    The RDAGNs are selected by at least 0.7\,dex stronger radio emission than that predicated by their SFR \citep[e.g.][]{2023MNRAS.523.1729B,2025A&A...694A.309J}.
    NLAGNs are required to have $\rm H\alpha$ equivalent width larger than 6 \citep{2010MNRAS.403.1036C}, and $P1>-0.5$. 
    P1 is defined as $\rm 0.63\,log([NII]/H\alpha)+0.51\,log([SII]/H\alpha)+0.59\,log([OIII]/H\beta)$, which is proposed in \citet{2020MNRAS.499.5749J} for MaNGA NLAGN classification.
    IRAGNs are selected based on their W1-W2 colour, and the BLAGNs are defined by the existence of broad Balmer lines. The $\sigma_{\rm H\alpha}$ in the lower right panel represents the velocity dispersion of the broad-line component for BLAGNs and of the narrow-line component for other galaxies; consequently, BLAGNs occupy a distinct region.
    If an AGN is classified into multiple types, it will be plotted at the same location using the corresponding symbols.
    }
    \label{fig:agnclass}
\end{figure*}

Based on the multi-wavelength photometric and spectroscopic data described in Section~\ref{sec:data}, we classify the AGNs into four populations: Broad-line AGNs with broad Balmer lines (hereafter BLAGN), narrow line AGNs with characteristic narrow emission lines (hereafter NLAGNs), infrared AGNs with red mid-infrared colour (hereafter IRAGNs), and radio AGNs showing excess radio emission (hereafter RDAGNs).
These four AGN populations are well known to represent different AGN structures, i.e., the broad-line region (BLR), narrow-line region (NLR), dust torus, and radio jets \citep[e.g.][]{2014ARA&A..52..589H,2015ARA&A..53..365N}.
It should be noted that these classifications are not mutually exclusive, and a single source may fall into multiple categories simultaneously.
For instance, a source classified as both an RDAGN and an BLAGN implies the coexistence of the radio jet and broad-line region.

BLAGNs are characterized by the broad Balmer emission lines in their optical spectra at MaNGA redshifts.
These lines are emitted from dense, ionised gas clouds close to the accreting SMBH \citep{1986ARA&A..24..171O}, which form the broad-line region (BLR).
Here, our BLAGN sample is defined as galaxies exhibiting a broad component (FWHM $> 700 \rm km\,s^{-1}$) in their Balmer emission lines.
We use the results of \citet{2023MNRAS.524.5827F}, who provided a complete catalogue of broad-line AGNs (BLAGNs) in the full MaNGA sample.
We also expand the BLAGN sample by fitting the nuclear spectra of all MaNGA galaxies with the same method.
In total we identify 130 BLAGNs in our parent sample, which are plotted in blue throughout the paper.

The narrow line regions (NLRs) of AGNs are created by low-density gas clouds which are photoionised by the AGN radiation.
The harder ionising spectrum from an AGN can excite gas to higher ionisation states compared to star formation, resulting in distinct emission-line ratios.
The Baldwin, Philips, and Terlevich (BPT) diagram \citep{1981PASP...93....5B} and Veilleux \& Osterbrock (VO) diagrams \citep{1987ApJS...63..295V}
are widely used to separate AGNs from star-forming galaxies.
\citet{2020MNRAS.499.5749J} reproject these diagrams into a 3D space,
and use the line ratios of [\ion{O}{III}]$\lambda 5008$/$\rm H\beta$ (R3), [\ion{N}{II}]$\lambda 6585$/$\rm H\alpha$ (N2), and [\ion{S}{II}]$\lambda\lambda 6718,6733$/$\rm H\alpha$ (S2)
to define a cleaner parameter $P_1$ ($P_1 \rm =0.63\,N2+0.51\,S2+0.59\,R3$) for separating NLAGNs from star-forming galaxies.
Here we combine the $P_1$ parameter and the low-ionisation emission line region (LINER) classification proposed by \citet{2010MNRAS.403.1036C} to select our NLAGN sample.
We require the NLAGNs to have $P_1>-0.5$ and $\rm H\alpha$ equivalent width (EW) $> 6\rm$\AA \ without broad emission lines, as shown in the top right panel of Figure~\ref{fig:agnclass}.
While a lower H$\alpha$ EW cut of 3,\AA\ is sometimes adopted \citep[e.g.][]{2010MNRAS.403.1036C}, 
our analysis indicates that varying the limit between 3 and 6,\AA\ does not significantly alter the general properties of the NLAGN sample.
Both $P_1$ and $\rm H\alpha$ EW are measured from the nuclear 3\arcsec \ spectra to avoid contamination from the host galaxies.
This way we classify 402 NLAGNs and plot them in green throughout the paper.

IRAGNs are classified by a red W1-W2 colour in the WISE bands, because the hot dust in the AGN torus will emit strongly around the W2 wavelength \citep[e.g.][]{2010AJ....140.1868W}.
We adopt the colour criterion of W1-W2 $> 0.35$ (Vega magnitudes) to define our IRAGN sample, as shown in the bottom left panel of Figure~\ref{fig:agnclass}.
This criterion was proposed by \citet{2024ApJ...977..102P} for MaNGA galaxies.
The 221 IRAGNs are shown in orange throughout the paper.

In normal star-forming galaxies without AGN activity, the radio luminosity is tightly correlated with the SFR \citep[e.g.][]{2022A&A...664A..83H,2024MNRAS.531..977D}.
The synchrotron emission from radio jets can produce excess radio luminosity relative to that expected from star formation, and can thus be used to identify RDAGNs \citep[e.g.][]{2025MNRAS.539.1856H}.
This approach is facilitated by the enhanced sensitivity of LoTSS over legacy surveys and has been demonstrated to be effective for classifying RDAGNs and estimating their jet energetics \citep[e.g.][]{2021MNRAS.506.5888M,2024MNRAS.529.3939Y}.
We require the RDAGNs to have a radio luminosity ($L_{\rm 144MHz}$) that is at least 0.7 dex higher than the expected value from the radio-SFR relation in \citet{2025A&A...694A.309J}.
This criterion has proven effective in selecting radio AGNs in MaNGA galaxies due to accurate SFR measurements based on the IFU data \citep{2025A&A...694A.309J}, as well as in other photometric samples \citep[e.g.][]{2023MNRAS.523.1729B}.
The RDAGN subsample contains 1354 galaxies, which are shown in red throughout the paper.

\begin{figure}
	\includegraphics[width=\columnwidth]{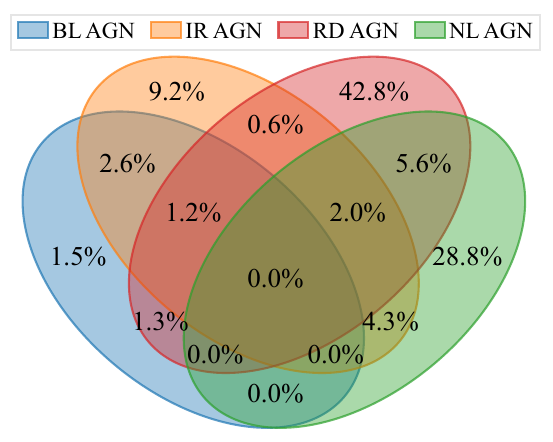}
    \caption{The Venn diagram showing the overlap between different AGN populations.
    IRAGNs, BLAGNs, NLAGNs, and RDAGNs are shown in orange, blue, green, and red, respectively.
    The fraction listed in each section is the percentage of the AGNs in that section relative to the total AGN sample, and is corrected by the volume weights. RDAGNs are the most populous AGN population, and most of them are not classified as other AGN types, while BLAGNs are relatively rare and most of them are also classified as other AGN types. Four regions show zero occupancy, as NLAGNs and BLAGNs are mutually exclusive.
    }
    \label{fig:venn}
\end{figure}

In total, we identify 1813 individual AGNs in our parent sample using the above four methods.
Figure~\ref{fig:agnclass} illustrates how the
different classes of AGNs are selected.
AGN populations are not mutually exclusive (except for BLAGNs and NLAGNs).
The Venn diagram in Figure~\ref{fig:venn} shows the overlap between different AGN populations.
RDAGNs are the most populous AGN population, and 88\% of them are not classified as other AGN types.
BLAGNs are relatively rare, and 81\% of them are found to have radio excess or torus emission.

\subsection{AGN luminosity at different wavelengths}
\label{sec:lagn}

\begin{figure*}
	\includegraphics[width=1.95\columnwidth]{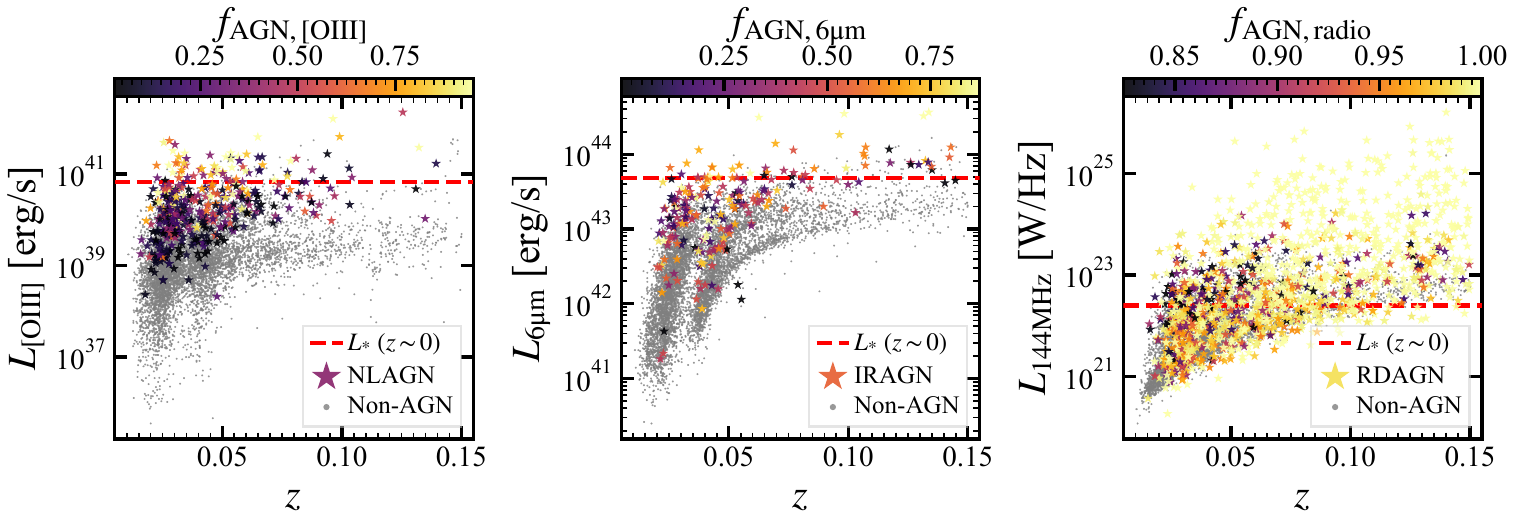}
    \caption{The luminosity of different AGN tracers across the MaNGA redshift range. Corresponding AGNs are shown as stars and colour-coded by their AGN fraction in each tracer. Non-AGNs are shown as grey dots. 
    We also plot the typical luminosity of the local luminosity functions in red dashed lines, which are derived from \citet{2016MNRAS.461.1076C} ([\ion{O}{III}]), \citet{2007ApJ...664..840H} (8$\mu$m, converted to 6$\mu$m), and \citet{2023MNRAS.523.6082C} (150\,MHz, converted to 144\, MHz), respectively.
    The AGN sample includes emission line AGNs with $L_{\rm [OIII]}$ larger than $10^{39}\rm \, erg\,s^{-1}$, IRAGNs with $L_{\rm 6\mu m}$ larger than $10^{42}\rm \, erg\,s^{-1}$, and RDAGNs with $L_{\rm RDAGN}$ larger than $10^{21}\rm \, W\,Hz^{-1}$.
    }
    \label{fig:lagn}
\end{figure*} 

The AGN classification methods described in Section~\ref{sec:agnclass} are designed to distinguish AGNs from normal galaxies.
The power of the AGN emission can vary by several orders of magnitude.
In the low-luminosity regime, AGN features may be overwhelmed by host galaxy emission and thus missed by the classification methods.

In this section, we estimate the power of the BLR, NLR, dust torus, and radio jets for those AGNs bright enough to be separated from the host, and we provide upper limits for other galaxies.

The broad Balmer lines can only be created by the dense gas close to and ionised by the accreting black hole; thus, the luminosity of the broad $\rm H\alpha$ can represent the power of the BLR.
This luminosity can be fitted using the {\tt ppxf} software \citep{2017MNRAS.466..798C} by modelling the Balmer lines with narrow and broad kinematic components.

The AGN contribution to the narrow emission lines can be estimated from the $P_1$ value (see Section~\ref{sec:agnclass}) based on the analysis in \citet{2020MNRAS.499.5749J}.
The fraction $f_{\rm NLAGN}$ can be calculated as $f_{\rm NLAGN}=0.14{P_1}^2 + 0.96P_1 + 0.47$, subject to the constraint $0\le f_{\rm NLAGN}\le 0.99$.
We note that this fraction is also used to correct the $\rm H\alpha$-derived SFR throughout the paper.
[\ion{O}{III}]$\lambda 5008$ is a strong forbidden line which is known as a good tracer of the NLR luminosity. 
It also correlates with the AGN bolometric luminosity in quasars \citep[$L_{\rm bol}$, e.g.][]{2004ApJ...613..109H}.
We thus use the attenuation-corrected nuclear $L_{\rm [OIII]}$ multiplied by $f_{\rm NLAGN}$ to represent the NLR power.
We assume a 10\% uncertainty for this conversion.
For non-NLAGNs, we take the measured nuclear $L_{\rm [OIII]}$ as an upper limit.

For the dust torus of IRAGNs, we need to separate the AGN emission from the host galaxy light in the mid-infrared bands using the spectral energy distribution (SED) fitting technique.
We follow the approach in \citet{2024ApJ...977..102P} to fit the SED from UV to mid-infrared wavelengths using the {\tt kcorrect} software \citep[][ version 5.1.8]{2007AJ....133..734B}.
For MaNGA galaxies, \citet{2024ApJ...977..102P} created a set of templates including both the host galaxy and AGN SEDs.
These templates are constructed from the Flexible Stellar Population
Synthesis code \citep{2010ApJ...712..833C} with dust attenuation and re-emission,
and the {\tt CLUMPY} AGN torus models \citep{2008ApJ...685..160N}.
A total of 11 bands of photometry from FUV to W4, as described in Section~\ref{sec:data}, are used to fit the SED.
All galaxies have at least 7 reliable bands ($ugriz$, W1, and W2) with $\rm S/N > 3$.
We then use the best-fit AGN SED to calculate the AGN luminosity at 6$\rm \mu m$ ($L_{\rm 6\mu m}$), a well-known tracer of the AGN torus power \citep[e.g.][]{2015MNRAS.449.1422M}. 
We add an additional 10\% error to the SED-fitting results, according to the scatter found among different fitting methods or models \citep[e.g.][]{2015A&A...576A..10C}
For non-IRAGNs, we take the $L_{\rm 6\mu m}$ derived from the best-fit AGN component as an upper limit.

The radio jet luminosity is calculated from the excess 144\,MHz luminosity above that expected from the host's SFR, following the method of \citet{2025A&A...694A.309J}.
The jet luminosity at 144\,MHz ($L_{\rm RDAGN}$) is defined as: $L_{\rm RDAGN}=L_{\rm 144MHz}^{\rm observed} - L_{\rm 144MHz}^{\rm SFR}$,
where $L_{\rm 144MHz}^{\rm SFR}$ is calculated from the SFR using the radio-SFR relation \citep[Eq. 1 in][]{2025A&A...694A.309J} with a 0.23 dex intrinsic scatter \citep[e.g.][]{2023MNRAS.523.1729B}:
\begin{equation}\label{eq:l144-sfr}
{\rm log}(\frac{L^{\rm SFR}_{\rm 144MHz}}{\rm W\,Hz^{-1}})=1.16\times {\rm log}(\frac{\rm SFR}{\rm M_{\odot}\,yr^{-1}}) + 21.99
\end{equation}

Figure~\ref{fig:lagn} shows the observed $L_{\rm [OIII]}$, $L_{\rm 6\mu m}$, and $L_{\rm 144MHz}$ (or upper limits) as a function of redshift.
Corresponding AGNs are shown as stars and colour-coded by their AGN fraction in each tracer.
Our AGN sample includes NLAGNs with $L_{\rm [OIII]}$ from $10^{39}$ to $10^{42}\rm erg\,s^{-1}$, IRAGNs with $L_{\rm 6\mu m}$ from $10^{42}$ to $10^{45}\rm erg\,s^{-1}$, and RDAGNs with $L_{\rm 144MHz}$ from $10^{21}$ to $10^{26}\rm \ W\,Hz^{-1}$.
The typical luminosities ($L_{*}$) derived from the local luminosity functions are 
$\rm 10^{40.8}\,erg\,s^{-1}$ for $L_{\rm [OIII]}$ \citep{2016MNRAS.461.1076C}, 
$\rm 10^{43.7}\,erg\,s^{-1}$ for $L_{\rm 6\mu m}$ \citep{2007ApJ...664..840H}, 
and $\rm 10^{22.4}\,W\,Hz^{-1}$ for $L_{\rm 144MHz}$ \citep{2023MNRAS.523.6082C}.
Therefore, the detection thresholds in the optical, infrared, and radio bands extend below typical luminosities, allowing for the inclusion of low-luminosity AGNs whose luminosity is comparable to that of their host galaxies.

\section{Host properties of different AGN populations}
\label{sec:results}

\subsection{The incidence of AGN}
\label{sec:fagn}
\begin{figure*}
	\includegraphics[width=2\columnwidth]{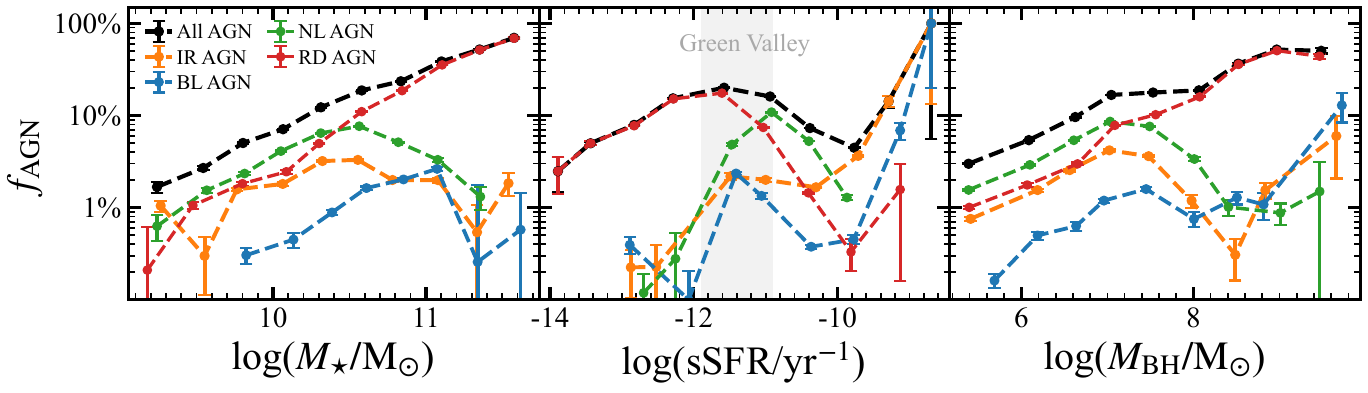}
    \caption{Fraction of galaxies hosting different AGN populations as a function of stellar mass, specific star formation rate (sSFR), and black hole mass. The fraction is already corrected by the volume weights. IRAGNs, BLAGNs, NLAGNs, and RDAGNs are shown in orange, blue, green, and red, respectively, and the total AGN fraction is shown in black. The AGN fraction increases with $M_{\star}$ and $M_{\rm BH}$, and is contributed by different AGN populations at different mass ranges.
    Along the sSFR axis, the AGN fraction shows a peak at the `green valley' region (the grey shaded area), and becomes dominant in the most strongly starbursting galaxies.}
    \label{fig:fagn}
\end{figure*}
Observed AGN incidence is related to triggering and the duty cycle of AGN activity \citep[e.g.,][]{2024A&A...686A..43I}, and is an important quantity to test the AGN feedback models \citep[e.g.,][]{2017MNRAS.465.3291W,2025A&A...697A.196I}.
In this section, we investigate the incidence of different AGN populations as a function of stellar mass ($M_{\star}$), specific star formation rate (sSFR), and black hole mass ($M_{\rm BH}$).
The AGN fraction is defined as the ratio between the sum of volume weights of AGNs and that of the parent sample in each parameter bin.
The results are shown in Figure~\ref{fig:fagn}, where the BLAGNs, NLAGNs, IRAGNs, and RDAGNs are shown in blue, green, orange, and red, respectively, and the total AGN fraction is shown in black.

In the left panel of Figure~\ref{fig:fagn}, the AGN fraction increases with $M_{\star}$, from $\sim 2\%$ at $10^{9.3}\rm \, M_{\odot}$ to $\sim 70$\% at $10^{11.6}\rm \, M_{\odot}$,
following an approximate power law ($f_{\rm AGN}\propto M_{\star}^{0.7}$).
This trend is mainly driven by the dominance of RDAGNs at the high mass end ($>10^{11}\rm \, M_{\odot}$) and NLAGNs at the low mass end ($<10^{10.5}\rm \, M_{\odot}$).

In the middle panel, the AGN fraction shows a peak at sSFR$\sim 10^{-11.5}\rm yr^{-1}$, between the star formation main sequence (SFMS, sSFR$\sim 10^{-10}\rm \, yr^{-1}$) and the quiescent population (sSFR$<10^{-12}\rm \, yr^{-1}$).
This region is often referred to as the `green valley', which is thought to be a transition region where galaxies are rapidly quenching their star formation \citep[e.g.,][]{2009ARA&A..47..159B}.
This peak supports the scenario that AGNs may play a role in the quenching process \citep[e.g.,][]{2014MNRAS.440..889S,2014ARA&A..52..589H}. 
We note that the overall AGN fraction is still low ($\sim 20$\%) in the green valley, indicating that the AGN duty cycle could be shorter than the quenching timescale, and/or other mechanisms are also at play in the quenching process, such as ram-pressure stripping \citep[e.g.][]{2023MNRAS.525.5359M}.
The peak in AGN content in the green valley is contributed by two populations:
(1) NLAGNs reach maximum fraction and contribute most at sSFR$\sim 10^{-10.8}\rm \, yr^{-1}$, (2) RDAGNs become dominant at sSFR$< 10^{-11.4}\rm \, yr^{-1}$. 
This supports a hypothesis that the structures of AGNs may transition during the quenching of their host galaxies.

Another feature is that the most strongly starbursting galaxies (sSFR$>10^{-9}\rm yr^{-1}$) are almost all hosting IRAGNs/BLAGNs,
indicating a common triggering mechanism \citep[e.g., gas inflow and galaxy mergers, ][]{2008ApJS..175..356H} for both starburst and high accretion rate AGNs.
This is also consistent with the AGN-starburst connection found in previous studies on luminous quasars \citep[e.g.,][]{2021ApJ...906...38Z}, and IRAGN samples \citep[e.g.,][]{2018MNRAS.478.4238D} which suggested that luminous IRAGNs could be the earlier phases of AGN evolution than the optical or radio AGNs.

The right panel shows the AGN fraction as a function of $M_{\rm BH}$. 
The AGN fraction also increases with $M_{\rm BH}$.
This trend is similar to that with $M_{\star}$ but with a shallower slope, which is also observed in the LoTSS radio AGN population \citep{2019A&A...622A..17S}.
Lower mass black holes ($<10^{7}\rm M_{\odot}$) are more likely to be NLAGNs,
while higher mass black holes ($>10^{7.5}\rm M_{\odot}$) are more likely to be RDAGNs.
The trends of different AGN fraction as a function of $M_{\rm BH}$ indicate a possible critical black hole mass around $\rm 10^{7.5}M_{\odot}$, 
where the AGN population sees a transition from IR/BL/NLAGNs to RDAGNs.

\begin{figure*}
	\includegraphics[width=1.7\columnwidth]{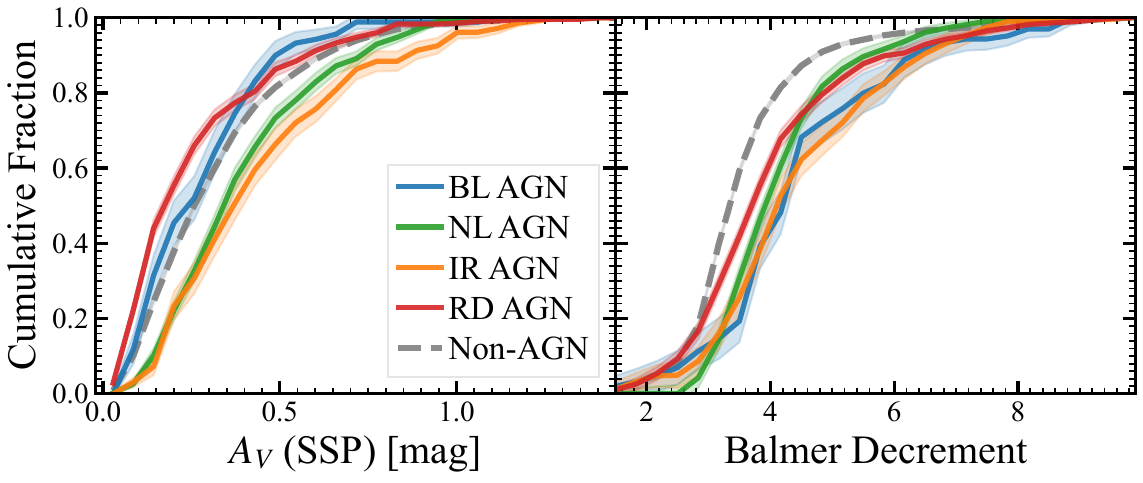}
    \caption{Cumulative distribution functions of the nuclear $A_V$ from stellar population synthesis and Balmer decrement ($\rm H\alpha/H\beta$) for different AGN populations. Volume weights are applied. The distributions are shown for IRAGNs (orange), BLAGNs (blue), NLAGNs (green), and RDAGNs (red). The non-AGN population is shown in grey dashed lines. The $A_V$ from stellar population synthesis shows that IRAGNs and NLAGNs have higher $A_V$ values than normal galaxies, while BLAGNs and RDAGNs have slightly lower $A_V$ values (K-S test, $p<0.01$). The Balmer decrement shows that all AGN populations have larger $\rm H\alpha/H\beta$ values than normal galaxies, likely due to higher electron density and temperature of the gas clouds around the SMBHs.}
    \label{fig:av}
\end{figure*}

To examine whether different AGN populations are the result of differing dust obscuration of the central source, 
we plot the cumulative distribution function (CDF) of the nuclear $A_V$ derived  from stellar population synthesis modelling of the SED and the  Balmer decrement ($\rm H\alpha/H\beta$) for different AGN populations in Figure~\ref{fig:av}.
The CDFs are built accounting for the volume weights, and the errors are estimated using the bootstrap resampling method.
IRAGNs, BLAGNs, NLAGNs, and RDAGNs are shown in orange, blue, green, and red, respectively, and the non-AGN population is shown in grey dashed lines.

In the left panel, we see that IRAGNs and NLAGNs have the higher $A_V$ values than the normal galaxies, while BLAGNs and RDAGNs have slightly lower $A_V$ values.
This is also supported by the Kolmogorov-Smirnov (K-S) test result ($p<0.01$).
This is consistent with the scenario that IRAGNs and NLAGNs are more likely to be obscured AGNs, while BLAGNs are unobscured \citep[e.g.,][]{2015ARA&A..53..365N}.
The low $A_V$ of RDAGNs is likely because their host galaxies are mostly quiescent galaxies lacking cold gas and dust (see the section below).

In the right panel, all AGN populations show larger $\rm H\alpha/H\beta$ values than the normal galaxies.
This is not necessarily caused by dust attenuation, but may instead reflect the different physical conditions of the gas clouds around the SMBHs.
For example, the gas in the narrow or broad line regions may have higher electron density and temperature,
thus the theoretical $\rm H\alpha / H\beta$ ratio can be higher than the standard Case B recombination value of 2.86 \citep[e.g.,][]{2008MNRAS.383..581D,2016MNRAS.461.4227H}.
Among different AGN populations, RDAGNs show slightly lower $\rm H\alpha/H\beta$ values, consistent with their low $A_V$ values. 
The other AGN populations all show similar $\rm H\alpha/H\beta$ distributions. 
This indicates that the Balmer decrement may be affected by both dust attenuation and the physical conditions of the gas clouds in AGNs, and should be used with caution when estimating the dust attenuation in AGN systems.

\subsection{Global star formation of AGN hosts}
\label{sec:global}

\begin{figure*}
	\includegraphics[width=2\columnwidth]{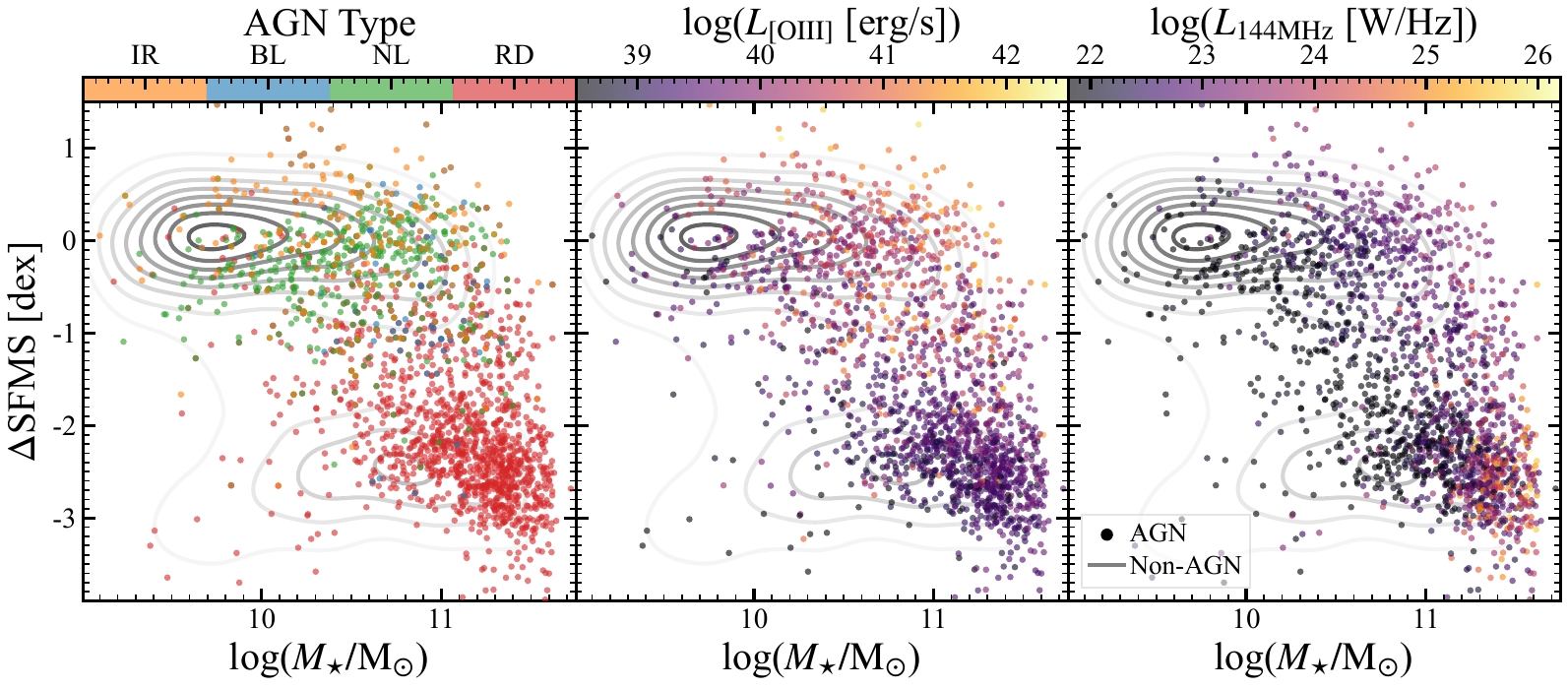}
    \caption{The distance to the star formation main sequence ($\Delta {\rm SFMS}$) versus stellar mass ($M_{\star}$) for different AGN populations as well as normal galaxies in our parent sample. The grey contours in each panel show the distribution of normal galaxies. AGN hosts are shown in coloured stars. The panels, from left to right, show AGN hosts coloured by their AGN types, nuclear $L_{\rm [OIII]}$, and $L_{\rm 144MHz}$, respectively.}
    \label{fig:sfms}
\end{figure*}

Star-forming galaxies follow a tight correlation in the SFR-$M_{\star}$ plane, known as the star formation main sequence (SFMS) \citep[e.g.,][]{2007ApJ...660L..43N},
while quiescent galaxies lie below the SFMS with much lower SFRs.
This bimodal distribution indicates that galaxies may experience a rapid quenching process during their evolution \citep[e.g.,][]{2009ARA&A..47..159B}.
AGNs are often proposed to play an important role in the quenching process \citep[e.g.,][]{2014MNRAS.440..889S,2014ARA&A..52..589H};
thus, it is interesting to see where different AGN populations lie in the SFR-$M_{\star}$ plane.
The distance from the SFMS can be defined as $\Delta {\rm SFMS} = \log({\rm SFR} / {\rm SFR}_{\rm SFMS}(M_{\star}))$,
where a more positive $\Delta {\rm SFMS}$ indicates a starbursting phase, while a more negative $\Delta {\rm SFMS}$ indicates a more quiescent phase.
Here, the SFMS fitted to MaNGA star-forming galaxies can be expressed as: $\log({\rm SFR_{SFMS}})=0.84 \log (M_{\star}/\rm M_{\odot}) - 8.59$.

Figure~\ref{fig:sfms} shows the $\Delta {\rm SFMS}$ versus $M_{\star}$ for different AGN populations as well as the normal galaxies in our parent sample.
The grey contours in each panel show the distribution of normal galaxies.
AGN hosts are shown in coloured stars.
The panels, from left to right, show AGN hosts coloured by their AGN types, nuclear $L_{\rm [OIII]}$, and $L_{\rm 144MHz}$, respectively.

In the left panel,  we see that as we go from  from lower-mass starburst hosts (upper left) to higher-mass quiescent hosts (lower right), the dominant AGN type shows a transition from IRAGN to NLAGN/BLAGN, and then to RDAGN.
Despite the different AGN selection methods between this work and \citet{2020ApJ...901..159C}, our results about the global star formation in AGN hosts are consistent with each other.
The global star formation conditions of different AGN hosts support the scenario that radiative-mode AGNs (such as IRAGNs, NLAGNs, and BLAGNs) prefer star-forming galaxies, while kinetic-mode AGNs (such as RDAGNs) prefer quiescent galaxies \citep[e.g.,][]{2014ARA&A..52..589H}.

$L_{\rm [OIII]}$ and $L_{\rm 144MHz}$ are roughly correlated with the radiative and kinetic power of the AGNs, respectively (see Section~\ref{sec:lagn}).
The middle panel shows that AGNs with high $L_{\rm [OIII]}$ are mostly found among the star-forming and green valley galaxies, while low $L_{\rm [OIII]}$ AGNs are mostly found in quiescent galaxies.
This suggests that higher SMBH accretion rates tend to occur in systems with higher $\Delta {\rm SFMS}$, whereas powerful radio jets are preferentially found in more massive galaxies.

\subsection{Radial gradients of star formation and gas ionisation}
\label{sec:rp}
\begin{figure*}
    \includegraphics[width=2\columnwidth]{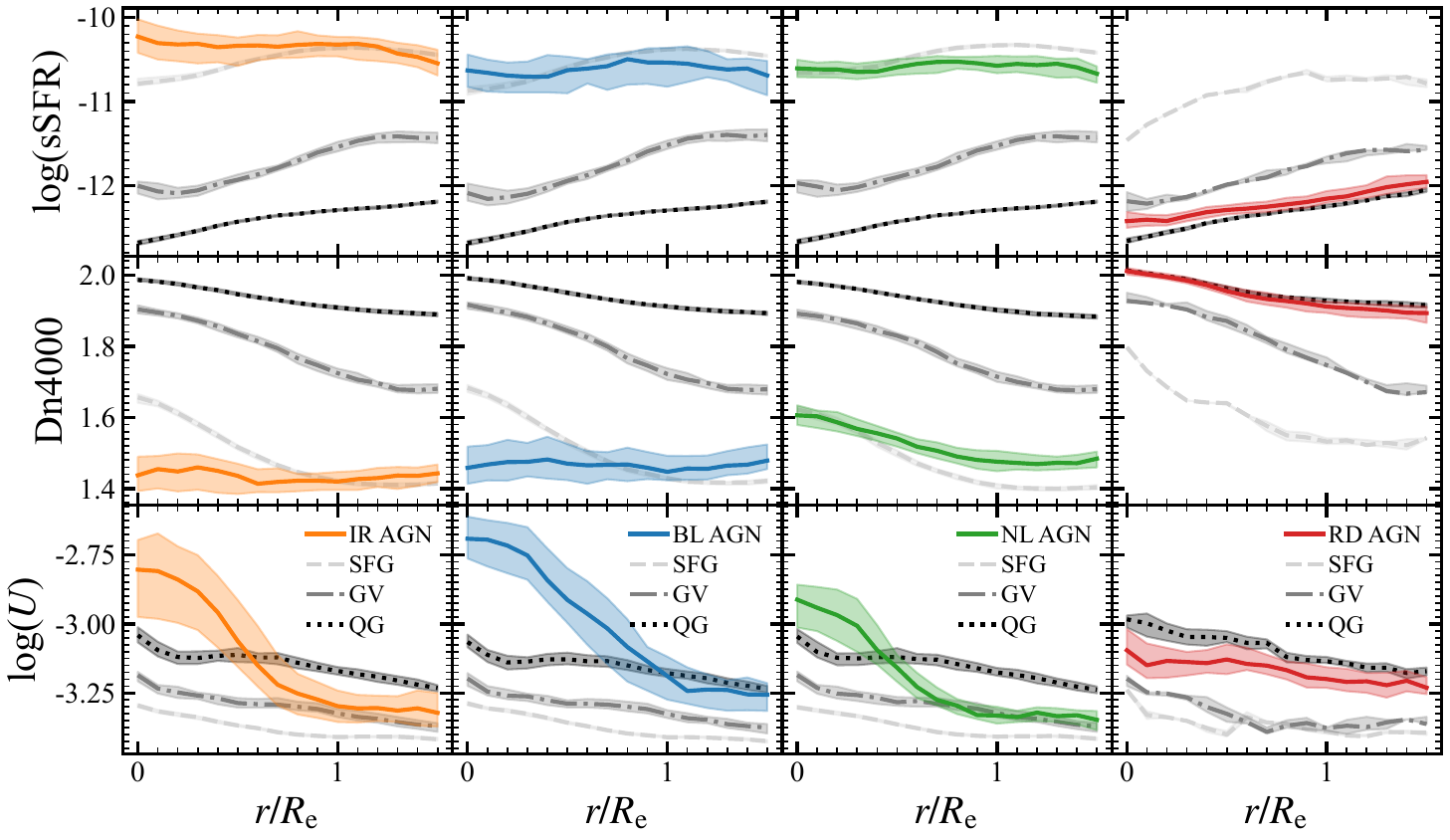}
    \caption{Radial profiles of sSFR, Dn4000, and the ionisation parameter for different AGN populations, compared to their mass-matched control samples.
    The control samples are divided into star-forming (light grey dashed lines), green valley (grey dash–dotted lines), and quiescent (black dotted lines) galaxies.
    The panels, from left to right, show IRAGNs, BLAGNs, NLAGNs, and RDAGNs, respectively, shown in orange, blue, green, and red lines.
    sSFR (in units of $\rm yr^{-1}$), Dn4000, and the ionisation parameter are shown from top to bottom panels and indicate the current star formation, average age of the stellar population (long-term star formation), and gas ionisation state, respectively.
    Shaded regions indicate the 95\% confidence intervals estimated through bootstrapping.
    }
    \label{fig:rp}
\end{figure*}

Galaxies are known to have radial gradients in their properties.
On average, the stellar mass is built up from the centre to the outskirts \citep[e.g.,][]{2016MNRAS.463.2799I}. 
This inside-out growth results in radial gradients of various properties which can be measured from the IFU data.
Here we focus on key properties such as the current star formation, the average age of the stellar population, and the gas ionisation state,
which may indicate the presence of AGN feedback.
We aim to understand how these properties are affected by AGN activity.

The star formation condition is traced by the specific star formation rate (sSFR, defined as $\rm H\alpha$-based SFR per unit stellar mass).
The average age of the stellar population, or long-timescale average star formation, is traced by the 4000\AA \ break strength \citep[][]{1999ApJ...527...54B}.
The ionisation parameter $U$, defined as the ratio between the ionising photon density and the hydrogen density, is estimated from the attenuation-corrected [\ion{O}{III}]/[\ion{O}{II}] ratio (O32), following the equation given by \citet{2002ApJS..142...35K} and assuming solar metallicity:
$\log(U)=-2.78+0.91\log \rm O32$.

These properties are known to strongly correlate with the stellar mass \citep[e.g.,][]{2022A&A...659A.160B};
thus, we compare the radial gradients in different AGN populations with their mass-matched control samples to minimize the mass effects.
For each AGN, we select non-AGN galaxies within $\Delta \log(M_{\star}/M_{\star}^{\rm AGN})<0.15$ as the control candidates.
These candidates are further median-stacked according to their volume weights as a control for the corresponding AGN.

We construct the radial profiles of these properties for each AGN and its control sample.
The radial distance is normalized by the effective radius ($R_{\rm e}$) of each galaxy to account for the size differences.
We limit the radial range to $r<1.5R_{\rm e}$, where most MaNGA galaxies have good coverage \citep{2015ApJ...798....7B}.

Figure~\ref{fig:rp} shows the median radial profiles of log(sSFR), Dn4000, and log($U$) for different AGN populations, compared to their mass-matched control samples.
The AGNs are shown in coloured lines (orange for IRAGNs, blue for BLAGNs, green for NLAGNs, and red for RDAGNs), which are built by stacking profiles of all AGNs in each population.
The non-AGN control samples are divided into star-forming (SFG, light grey dashed lines), green valley (GV, grey dash–dotted lines), and quiescent (QG, black dotted lines) galaxies according to their global $\Delta$SFMS.
The empirical dividing lines are $\Delta {\rm SFMS}=-1.8$ for QG and GV, and $-0.8$ for GV and SFG.
The shaded regions indicate the 95\% confidence intervals of each profile, estimated through bootstrapping on the subsamples.

In the top panels of Figure~\ref{fig:rp}, the sSFR profiles show flat gradients for IR, BL, and NL AGNs hosts, while RDAGN hosts show a decreasing sSFR trend towards the centre.
Compared to their  control galaxies, IRAGNs show sSFR profiles similar to SFGs in the outskirts ($r>R_{\rm e}$), but significantly enhanced sSFR towards the nuclear region ($r<0.5R_{\rm e}$). The ionization profile is also peaked towards the central region in IRAGNs.
This indicates that the SMBH accretion and nuclear star formation are both enhanced in these systems, consistent with a `positive AGN feedback' scenario in which a common gas inflow fuels both the AGNs and the nuclear starburst \citep[e.g.,][]{2008ApJS..175..356H}.
The profiles suggest that this AGN + starburst phase is mainly found in IRAGNs, with a $\sim 0.6$ dex nuclear sSFR enhancement, and this impact is only prominent within the effective radius (typically $\sim 5$ kpc).
This finding strongly supports our observation in Section 4.1. 
The centrally-enhanced star formation in IRAGN hosts explains their high global $\Delta$SFMS (starburst).
The BLAGN and NLAGN hosts also have flatter sSFR profiles than their SFG controls, indicating that they have a mild  or no nuclear sSFR enhancement.
The sSFR profiles of RDAGN hosts lie between those of the GV and QG controls, and show a decreasing trend towards the centre,  similar to that of their quiescent controls.

The Dn4000 profiles in the middle panels are broadly consistent with the sSFR profiles, with lower Dn4000 corresponding to higher sSFR.
IRAGN and BLAGN hosts show significantly lower Dn4000 than their SFG controls in the central region, indicating the existence of younger stellar populations within the effective radius.
NLAGN hosts show Dn4000 profiles similar to their SFG controls, with a mildly older stellar population in the outskirts.
RDAGN hosts show Dn4000 profiles similar to their QG controls, indicating that they have stopped forming stars for a long period.

In the bottom panels, AGN hosts show a clear  trend of increasing ionisation parameter $U$ towards the centre.
From BLAGNs, IRAGNs, NLAGNs to RDAGNs, the central $U$ decreases systematically, indicating a decreasing level of gas ionisation.
Compared to their control galaxies, IRAGNs, BLAGNs, and NLAGNs show significantly higher $U$  in their nuclear regions.
The trend of $U$   with radius is steeper within $\sim R_{\rm e}$, and then  flattens out at larger radii,  where the $U$ values become similar to their controls.
Our results indicate  that the AGN radiation can affect the gas ionisation on kpc scales.
The RDAGN hosts show $U$ profiles similar to their quiescent controls, 
indicating that most RDAGNs have low accretion rates and that low luminosity jets do not significantly affect the gas ionisation of their hosts.

The radial profiles presented here characterise the typical behaviour of the subsamples due to the stacking analysis. We caution that this statistical approach averages over individual variations, and specific sources may possess unique structural features that deviate from the stacked median.

\begin{figure*}
    \includegraphics[width=2\columnwidth]{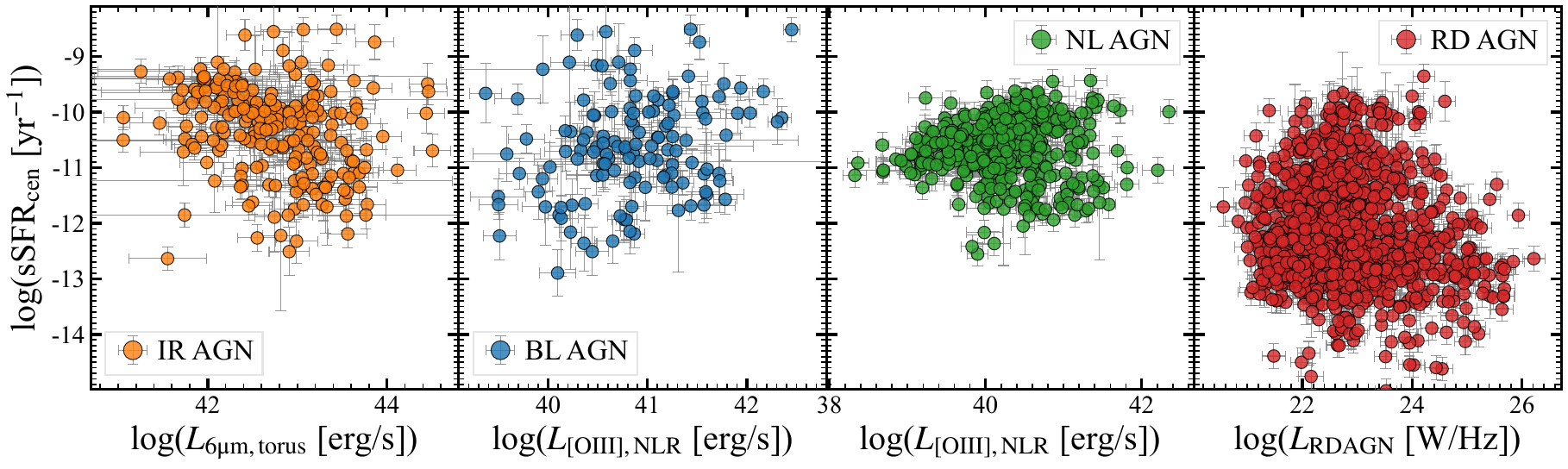}
    \caption{The central sSFR (within 0.5 $R_{\rm e}$) versus AGN luminosities at different wavelengths.
    From left to right, the panels show the central sSFR versus $L_{\rm 6\mu m}$, $L_{\rm [OIII]}$, $L_{\rm [OIII]}$, and $L_{\rm RDAGN}$ for IRAGNs, BLAGNs, NLAGNs, and RDAGNs, respectively.
    AGN populations show large scatters in their central sSFR, and the low Spearman coefficients indicate no significant correlations between the central sSFR and any of the AGN luminosities.
    }
    \label{fig:ssfr_lagn}
\end{figure*}

Instead of the types of the AGNs, we also investigate how the power of AGNs at different wavelengths affects their host star formation.
\citet{2024ApJ...977..194S}, using a MaNGA radio AGN sample, has reported that there is no correlation between sSFR and the radio AGN activities.
Since the AGNs tend to affect the innermost region of their host galaxies, 
we further compare the central (within 0.5 $R_{\rm e}$) sSFR with AGN luminosities at different wavelengths defined in Section~\ref{sec:lagn}.
Figure~\ref{fig:ssfr_lagn} shows the central sSFR versus $L_{\rm 6\mu m}$, $L_{\rm [OIII]}$, and $L_{\rm RDAGN}$ for IRAGNs, BLAGNs, NLAGNs, and RDAGNs.
AGN populations show large scatters in their central sSFR, 
and we do not find significant correlations between the central sSFR and any of the AGN luminosities.
The Spearman correlation coefficients are all within $\pm 0.3$,
with only marginal positive correlation between the central sSFR and $L_{\rm [OIII]}$ for BLAGNs (coefficient = 0.2, p-value = 0.013).
We also check the correlation between the outskirts sSFR and the $L_{\rm [OIII]}$ for BLAGNs, 
and the weak correlation found for the nuclear region disappears (coefficient = 0.01, p-value = 0.95).
The lack of correlation suggests that the AGN host galaxies can have a wide range of central star formation conditions, and most low-to-moderate luminosity AGNs may not have a immediate strong impact on the central star formation of their hosts, or vice versa.
Since the duty cycle of AGN activity is typically shorter than the quenching timescale \citep[e.g.,][]{2015MNRAS.451.2517S},
however, the feedback on star formation may still be effective on longer timescales with cumulative AGN activities, as seen in simulations \citep[e.g.,][]{2017MNRAS.472..949B}.

\subsection{Ionized outflows traced by [\ion{O}{III}]}
\label{sec:outflow}

\begin{figure}
	\includegraphics[width=\columnwidth]{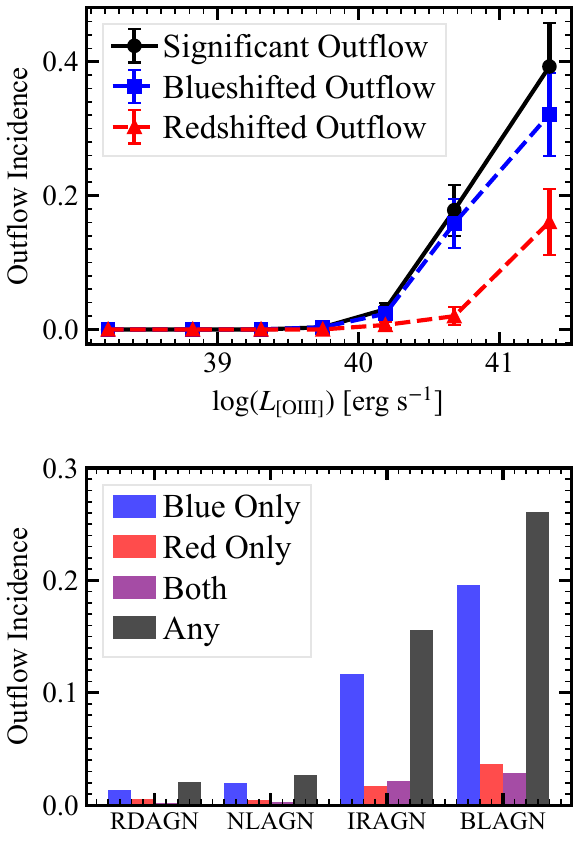}
    \caption{Upper panel: The incidence of ionised outflows as a function of $L_{\rm [OIII]}$. 
    Blue and red lines show the incidence of blueshifted and redshifted outflows, respectively, while the black line shows the total outflow incidence.
    The outflow start to appear at $L_{\rm [OIII]} \sim 10^{40}\rm \ erg\, s^{-1}$, and the incidence clearly increases with $L_{\rm [OIII]}$. 
    Lower panel: The outflow incidence in different AGN populations. IRAGNs and BLAGNs show the high outflow incidence ($\sim 16\%$ and $\sim 26\%$), NLAGNs ($\sim 4\%$) and RDAGNs ($\sim 2\%$) rarely have outflows.
    }
    \label{fig:foutflow}
\end{figure}

Ionised outflows are often observed in all types of AGN host galaxies,
and is considered as an important feedback mechanism to ionise and expel the interstellar medium \citep[e.g.,][]{2011MNRAS.415.2359N,2013MNRAS.433..622M,2016ApJ...817..108W,2025MNRAS.536.1166E}.
Several MaNGA studies have reported the detection of ionised outflows in different AGN host galaxies through their [\ion{O}{III}] properties \citep[e.g.,][]{2020MNRAS.492.4680W,2024A&A...691A.124A,2025A&A...698A..99K}.
Here we systematically investigate the ionised outflow features in different AGN populations in our sample.

We use the {\tt ppxf} software \citep{2017MNRAS.466..798C} to fit the spectra of all AGN host galaxies in our sample.
Following the standard procedure, we allow a broad line kinematic component to fit the Balmer lines and a narrow line component to fit all emission lines.
To identify the outflow features, we focus on the [\ion{O}{III}]$\lambda 5008$ line, which is a strong forbidden line and a good tracer of the ionised outflow.
In addition to the normal narrow line component, we allow a blueshifted and a redshifted kinematic component to fit the [\ion{O}{III}] doublets, serving as a proxy for the outflow.
The ratio of the [\ion{O}{III}] doublets ([\ion{O}{III}]$\lambda 5008$/[\ion{O}{III}]$\lambda 4959$) is fixed to the theoretical value of 2.98 \citep{2006agna.book.....O}.
We set several criteria for the outflow components to ensure their robustness.
Firstly, the redshifted and blueshifted components must have central velocities at least $200\rm \, km\, s^{-1}$ different from that of the narrow line component.
Secondly, we require that the velocity dispersion of these components must be at least $200\rm \, km\, s^{-1}$ (about $3\sigma$ of the spectral resolution).
Finally, the flux of these components must be detected with an amplitude-to-noise ratio greater than 10.
This approach allows us to also account for asymmetric and non-Gaussian features characteristic of outflows, especially when the emission lines show both redshifted and blueshifted wings.

In total, we identify 51 outflow candidates from 1813 AGN host galaxies (2.8\%$\pm$0.4\%), with 43 showing blueshifted outflow components and 13 showing redshifted outflow components.
Blueshifted outflows are much more common than redshifted outflows, as expected that the dust attenuation of the receding outflowing gas on the far side of the galaxy.
We then investigate the dependence of outflow incidence on AGN luminosity and type, shown in Figure~\ref{fig:foutflow}.
The upper panel shows that the outflow starts to appear at $L_{\rm [OIII]} \sim 10^{40}\rm \, erg\, s^{-1}$, and the incidence increases with $L_{\rm [OIII]}$, from $\sim 3\% \pm 1\%$ at $10^{40.2}\rm \, erg\, s^{-1}$ to $\sim 39\% \pm 7\%$ at $10^{41.4}\rm \, erg\, s^{-1}$.
The overall outflow incidence and its dependence on $L_{\rm [OIII]}$ are consistent with the result in a infrared-selected AGN sample at $z\sim0.5$ \citep{2024ApJ...962..146O}.
The lower panel shows the outflow incidence in different AGN populations.
IRAGNs and BLAGNs show high outflow incidence ($\sim 16\% \pm 3\%$ and $\sim 26\% \pm 5\%$),
while NLAGNs ($\sim 3\% \pm 1\%$) and RDAGNs ($\sim 2\% \pm 1\%$) rarely have outflows.

\begin{figure*}
	\includegraphics[width=2\columnwidth]{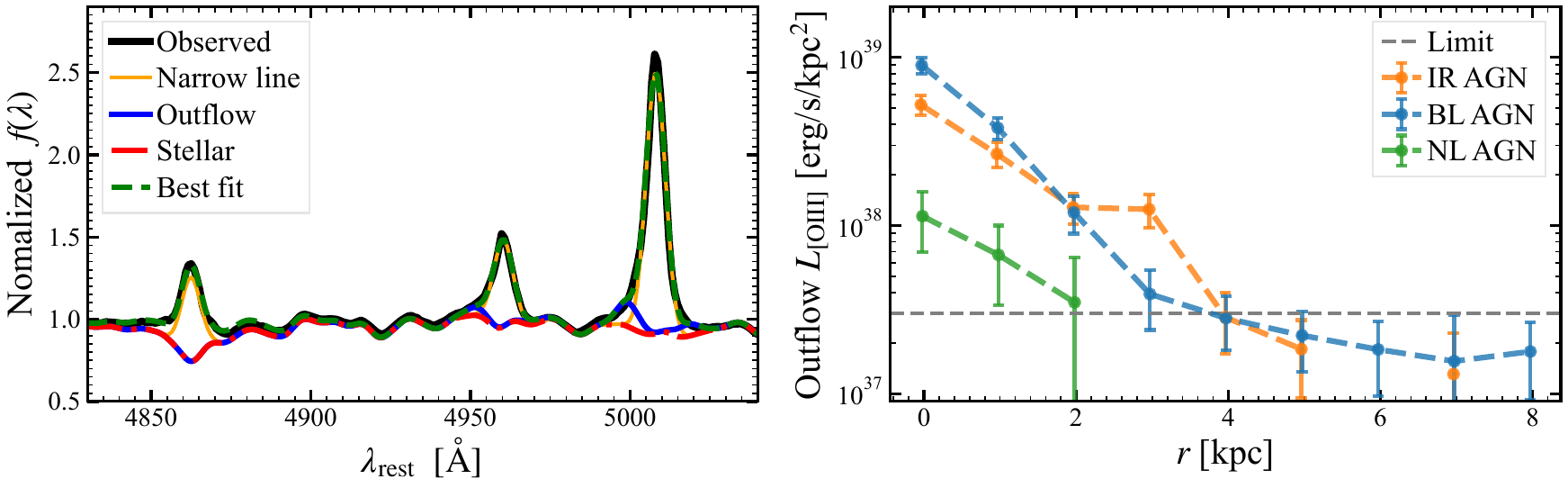}
    \caption{Left panel: An example of the outflow fitting for the [\ion{O}{III}] line using {\tt ppxf}. The observed spectrum (black line), which is the median stack of the nuclear spectra from all IRAGNs, is shown. Different components of the best-fit model are shown in different colours. The outflow component (blue line) is clearly identified. Right panel: The radial profiles of the outflow power in different AGN populations. IRAGNs and BLAGNs show the strongest outflow, especially in the inner 2 kpc region of the host galaxies. NLAGNs have weaker outflow in the nuclear region, but the outflow level at $r>3$kpc is similar to those of BLAGNs and IRAGNs. RDAGNs show negligible outflow features.
    }
    \label{fig:outflow}
\end{figure*}

With IFU data, the spatial distribution of outflows can be mapped.
Here we investigate the radial distribution of outflow features in different AGN populations.
To achieve higher S/N, we stack the spectra of AGN host galaxies in different radial bins,
and then stack them for each subsample of different AGN types.
The left panel of Figure~\ref{fig:outflow} shows an example fit of the stacked nuclear spectra of all BLAGNs,
where the outflow component (blue line) is clearly identified.

The right panel of Figure~\ref{fig:outflow} shows the radial profiles of the median outflow power in different AGN populations.
The stacking and fitting procedures allow us to detect weak outflow features down to a surface brightness level of $\sim 3\times 10^{37}\ \rm erg\, s^{-1}\, kpc^{-2}$.
The outflow power is traced by the surface brightness of the [\ion{O}{III}] outflow component (blueshifted + redshifted).
IRAGNs and BLAGNs show the strongest outflow, up to $\sim 10^{39}\ \rm erg\, s^{-1}\, kpc^{-2}$, especially in the inner 2 kpc region of the host galaxies.
NLAGNs have weaker outflow in the nuclear region ($\sim 10^{38}\ \rm erg\, s^{-1}\, kpc^{-2}$), and the outflow region is also smaller ($\sim 1\, $kpc) than that of BLAGNs and IRAGNs.
RDAGNs do not show significant outflow features at any radius, thus are not plotted in the figure.
These results suggest that ionised outflows are commonly found in IRAGN, BLAGN and NLAGN hosts, and can extend to several kpc scales,
suggesting they can be an important channel of AGN feedback and affect the host galaxy on a global scale.
RDAGNs, on average, show no outflow features and may have different feedback mechanisms other than ionised outflows.
The different incidences and strengths of outflow features between RDAGNs and other types of AGNs also suggest that these [\ion{O}{III}] outflows are mainly driven by radiative winds rather than radio jets.

\section{Discussion}
\label{sec:discussion}

\subsection{An evolutionary sequence of AGNs and host galaxies}
\label{sec:sequence}

Our results in Section~\ref{sec:results} indicate a possible evolutionary sequence among different AGN populations and their host galaxies.
In this sequence, AGNs with strong emission lines or a dusty torus represent an early phase where both the SMBHs and the host galaxies are rapidly growing through strong accretion and nuclear starbursts, respectively,
while most AGNs with jets represent a later phase where the SMBHs are accreting inefficiently and the host galaxies are quenched.

The bimodal distribution of AGN host galaxies in the SFR/$M_*$
versus $M_*$ plane raises a natural question: is galaxy quenching related to the transition of AGN types?
The overlap sample between different AGN types could
be a test of this transition hypothesis, especially those showing both optical/IR and radio classifications.
To examine this, we select the AGNs with both RDAGN classification and IR/BL/NLAGN classification as a `mixed AGN' sample,
and compare them with AGNs with only RDAGN classification (hereafter `radio-only AGNs') and AGNs with only IR/BL/NLAGN classification (hereafter `Optical/IR-only AGNs').
These three populations make up 11\%, 46\%, and 43\% of the full AGN sample, respectively.

\begin{figure*}
	\includegraphics[width=\textwidth]{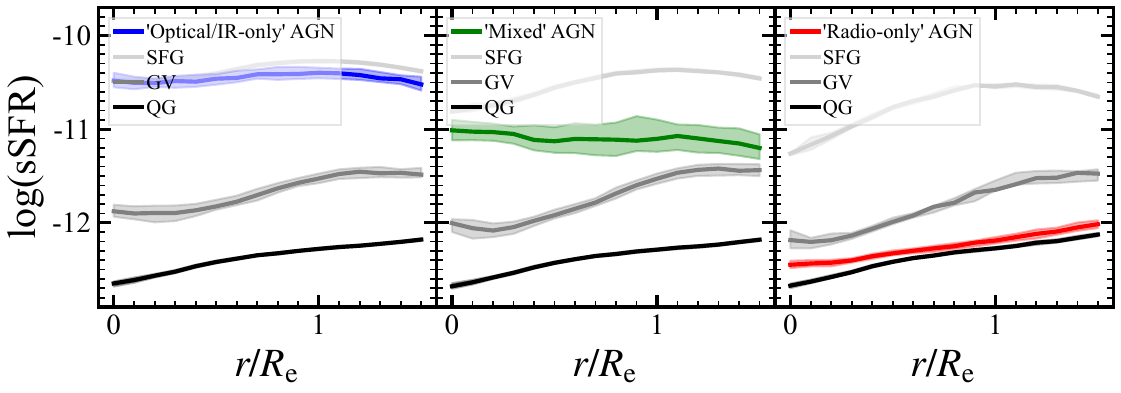}
    \caption{The median sSFR radial profile of the mixed AGNs (green), optical/IR-only AGNs (blue), and radio-only AGNs (red) populations, along with their mass-matched control samples of star-forming (light grey), green valley (grey), and quiescent galaxies (dark grey). The profiles are built following the same procedure as in Section~\ref{sec:rp}. The mixed AGNs show sSFR profiles between optical/IR-only AGNs and radio-only AGNs, and also between the main sequence and green valley control galaxies, which supports the scenario that the host galaxies of mixed AGNs are likely in a transition phase from star-forming to quiescent.}
    \label{fig:profilemix}
\end{figure*}

Figure~\ref{fig:profilemix} shows the median sSFR radial profiles of these three populations, and the comparisons with their stellar mass-matched control samples of normal galaxies.
The profiles are built following the same procedure as in Section~\ref{sec:rp}.
The mixed AGNs exhibit sSFR profiles that are clearly intermediate between those of optical/IR-only and radio-only AGNs, positioning them between the star-forming main sequence and the green valley control galaxies. This suggests that as AGNs evolve from IR/BL/NLAGNs to RDAGNs, their host galaxies are likely undergoing a transition from a star-forming to a quiescent phase.

The median dust attenuation values ($A_V$) measured from stellar population synthesis are 0.44, 0.34, and 0.26 for optical/IR-only AGNs, mixed AGNs, and radio-only AGNs, respectively.
This indicates that the AGNs become less obscured along the sequence.

\begin{figure}
	\includegraphics[width=\columnwidth]{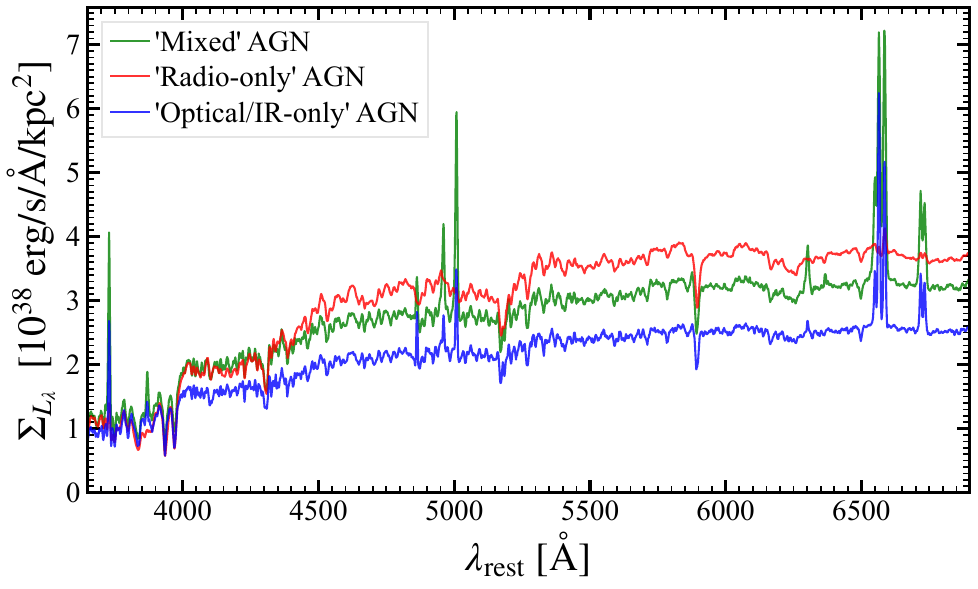}
    \caption{The stacked nuclear spectra of the surface luminosity density for different AGN subsamples, in units of $10^{38} \, \text{erg} \, \text{s}^{-1} \, \text{kpc}^{-2} \, \text{\AA}^{-1}$. Mixed AGNs, optical/IR-only AGNs, and radio-only AGNs are shown in green, blue, and red, respectively. The mixed AGNs show stronger emission lines (especially the high-ionisation lines like [\ion{O}{III}]) but a larger 4000\AA \ break strength than the optical/IR-only AGNs, supporting that the host galaxies of mixed AGNs are likely in a transition phase from star-forming to quiescent while maintaining strong SMBH accretion.}
    \label{fig:specmix}
\end{figure}

We further stack the nuclear spectra of these three populations to compare their spectral features, as shown in Figure~\ref{fig:specmix}.
The spectra are presented as the surface luminosity density (in units of $10^{38} \, \text{erg} \, \text{s}^{-1} \, \text{kpc}^{-2} \, \text{\AA}^{-1}$).
Mixed AGNs, optical/IR-only AGNs, and radio-only AGNs are shown in green, blue, and red, respectively.
The mixed AGNs show stronger emission lines (especially the high-ionisation lines like [\ion{O}{III}]) but a larger 4000\AA \ break strength than the optical/IR-only AGNs.
Considering that they also have a radio excess, the mixed AGNs could be a population where the host galaxies are starting to quench while the accreting SMBHs start to launch radio jets.

Smaller radio sizes are an additional  piece of evidence supporting 
our scenario that the jets in mixed AGNs are relatively young.
We have compared the sizes of the radio emission with those of the optical host galaxies for these three populations.
The relative size of the radio emission to the host galaxy can be measured by the ratio of the radio semi-major axis and the optical effective radius (hereafter referred to as $a/R_{\rm e}$).
The median (and mean) values of $a/R_{\rm e}$ are 1.0 (1.4), 1.0 (1.5), and 1.2 (2.9) for optical/IR-only AGNs, mixed AGNs, and radio-only AGNs, respectively.
We note that the resolution of LoTSS images is 6 arcsec, similar with the typical $r$-band effective radius (5.3 arcsec).
The radio emission in optical/IR-only AGNs therefore are mostly unresolved and does not extend much beyond the host galaxy, indicating that the radio emission in this AGN population is likely from star formation activity.
The mixed AGNs have radio excess by classification, but their radio size is still comparable to the host galaxy size.
This indicates that the radio jets in mixed AGNs are still compact and confined within the host galaxy.
In contrast, the radio-only AGNs have more extended radio emission than their host galaxies, indicating that the radio jets in this population are more developed and can extend beyond the host galaxy.
Another explanation for the small radio sizes in mixed AGNs is that the radio excess in these sources is emitted by radiatively driven winds instead of jets, such as those found in supernova remnants,
as proposed by \citet{2013MNRAS.436.2576L} and \citet{2014MNRAS.442..784Z}.

A similar evolutionary sequence has been proposed for quasar populations \citep{2019MNRAS.488.3109K},
in which young red quasars with dust torus classification are found to have strong winds and weak or compact jets,
while older blue quasars are found to have weaker winds and more powerful, extended jets.
Our results are consistent with this scenario and extend the sequence to the more numerous low-luminosity AGN populations.
In addition, we find that the spatially resolved stellar populations  of the AGN hosts are also evolving along this sequence, 
with younger AGNs (e.g., IR/BLAGNs) typically residing in actively star-forming galaxies with a nuclear starburst, while older AGNs (e.g., RDAGNs) are found in globally quiescent systems.

This evolutionary scenario also matches the classification of radiative-mode and kinetic-mode AGNs \citep[e.g.,][]{2014ARA&A..52..589H},
where the radiative-mode AGNs (IRAGNs, BLAGNs, and NLAGNs) represent the early phase with a high Eddington ratio and efficient accretion,
while the kinetic-mode AGNs (RDAGNs) represent the later phase with a low Eddington ratio and inefficient accretion.
In semi-analytic models of galaxy evolution, radiative-mode AGNs are usually fuelled by cold gas accretion, whereas kinetic-mode AGNs are usually fuelled by gas cooling from a hot-gas halo \citep[e.g.,][]{2000MNRAS.311..576K,2008MNRAS.391..481S}.
We check the AGNs in our sample that are observed by the \ion{H}{I}-MaNGA survey \citep{2019MNRAS.488.3396M} to see if there is a difference in their cold gas content.
Among the 1813 AGNs in our sample, 762 have \ion{H}{I} observations, and 310 are detected in \ion{H}{I}.
Taking into account the upper limits of the  non-detections,
we find that the median \ion{H}{I} gas fraction ($M_{\rm HI}/M_{\star}$) is around 26\%, 12\%, and 6\% for optical/IR-only AGNs, mixed AGNs, and radio-only AGNs, respectively (30\%, 19\%, and 15\% if we exclude the non-detections),
which demonstrates that the neutral gas content is decreasing along the AGN evolutionary sequence.
A more complete view of the fuelling mechanism for different AGN populations will require molecular gas images from CO observations and hot gas images from X-ray observations in future studies.

The depletion of cold gas supply could be one reason of this evolutionary sequence, which can explain that the accretion rate of AGNs and the star formation rate of their hosts are both decreasing along the sequence.
In addition, dramatic events such as major mergers may also be related to the different phenomena of the AGN types.
For example, luminous infrared AGNs are found to prefer merger systems \citep[e.g.][]{2017MNRAS.464.3882W}.
By cross-matching with merger classifications for MaNGA galaxies \citep{2021ApJ...923....6J}, we found that the fraction of interacting systems is higher in IRAGNs ($\sim$15\%) and BLAGNs ($\sim$12\%) than in NLAGNs ($\sim$7\%) and RDAGNs ($\sim$5\%).

We note that the quenching sequence will only apply in a population-averaged sense, 
because the duty cycle of AGN activity is typically shorter (on the order of millions of years) than the quenching timescale (on the order of Gyr) of galaxies \citep[e.g.,][]{2015MNRAS.451.2517S}. 
For example, a single galaxy may undergo multiple cycles of AGN activity during its growth and quenching process, and the type of AGN activity at different stages follows the proposed sequence.

\subsection{The growth of SMBHs and different host regions}
\label{sec:bhar}
\begin{figure*}
    \includegraphics[width=2\columnwidth]{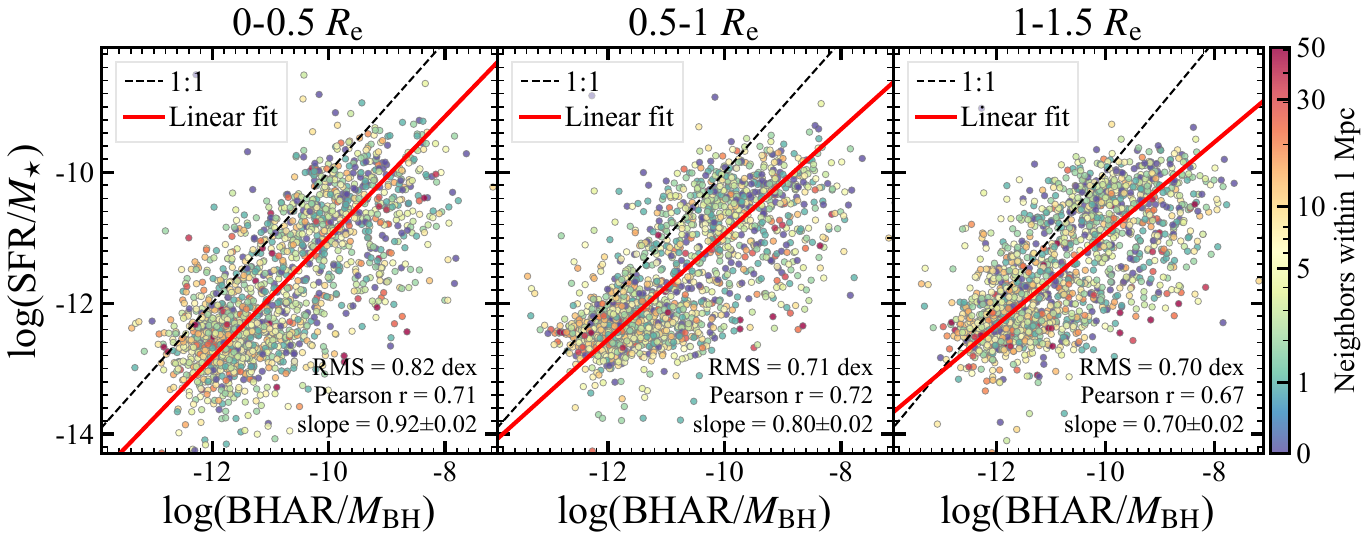}
    \caption{Comparison between the specific black hole accretion rate (sBHAR) and specific star formation rate (sSFR) in different radial bins.
    The panels, from left to right, show $r<0.5R_{\rm e}$, $0.5R_{\rm e}<r<R_{\rm e}$, and $R_{\rm e}<r<1.5R_{\rm e}$, respectively. The dashed black line indicates the one-to-one relation expected from the co-evolution scenario. The solid red lines indicate the best-fit linear relations. The slopes, Pearson correlation coefficients, and the root-mean-square errors of the fitting are listed in each panel. The galaxies are colour-coded by the number of their neighbours within 1 Mpc.
    Clear positive correlations between ${\rm BHAR}/M_{\rm BH}$ and ${\rm SFR}/M_{\star}$ are observed in all radial bins, with the slope decreasing from inner to outer bins.
    The numbers of neighbours within 1 Mpc \citep{2015A&A...578A.110A} are shown in different colours, but no clear dependence is observed.
    We note that typical errors on the x and y axes are about 0.6 dex and 0.3 dex, respectively.
    }
    \label{fig:sbhar}
\end{figure*}

It is well known that there are tight correlations between the mass of SMBHs and their host galaxies \citep[e.g.,][]{2003ApJ...589L..21M,2013ARA&A..51..511K}.
This suggests a co-evolution scenario where the growth of SMBHs and their host galaxies are globally linked.
However, the growth rates of different galactic regions can be different, as shown by the radial gradients of sSFR in normal galaxies (Figure~\ref{fig:rp}).
In this section, we investigate the relation between the SMBH growth and the growth of different galactic regions.

Assuming the $M_{\rm BH}$-$M_{\star}$ correlation holds at all times during co-evolution,
we expect the specific black hole accretion rate (sBHAR) to equal the specific star formation rate (sSFR), i.e., ${\rm BHAR}/M_{\rm BH} = {\rm SFR}/M_{\star}$.
Here we use the sum of the bolometric and kinetic luminosities to calculate the total black hole accretion rate:
\begin{equation}
{\rm BHAR}=\frac{(L_{\rm Bol}+L_{\rm Kin})}{\eta c^{2}},
\end{equation}
where $L_{\rm Bol}$ and $L_{\rm Kin}$ are defined in Section~\ref{sec:param}; $\eta$ is assumed to be 0.1; and $c$ is the speed of light.

In Figure~\ref{fig:sbhar}, we plot the comparisons between ${\rm BHAR}/M_{\rm BH}$ and ${\rm SFR}/M_{\star}$ in three radial bins, i.e., $r<0.5R_{\rm e}$, $0.5R_{\rm e}<r<R_{\rm e}$, and $R_{\rm e}<r<1.5R_{\rm e}$.
In each panel, the dashed black line indicates the one-to-one relation expected from equal growth rates, and the solid red line indicates the best-fit linear relation.
The linear fitting is performed in the logarithmic space using the {\tt LtsFit} package \citep{2013MNRAS.432.1709C}.
We assume a 0.6 dex intrinsic scatter in ${\rm BHAR}/M_{\rm BH}$ by combining the uncertainties on $M_{\rm BH}$ \citep{2013ApJ...764..184M}, $L_{\rm Bol}$ \citep{2010ApJ...720..786L}, and $L_{\rm Kin}$ \citep{2010ApJ...720.1066C}.
The typical uncertainties on ${\rm SFR}/M_{\star}$ are assumed to be 0.3 dex, according to the scatter among different measurements.
We note that the intercepts of the relations depend on several assumed parameters with large uncertainties, such as the bolometric correction for the [\ion{O}{III}] luminosity, the kinetic power-radio luminosity conversion factor, and accretion efficiency;
thus, we mainly focus on the slope of the relation between the ${\rm BHAR}/M_{\rm BH}$ and ${\rm SFR}/M_{\star}$.

From inner to outer radial bins, i.e., left to right panels, the best-fit slopes are 0.92, 0.80, and 0.70, respectively, with typical uncertainties of $\sim 0.02$.
The Pearson correlation coefficients are all around 0.7, indicating clear positive correlations between ${\rm BHAR}/M_{\rm BH}$ and ${\rm SFR}/M_{\star}$ in all radial bins.
The Spearman correlation coefficients are also around 0.7, indicating that the relationship is not only monotonic but primarily linear.
Considering that our sample covers wide ranges of sSFR and specific BHAR,
the linear correlations suggest that the SMBH growth and the galaxy growth are globally linked across their evolutionary stages.
This suggests that the fuel for the central SMBH is related to the galactic-scale star formation supply,
and the mechanism that quenches star formation may also suppress the SMBH accretion.

On the other hand, the slopes of the relations decrease from inner to outer radial bins, suggesting that this black hole accretion-star formation connection may vary with galactic radius.
This is expected, since SMBHs are located in the nuclear regions.
The decreasing slopes are mainly driven by the low accretion rate sources,
where the low-specific-BHAR AGN hosts still maintain relatively high sSFRs in the outskirts compared to their nuclear regions.
This may be the reason why the SMBH mass is found to correlate better with the bulge mass than the total stellar mass of the host galaxies \citep[e.g.,][]{2003ApJ...589L..21M}.
Alternatively, the decreasing slopes may be due to feedback from the SMBH that first affects the nuclear star formation and then gradually affects the outer regions.
The slopes are all less than unity, which is not consistent with the equal growth rates scenario.
One possible explanation for this discrepancy is that this AGN sample is biased towards high BHAR sources because they are more likely to be classified,
thus some of the low-sBHAR, high-sSFR sources are missing from the sample.

In the literature, the cosmic environment is often considered a key factor in the co-evolution of SMBHs, galaxies, and halos.
This environmental dependence is found in some observations and simulations \citep[e.g.,][]{2013MNRAS.436.2708M,2024MNRAS.528.4891G,2025MNRAS.536..777V}.
In  Figure~\ref{fig:sbhar}, we colour-code the AGNs by the number of their neighbours within 1 Mpc to indicate the local environment.
The number of neighbours is calculated using the SDSS spectroscopic catalogue within a velocity window of $\pm 1000\rm \, km\, s^{-1}$,
as described in \citet{2015A&A...578A.110A}.
We do not find a significant dependence of the ${\rm BHAR}/M_{\rm BH}-{\rm SFR}/M_{\star}$ relation on the number of neighbours,
although more reliable environmental indicators, beyond just the number of neighbours, are needed to perform a more detailed analysis.

\section{Conclusions}
\label{sec:conclusion}
In this work, we combine the latest LoTSS observations with the full MaNGA sample.
Together with the infrared photometry from the WISE all-sky survey,
we perform AGN classifications in optical, infrared, and radio wavelengths for this largest volume-limited IFU sample.
We identify 1813 AGNs classified as infrared AGNs, broad-line AGNs, narrow-line AGNs, and/or radio AGNs.
Besides the classification, we also estimate the luminosities or upper limits for different AGN structures: the $L_{\rm 6\mu m}$ of the dust torus based on SED fitting, broad-line region and narrow-line region luminosities from IFU spectral fitting, and jet luminosities from the excess radio luminosities.

With this large sample, we investigate the global and spatially resolved star formation in different AGN populations and compare them with their star-forming, green valley, and quiescent counterparts without AGN detections. The AGN incidence as a function of stellar
mass, specific star formation rate and black hole mass is presented and 
we also study the radial gradients of the gas ionisation and ionised outflows in different AGN host galaxies.
We discuss how the observational results can be explained by an evolutionary sequence of AGNs and their host galaxies
and investigate the connection between SMBH growth and stellar mass growth in  different galactic regions.

The main conclusions are summarized as follows:
\begin{itemize}
    \item The incidence of different AGN populations strongly depends on stellar mass, specific star formation rate, and black hole mass. The total AGN fraction increases with $M_{\star}$ following a power-law, with RDAGNs dominant at high $M_{\star}$ and NLAGNs at low $M_{\star}$.
    \item Along the sSFR axis, the AGN fraction shows a peak in the `green valley' region, and becomes dominant in the most strongly starbursting galaxies. This indicates that AGNs may have underlying connections to both the quenching process and the nuclear starburst activity.
    \item  The dust attenuation estimated from SED fitting using stellar population synthesis 
    models indicates that IRAGN and NLAGN hosts are more dusty than BLAGN and RDAGN hosts.
    In contrast, for Balmer decrements, there are no significant differences among different AGN populations, and AGN hosts generally have higher Balmer decrements than normal galaxies. This is consistent with the literature finding that gas clouds around SMBHs can have higher electron densities and/or temperatures.
    Thus, Balmer decrements may not be a good tracer of the dust attenuation in AGN hosts.
    \item The host galaxies of IRAGNs, BLAGNs, and NLAGNs are on average globally similar to star-forming galaxies, with a large scatter in SFR, while RDAGN hosts are similar to quiescent galaxies.
    \item Radial gradients of sSFR show that IRAGN hosts have significant nuclear sSFR enhancement within the effective radius compared to their
    non-AGN controls, consistent with a `positive AGN feedback' scenario. BLAGN and NLAGN hosts also show mild nuclear sSFR enhancement, while RDAGN hosts show decreasing sSFR towards the centre similar to their quiescent controls.
    \item Radial gradients of the gas ionisation parameter $U$ show that IRAGN, BLAGN, and NLAGN hosts have significantly higher $U$ in the nuclear region compared to their mass controls, indicating that the AGN radiation can affect the gas ionisation out to several kpc. RDAGN hosts show similar $U$ profiles to their quiescent controls.
    \item AGN hosts show a wide range of nuclear sSFR, and the nuclear sSFR is not strongly correlated with the AGN luminosities traced by $L_{\rm 6\mu m}$, $L_{\rm [OIII]}$, or $L_{\rm RDAGN}$.
    \item Ionised outflows traced by the [\ion{O}{III}] line are mainly found in IRAGN and BLAGN hosts, with blueshifted outflows being much more common than redshifted outflows. The outflow incidence increases with $L_{\rm [OIII]}$, from $\sim 3\%$ at $10^{40.2}\rm \, erg\, s^{-1}$ to $\sim 39\%$ at $10^{41.4}\rm \, erg\, s^{-1}$. 
    The outflow features on average extend to $\sim 2$ kpc scales for BLAGNs and IRAGNs. 
    The majority of RDAGNs do not show outflow features, indicating the observed [\ion{O}{III}] outflows are mainly driven radiatively rather than by jets.
    \item Clear positive linear correlations between ${\rm BHAR}/M_{\rm BH}$ and ${\rm SFR}/M_{\star}$ are observed in all radial bins, with the slope decreasing from inner to outer bins. The strong correlations suggest that the SMBH growth and the galaxy growth are globally linked across their evolutionary stages, possibly by the large-scale gas supply. The decreasing slopes with radius hint at underlying feedback from the SMBH, which first affects the nuclear star formation and then gradually affects the outer regions.
    \item We propose an evolutionary sequence for AGNs and their hosts from IR/BL/NLAGNs to RDAGNs, connected by a `mixed AGN' population (showing both BLR/NLR/torus and radio jet features). 
    These transitional AGNs exhibit intermediate properties: their sSFR profiles lie between star-forming and quiescent; the obscuration levels are intermediate; and their stacked spectra show both strong ionisation lines and a significant 4000\AA\ break.
    \item This evolutionary sequence is also supported by their radio morphology and atomic gas content. From IR/BL/NLAGNs to RDAGNs, the radio emission becomes more extended and the atomic gas fraction becomes lower, consistent with a scenario where AGNs evolve from the cold-gas-fuelled radiative mode to the jet-dominated kinetic mode as their host galaxies grow and the star formation ceases.
    This sequence could be driven by the depletion of cold gas supply and specific events such as major mergers, and confirmation of this scenario requires future high-resolution gas observations.
\end{itemize}

To connect the observational results to the underlying physical processes in the evolution of AGNs and their host galaxies,
we still need more detailed multi-wavelength observations of different AGN populations,
especially high-quality gas observations which can be compared with current galaxy evolution models and hydrodynamic simulations.
We expect future high-resolution observations of molecular gas at sub-millimetre wavelengths and hot gas distributions from X-ray observations,
to improve our understanding  the fuelling and feedback mechanisms during the AGN fuelling cycle.
This will help us to build a more complete picture of the secular evolution of AGNs and the co-evolution of SMBHs and their host galaxies, which can improve our understanding of the diverse observational features in different AGN populations.

\section*{Acknowledgements}

GJ would like to thank Michael Blanton, Bo Peng, Catarina Aydar, and Iker Millan Irigoyen for their helpful discussions on the development of this paper.
Part of this work was supported by the German
\emph{Deut\-sche For\-schungs\-ge\-mein\-schaft, DFG\/} project number Ts\ 17/2--1.
MJH thanks the UK STFC for support [ST/V000624/1, ST/Y001249/1].
Y.D. would like to thank the support from National Key R\&D Program of China (MOST) with grant
No. 2022YFA1605300, and the National Natural Science Foundation of China (NSFC, grant No. 12273051).
    
    Funding for the Sloan Digital Sky 
    Survey IV has been provided by the 
    Alfred P. Sloan Foundation, the U.S. 
    Department of Energy Office of 
    Science, and the Participating 
    Institutions. 
    
    SDSS-IV acknowledges support and 
    resources from the Center for High 
    Performance Computing  at the 
    University of Utah. The SDSS 
    website is www.sdss4.org.
    
    SDSS-IV is managed by the 
    Astrophysical Research Consortium 
    for the Participating Institutions 
    of the SDSS Collaboration including 
    the Brazilian Participation Group, 
    the Carnegie Institution for Science, 
    Carnegie Mellon University, Center for 
    Astrophysics | Harvard \& 
    Smithsonian, the Chilean Participation 
    Group, the French Participation Group, 
    Instituto de Astrof\'isica de 
    Canarias, The Johns Hopkins 
    University, Kavli Institute for the 
    Physics and Mathematics of the 
    Universe (IPMU) / University of 
    Tokyo, the Korean Participation Group, 
    Lawrence Berkeley National Laboratory, 
    Leibniz Institut f\"ur Astrophysik 
    Potsdam (AIP),  Max-Planck-Institut 
    f\"ur Astronomie (MPIA Heidelberg), 
    Max-Planck-Institut f\"ur 
    Astrophysik (MPA Garching), 
    Max-Planck-Institut f\"ur 
    Extraterrestrische Physik (MPE), 
    National Astronomical Observatories of 
    China, New Mexico State University, 
    New York University, University of 
    Notre Dame, Observat\'ario 
    Nacional / MCTI, The Ohio State 
    University, Pennsylvania State 
    University, Shanghai 
    Astronomical Observatory, United 
    Kingdom Participation Group, 
    Universidad Nacional Aut\'onoma 
    de M\'exico, University of Arizona, 
    University of Colorado Boulder, 
    University of Oxford, University of 
    Portsmouth, University of Utah, 
    University of Virginia, University 
    of Washington, University of 
    Wisconsin, Vanderbilt University, 
    and Yale University.

    LOFAR \citep{2013A&A...556A...2V}
    is the Low Frequency Array designed and constructed by 
    ASTRON. It has observing, data processing, and data storage facilities in several countries, 
    which are owned by various parties (each with their own funding sources), and that are 
    collectively operated by the ILT foundation under a joint scientific policy. The ILT resources 
    have benefited from the following recent major funding sources: CNRS-INSU, Observatoire de 
    Paris and Université d'Orléans, France; BMBF, MIWF-NRW, MPG, Germany; Science 
    Foundation Ireland (SFI), Department of Business, Enterprise and Innovation (DBEI), Ireland; 
    NWO, The Netherlands; The Science and Technology Facilities Council, UK; Ministry of 
    Science and Higher Education, Poland; The Istituto Nazionale di Astrofisica (INAF), Italy. 
    This research made use of the Dutch national e-infrastructure with support of the SURF 
    Cooperative (e-infra 180169) and the LOFAR e-infra group. The Jülich LOFAR Long Term 
    Archive and the German LOFAR network are both coordinated and operated by the Jülich 
    Supercomputing Centre (JSC), and computing resources on the supercomputer JUWELS at JSC 
    were provided by the Gauss Centre for Supercomputing e.V. (grant CHTB00) through the John 
    von Neumann Institute for Computing (NIC). 
    This research made use of the University of Hertfordshire high-performance computing facility
    and the LOFAR-UK computing facility located at the University of Hertfordshire and supported 
    by STFC [ST/P000096/1], and of the Italian LOFAR IT computing infrastructure supported and 
    operated by INAF, and by the Physics Department of Turin university (under an agreement with 
    Consorzio Interuniversitario per la Fisica Spaziale) at the C3S Supercomputing Centre, Italy. 

This project makes use of the MaNGA-Pipe3D dataproducts. We thank the IA-UNAM MaNGA team for creating this catalogue, and the Conacyt Project CB-285080 for supporting them.

\section*{Data Availability}
The LoTSS measurements will be made available upon the publication of the LoTSS DR3 paper (Shimwell et al., submitted). Other data underlying this article will be shared on reasonable request to the corresponding author.



\bibliographystyle{mnras}
\bibliography{article} 

@ARTICLE{2021MNRAS.506.5888M,
       author = {{Macfarlane}, C. and {Best}, P.~N. and {Sabater}, J. and {G{\"u}rkan}, G. and {Jarvis}, M.~J. and {R{\"o}ttgering}, H.~J.~A. and {Baldi}, R.~D. and {Calistro Rivera}, G. and {Duncan}, K.~J. and {Morabito}, L.~K. and {Prandoni}, I. and {Retana-Montenegro}, E.},
        title = "{The radio loudness of SDSS quasars from the LOFAR Two-metre Sky Survey: ubiquitous jet activity and constraints on star formation}",
      journal = {\mnras},
         year = 2021,
        month = oct,
       volume = {506},
       number = {4},
        pages = {5888-5907},
          doi = {10.1093/mnras/stab1998},
archivePrefix = {arXiv},
       eprint = {2107.09141},
 primaryClass = {astro-ph.GA},
       adsurl = {https://ui.adsabs.harvard.edu/abs/2021MNRAS.506.5888M}
}

@ARTICLE{2021ApJ...923....6J,
       author = {{Jin}, Gaoxiang and {Dai}, Y. Sophia and {Pan}, Hsi-An and {Lin}, Lihwai and {Li}, Cheng and {Hsieh}, Bau-Ching and {Shen}, Shiyin and {Yuan}, Fang-Ting and {Feng}, Shuai and {Cheng}, Cheng and {Xu}, Hai and {Huang}, Jia-Sheng and {Zhang}, Kai},
        title = "{An IFU View of the Active Galactic Nuclei in MaNGA Galaxy Pairs}",
      journal = {\apj},
         year = 2021,
        month = dec,
       volume = {923},
       number = {1},
          eid = {6},
        pages = {6},
          doi = {10.3847/1538-4357/ac2901},
archivePrefix = {arXiv},
       eprint = {2109.11084},
 primaryClass = {astro-ph.GA},
       adsurl = {https://ui.adsabs.harvard.edu/abs/2021ApJ...923....6J}
}

@ARTICLE{2017MNRAS.464.3882W,
       author = {{Weston}, Madalyn E. and {McIntosh}, Daniel H. and {Brodwin}, Mark and {Mann}, Justin and {Cooper}, Andrew and {McConnell}, Adam and {Nielsen}, Jennifer L.},
        title = "{Incidence of WISE -selected obscured AGNs in major mergers and interactions from the SDSS}",
      journal = {\mnras},
         year = 2017,
        month = feb,
       volume = {464},
       number = {4},
        pages = {3882-3906},
          doi = {10.1093/mnras/stw2620},
archivePrefix = {arXiv},
       eprint = {1609.04832},
 primaryClass = {astro-ph.GA},
       adsurl = {https://ui.adsabs.harvard.edu/abs/2017MNRAS.464.3882W}
}

@ARTICLE{2016MNRAS.461.1076C,
       author = {{Comparat}, Johan and {Zhu}, Guangtun and {Gonzalez-Perez}, Violeta and {Norberg}, Peder and {Newman}, Jeffrey and {Tresse}, Laurence and {Richard}, Johan and {Yepes}, Gustavo and {Kneib}, Jean-Paul and {Raichoor}, Anand and {Prada}, Francisco and {Maraston}, Claudia and {Y{\`e}che}, Christophe and {Delubac}, Timoth{\'e}e and {Jullo}, Eric},
        title = "{The evolution of the [O II], H {\ensuremath{\beta}} and [O III] emission line luminosity functions over the last nine billions years}",
      journal = {\mnras},
         year = 2016,
        month = sep,
       volume = {461},
       number = {1},
        pages = {1076-1087},
          doi = {10.1093/mnras/stw1393},
archivePrefix = {arXiv},
       eprint = {1605.02875},
 primaryClass = {astro-ph.GA},
       adsurl = {https://ui.adsabs.harvard.edu/abs/2016MNRAS.461.1076C}
}

@ARTICLE{2007ApJ...664..840H,
       author = {{Huang}, J.-S. and {Ashby}, M.~L.~N. and {Barmby}, P. and {Brodwin}, M. and {Brown}, M.~J.~I. and {Caldwell}, N. and {Cool}, R.~J. and {Eisenhardt}, P. and {Eisenstein}, D. and {Fazio}, G.~G. and {Le Floc'h}, E. and {Green}, P. and {Kochanek}, C.~S. and {Lu}, Nanyao and {Pahre}, M.~A. and {Rigopoulou}, D. and {Rosenberg}, J.~L. and {Smith}, H.~A. and {Wang}, Z. and {Willmer}, C.~N.~A. and {Willner}, S.~P.},
        title = "{The Local Galaxy 8 {\ensuremath{\mu}}m Luminosity Function}",
      journal = {\apj},
         year = 2007,
        month = aug,
       volume = {664},
       number = {2},
        pages = {840-849},
          doi = {10.1086/519241},
archivePrefix = {arXiv},
       eprint = {0704.3609},
 primaryClass = {astro-ph},
       adsurl = {https://ui.adsabs.harvard.edu/abs/2007ApJ...664..840H}
}

@ARTICLE{2023MNRAS.523.6082C,
       author = {{Cochrane}, R.~K. and {Kondapally}, R. and {Best}, P.~N. and {Sabater}, J. and {Duncan}, K.~J. and {Smith}, D.~J.~B. and {Hardcastle}, M.~J. and {R{\"o}ttgering}, H.~J.~A. and {Prandoni}, I. and {Haskell}, P. and {G{\"u}rkan}, G. and {Miley}, G.~K.},
        title = "{The LOFAR Two-metre Sky Survey: the radio view of the cosmic star formation history}",
      journal = {\mnras},
         year = 2023,
        month = aug,
       volume = {523},
       number = {4},
        pages = {6082-6102},
          doi = {10.1093/mnras/stad1602},
archivePrefix = {arXiv},
       eprint = {2305.15510},
 primaryClass = {astro-ph.GA},
       adsurl = {https://ui.adsabs.harvard.edu/abs/2023MNRAS.523.6082C}
}

@ARTICLE{2023MNRAS.525.5359M,
       author = {{Marasco}, A. and {Poggianti}, B.~M. and {Fritz}, J. and {Werle}, A. and {Vulcani}, B. and {Moretti}, A. and {Gullieuszik}, M. and {Kulier}, A.},
        title = "{The morphological transformation of ram pressure stripped galaxies: a pathway from late to early galaxy types}",
      journal = {\mnras},
         year = 2023,
        month = nov,
       volume = {525},
       number = {4},
        pages = {5359-5377},
          doi = {10.1093/mnras/stad2604},
archivePrefix = {arXiv},
       eprint = {2308.14791},
 primaryClass = {astro-ph.GA},
       adsurl = {https://ui.adsabs.harvard.edu/abs/2023MNRAS.525.5359M}
}

@ARTICLE{2019A&A...631A..46G,
       author = {{Gaur}, H. and {Gu}, M. and {Ramya}, S. and {Guo}, H.},
        title = "{Properties of radio-loud quasars in the Sloan Digital Sky Survey}",
      journal = {\aap},
         year = 2019,
        month = nov,
       volume = {631},
          eid = {A46},
        pages = {A46},
          doi = {10.1051/0004-6361/201935398},
archivePrefix = {arXiv},
       eprint = {1908.09489},
 primaryClass = {astro-ph.HE},
       adsurl = {https://ui.adsabs.harvard.edu/abs/2019A&A...631A..46G}
}

@ARTICLE{2017A&ARv..25....2P,
       author = {{Padovani}, P. and {Alexander}, D.~M. and {Assef}, R.~J. and {De Marco}, B. and {Giommi}, P. and {Hickox}, R.~C. and {Richards}, G.~T. and {Smol{\v{c}}i{\'c}}, V. and {Hatziminaoglou}, E. and {Mainieri}, V. and {Salvato}, M.},
        title = "{Active galactic nuclei: what's in a name?}",
      journal = {\aapr},
         year = 2017,
        month = aug,
       volume = {25},
       number = {1},
          eid = {2},
        pages = {2},
          doi = {10.1007/s00159-017-0102-9},
archivePrefix = {arXiv},
       eprint = {1707.07134},
 primaryClass = {astro-ph.GA},
       adsurl = {https://ui.adsabs.harvard.edu/abs/2017A&ARv..25....2P}
}

@ARTICLE{2020NewAR..8801539H,
       author = {{Hardcastle}, M.~J. and {Croston}, J.~H.},
        title = "{Radio galaxies and feedback from AGN jets}",
      journal = {\nar},
         year = 2020,
        month = jun,
       volume = {88},
          eid = {101539},
        pages = {101539},
          doi = {10.1016/j.newar.2020.101539},
archivePrefix = {arXiv},
       eprint = {2003.06137},
 primaryClass = {astro-ph.HE},
       adsurl = {https://ui.adsabs.harvard.edu/abs/2020NewAR..8801539H}
}

@ARTICLE{2024ApJ...962..146O,
       author = {{Oio}, Gabriel A. and {Dai}, Y. Sophia and {Bornancini}, C.~G. and {Li}, Zi-Jian},
        title = "{Host Galaxy and Nuclear Properties of IR-selected AGNs with and without Outflow Signatures}",
      journal = {\apj},
         year = 2024,
        month = feb,
       volume = {962},
       number = {2},
          eid = {146},
        pages = {146},
          doi = {10.3847/1538-4357/ad18a5},
archivePrefix = {arXiv},
       eprint = {2312.15957},
 primaryClass = {astro-ph.GA},
       adsurl = {https://ui.adsabs.harvard.edu/abs/2024ApJ...962..146O}
}

@ARTICLE{2012MNRAS.427.2275B,
       author = {{Banerji}, Manda and {McMahon}, Richard G. and {Hewett}, Paul C. and {Alaghband-Zadeh}, Susannah and {Gonzalez-Solares}, Eduardo and {Venemans}, Bram P. and {Hawthorn}, Melanie J.},
        title = "{Heavily reddened quasars at z {\ensuremath{\sim}} 2 in the UKIDSS Large Area Survey: a transitional phase in AGN evolution}",
      journal = {\mnras},
         year = 2012,
        month = dec,
       volume = {427},
       number = {3},
        pages = {2275-2291},
          doi = {10.1111/j.1365-2966.2012.22099.x},
archivePrefix = {arXiv},
       eprint = {1203.5530},
 primaryClass = {astro-ph.CO},
       adsurl = {https://ui.adsabs.harvard.edu/abs/2012MNRAS.427.2275B}
}

@ARTICLE{2024ApJ...977..194S,
       author = {{Suresh}, Arjun and {Blanton}, Michael R.},
        title = "{Radio Active Galactic Nuclei Activity in Low Redshift Galaxies Is Not Directly Related to Star Formation Rates}",
      journal = {\apj},
         year = 2024,
        month = dec,
       volume = {977},
       number = {2},
          eid = {194},
        pages = {194},
          doi = {10.3847/1538-4357/ad8ac7},
archivePrefix = {arXiv},
       eprint = {2404.04780},
 primaryClass = {astro-ph.GA},
       adsurl = {https://ui.adsabs.harvard.edu/abs/2024ApJ...977..194S}
}

@ARTICLE{2024MNRAS.529.3939Y,
       author = {{Yue}, B.-H. and {Best}, P.~N. and {Duncan}, K.~J. and {Calistro-Rivera}, G. and {Morabito}, L.~K. and {Petley}, J.~W. and {Prandoni}, I. and {R{\"o}ttgering}, H.~J.~A. and {Smith}, D.~J.~B.},
        title = "{A novel Bayesian approach for decomposing the radio emission of quasars: I. Modelling the radio excess in red quasars}",
      journal = {\mnras},
         year = 2024,
        month = apr,
       volume = {529},
       number = {4},
        pages = {3939-3957},
          doi = {10.1093/mnras/stae725},
archivePrefix = {arXiv},
       eprint = {2403.07074},
 primaryClass = {astro-ph.GA},
       adsurl = {https://ui.adsabs.harvard.edu/abs/2024MNRAS.529.3939Y}
}

@ARTICLE{2025MNRAS.536.1166E,
       author = {{Escott}, Emmy L. and {Morabito}, Leah K. and {Scholtz}, Jan and {Hickox}, Ryan C. and {Harrison}, Chris M. and {Alexander}, David M. and {Arnaudova}, Marina I. and {Smith}, Daniel J.~B. and {Duncan}, Kenneth J. and {Petley}, James and et al.},
        title = "{Unveiling AGN outflows: [O III] outflow detection rates and correlation with low-frequency radio emission}",
      journal = {\mnras},
         year = 2025,
        month = jan,
       volume = {536},
       number = {2},
        pages = {1166-1179},
          doi = {10.1093/mnras/stae2645},
archivePrefix = {arXiv},
       eprint = {2411.19326},
 primaryClass = {astro-ph.GA},
       adsurl = {https://ui.adsabs.harvard.edu/abs/2025MNRAS.536.1166E}
}

@ARTICLE{2019MNRAS.488.3396M,
       author = {{Masters}, Karen L. and {Stark}, David V. and {Pace}, Zachary J. and {Phipps}, Frederika and {Rujopakarn}, Wiphu and {Samanso}, Nattida and {Harrington}, Emily and {S{\'a}nchez-Gallego}, Jos{\'e} R. and {Avila-Reese}, Vladimir and {Bershady}, Matthew and {Cherinka}, Brian and {Fielder}, Catherine E. and {Finnegan}, Daniel and {Riffel}, Rogemar A. and {Rowlands}, Kate and {Shamsi}, Shoaib and {Newnham}, Lucy and {Weijmans}, Anne-Marie and {Witherspoon}, Catherine A.},
        title = "{H I-MaNGA: H I follow-up for the MaNGA survey}",
      journal = {\mnras},
         year = 2019,
        month = sep,
       volume = {488},
       number = {3},
        pages = {3396-3405},
          doi = {10.1093/mnras/stz1889},
archivePrefix = {arXiv},
       eprint = {1901.05579},
 primaryClass = {astro-ph.GA},
       adsurl = {https://ui.adsabs.harvard.edu/abs/2019MNRAS.488.3396M}
}

@ARTICLE{2008MNRAS.391..481S,
       author = {{Somerville}, Rachel S. and {Hopkins}, Philip F. and {Cox}, Thomas J. and {Robertson}, Brant E. and {Hernquist}, Lars},
        title = "{A semi-analytic model for the co-evolution of galaxies, black holes and active galactic nuclei}",
      journal = {\mnras},
         year = 2008,
        month = dec,
       volume = {391},
       number = {2},
        pages = {481-506},
          doi = {10.1111/j.1365-2966.2008.13805.x},
archivePrefix = {arXiv},
       eprint = {0808.1227},
 primaryClass = {astro-ph},
       adsurl = {https://ui.adsabs.harvard.edu/abs/2008MNRAS.391..481S}
}

@ARTICLE{2000MNRAS.311..576K,
       author = {{Kauffmann}, Guinevere and {Haehnelt}, Martin},
        title = "{A unified model for the evolution of galaxies and quasars}",
      journal = {\mnras},
         year = 2000,
        month = jan,
       volume = {311},
       number = {3},
        pages = {576-588},
          doi = {10.1046/j.1365-8711.2000.03077.x},
archivePrefix = {arXiv},
       eprint = {astro-ph/9906493},
 primaryClass = {astro-ph},
       adsurl = {https://ui.adsabs.harvard.edu/abs/2000MNRAS.311..576K}
}

@ARTICLE{2019MNRAS.488.3109K,
       author = {{Klindt}, L. and {Alexander}, D.~M. and {Rosario}, D.~J. and {Lusso}, E. and {Fotopoulou}, S.},
        title = "{Fundamental differences in the radio properties of red and blue quasars: evolution strongly favoured over orientation}",
      journal = {\mnras},
         year = 2019,
        month = sep,
       volume = {488},
       number = {3},
        pages = {3109-3128},
          doi = {10.1093/mnras/stz1771},
archivePrefix = {arXiv},
       eprint = {1905.12108},
 primaryClass = {astro-ph.GA},
       adsurl = {https://ui.adsabs.harvard.edu/abs/2019MNRAS.488.3109K}
}

@BOOK{2006agna.book.....O,
       author = {{Osterbrock}, Donald E. and {Ferland}, Gary J.},
        title = "{Astrophysics of gaseous nebulae and active galactic nuclei}",
         year = 2006,
       publisher = {University Science Books}, 
       adsurl = {https://ui.adsabs.harvard.edu/abs/2006agna.book.....O}
}

@ARTICLE{2016MNRAS.461.4227H,
       author = {{Heard}, Clio Z.~P. and {Gaskell}, C. Martin},
        title = "{The location of the dust causing internal reddening of active galactic nuclei}",
      journal = {\mnras},
         year = 2016,
        month = oct,
       volume = {461},
       number = {4},
        pages = {4227-4232},
          doi = {10.1093/mnras/stw1616},
archivePrefix = {arXiv},
       eprint = {1606.08914},
 primaryClass = {astro-ph.GA},
       adsurl = {https://ui.adsabs.harvard.edu/abs/2016MNRAS.461.4227H}
}

@ARTICLE{2008MNRAS.383..581D,
       author = {{Dong}, Xiaobo and {Wang}, Tinggui and {Wang}, Jianguo and {Yuan}, Weimin and {Zhou}, Hongyan and {Dai}, Haifeng and {Zhang}, Kai},
        title = "{Broad-line Balmer decrements in blue active galactic nuclei}",
      journal = {\mnras},
         year = 2008,
        month = jan,
       volume = {383},
       number = {2},
        pages = {581-592},
          doi = {10.1111/j.1365-2966.2007.12560.x},
archivePrefix = {arXiv},
       eprint = {0710.1458},
 primaryClass = {astro-ph},
       adsurl = {https://ui.adsabs.harvard.edu/abs/2008MNRAS.383..581D}
}

@ARTICLE{2015A&A...576A..10C,
       author = {{Ciesla}, L. and {Charmandaris}, V. and {Georgakakis}, A. and {Bernhard}, E. and {Mitchell}, P.~D. and {Buat}, V. and {Elbaz}, D. and {LeFloc'h}, E. and {Lacey}, C.~G. and {Magdis}, G.~E. and {Xilouris}, M.},
        title = "{Constraining the properties of AGN host galaxies with spectral energy distribution modelling}",
      journal = {\aap},
         year = 2015,
        month = apr,
       volume = {576},
          eid = {A10},
        pages = {A10},
          doi = {10.1051/0004-6361/201425252},
archivePrefix = {arXiv},
       eprint = {1501.03672},
 primaryClass = {astro-ph.GA},
       adsurl = {https://ui.adsabs.harvard.edu/abs/2015A&A...576A..10C}
}

@ARTICLE{2013A&A...556A...2V,
       author = {{van Haarlem}, M.~P. and {Wise}, M.~W. and {Gunst}, A.~W. and {Heald}, G. and {McKean}, J.~P. and {Hessels}, J.~W.~T. and {de Bruyn}, A.~G. and {Nijboer}, R. and {Swinbank}, J. and {Fallows}, R. and {Brentjens}, M. and {Nelles}, A. and {Beck}, R. and {Falcke}, H. and {Fender}, R. and {H{\"o}randel}, J. and {Koopmans}, L.~V.~E. and {Mann}, G. and {Miley}, G. and {R{\"o}ttgering}, H. and {Stappers}, B.~W. and {Wijers}, R.~A.~M.~J. and {Zaroubi}, S. and {van den Akker}, M. and {Alexov}, A. and {Anderson}, J. and {Anderson}, K. and {van Ardenne}, A. and {Arts}, M. and {Asgekar}, A. and {Avruch}, I.~M. and {Batejat}, F. and {B{\"a}hren}, L. and {Bell}, M.~E. and {Bell}, M.~R. and {van Bemmel}, I. and {Bennema}, P. and {Bentum}, M.~J. and {Bernardi}, G. and {Best}, P. and {B{\^\i}rzan}, L. and {Bonafede}, A. and {Boonstra}, A. -J. and {Braun}, R. and {Bregman}, J. and {Breitling}, F. and {van de Brink}, R.~H. and {Broderick}, J. and {Broekema}, P.~C. and {Brouw}, W.~N. and {Br{\"u}ggen}, M. and {Butcher}, H.~R. and {van Cappellen}, W. and {Ciardi}, B. and {Coenen}, T. and {Conway}, J. and {Coolen}, A. and {Corstanje}, A. and {Damstra}, S. and {Davies}, O. and {Deller}, A.~T. and {Dettmar}, R. -J. and {van Diepen}, G. and {Dijkstra}, K. and {Donker}, P. and {Doorduin}, A. and {Dromer}, J. and {Drost}, M. and {van Duin}, A. and {Eisl{\"o}ffel}, J. and {van Enst}, J. and {Ferrari}, C. and {Frieswijk}, W. and {Gankema}, H. and {Garrett}, M.~A. and {de Gasperin}, F. and {Gerbers}, M. and {de Geus}, E. and {Grie{\ss}meier}, J. -M. and {Grit}, T. and {Gruppen}, P. and {Hamaker}, J.~P. and {Hassall}, T. and {Hoeft}, M. and {Holties}, H.~A. and {Horneffer}, A. and {van der Horst}, A. and {van Houwelingen}, A. and {Huijgen}, A. and {Iacobelli}, M. and {Intema}, H. and {Jackson}, N. and {Jelic}, V. and {de Jong}, A. and {Juette}, E. and {Kant}, D. and {Karastergiou}, A. and {Koers}, A. and {Kollen}, H. and {Kondratiev}, V.~I. and {Kooistra}, E. and {Koopman}, Y. and {Koster}, A. and {Kuniyoshi}, M. and {Kramer}, M. and {Kuper}, G. and {Lambropoulos}, P. and {Law}, C. and {van Leeuwen}, J. and {Lemaitre}, J. and {Loose}, M. and {Maat}, P. and {Macario}, G. and {Markoff}, S. and {Masters}, J. and {McFadden}, R.~A. and {McKay-Bukowski}, D. and {Meijering}, H. and {Meulman}, H. and {Mevius}, M. and {Middelberg}, E. and {Millenaar}, R. and {Miller-Jones}, J.~C.~A. and {Mohan}, R.~N. and {Mol}, J.~D. and {Morawietz}, J. and {Morganti}, R. and {Mulcahy}, D.~D. and {Mulder}, E. and {Munk}, H. and {Nieuwenhuis}, L. and {van Nieuwpoort}, R. and {Noordam}, J.~E. and {Norden}, M. and {Noutsos}, A. and {Offringa}, A.~R. and {Olofsson}, H. and {Omar}, A. and {Orr{\'u}}, E. and {Overeem}, R. and {Paas}, H. and {Pandey-Pommier}, M. and {Pandey}, V.~N. and {Pizzo}, R. and {Polatidis}, A. and {Rafferty}, D. and {Rawlings}, S. and {Reich}, W. and {de Reijer}, J. -P. and {Reitsma}, J. and {Renting}, G.~A. and {Riemers}, P. and {Rol}, E. and {Romein}, J.~W. and {Roosjen}, J. and {Ruiter}, M. and {Scaife}, A. and {van der Schaaf}, K. and {Scheers}, B. and {Schellart}, P. and {Schoenmakers}, A. and {Schoonderbeek}, G. and {Serylak}, M. and {Shulevski}, A. and {Sluman}, J. and {Smirnov}, O. and {Sobey}, C. and {Spreeuw}, H. and {Steinmetz}, M. and {Sterks}, C.~G.~M. and {Stiepel}, H. -J. and {Stuurwold}, K. and {Tagger}, M. and {Tang}, Y. and {Tasse}, C. and {Thomas}, I. and {Thoudam}, S. and {Toribio}, M.~C. and {van der Tol}, B. and {Usov}, O. and {van Veelen}, M. and {van der Veen}, A. -J. and {ter Veen}, S. and {Verbiest}, J.~P.~W. and {Vermeulen}, R. and {Vermaas}, N. and {Vocks}, C. and {Vogt}, C. and {de Vos}, M. and {van der Wal}, E. and {van Weeren}, R. and {Weggemans}, H. and {Weltevrede}, P. and {White}, S. and {Wijnholds}, S.~J. and {Wilhelmsson}, T. and {Wucknitz}, O. and {Yatawatta}, S. and {Zarka}, P. and {Zensus}, A.},
        title = "{LOFAR: The LOw-Frequency ARray}",
      journal = {\aap},
         year = 2013,
        month = aug,
       volume = {556},
          eid = {A2},
        pages = {A2},
          doi = {10.1051/0004-6361/201220873},
archivePrefix = {arXiv},
       eprint = {1305.3550},
 primaryClass = {astro-ph.IM},
       adsurl = {https://ui.adsabs.harvard.edu/abs/2013A&A...556A...2V}
}

@ARTICLE{2022A&A...659A...1S,
       author = {{Shimwell}, T.~W. and {Hardcastle}, M.~J. and {Tasse}, C. and {Best}, P.~N. and {R{\"o}ttgering}, H.~J.~A. and {Williams}, W.~L. and {Botteon}, A. and {Drabent}, A. and {Mechev}, A. and {Shulevski}, A. and {van Weeren}, R.~J. and {Bester}, L. and {Br{\"u}ggen}, M. and {Brunetti}, G. and {Callingham}, J.~R. and {Chy{\.z}y}, K.~T. and {Conway}, J.~E. and {Dijkema}, T.~J. and {Duncan}, K. and {de Gasperin}, F. and {Hale}, C.~L. and {Haverkorn}, M. and {Hugo}, B. and {Jackson}, N. and {Mevius}, M. and {Miley}, G.~K. and {Morabito}, L.~K. and {Morganti}, R. and {Offringa}, A. and {Oonk}, J.~B.~R. and {Rafferty}, D. and {Sabater}, J. and {Smith}, D.~J.~B. and {Schwarz}, D.~J. and {Smirnov}, O. and {O'Sullivan}, S.~P. and {Vedantham}, H. and {White}, G.~J. and {Albert}, J.~G. and {Alegre}, L. and {Asabere}, B. and {Bacon}, D.~J. and {Bonafede}, A. and {Bonnassieux}, E. and {Brienza}, M. and {Bilicki}, M. and {Bonato}, M. and {Calistro Rivera}, G. and {Cassano}, R. and {Cochrane}, R. and {Croston}, J.~H. and {Cuciti}, V. and {Dallacasa}, D. and {Danezi}, A. and {Dettmar}, R.~J. and {Di Gennaro}, G. and {Edler}, H.~W. and {En{\ss}lin}, T.~A. and {Emig}, K.~L. and {Franzen}, T.~M.~O. and {Garc{\'\i}a-Vergara}, C. and {Grange}, Y.~G. and {G{\"u}rkan}, G. and {Hajduk}, M. and {Heald}, G. and {Heesen}, V. and {Hoang}, D.~N. and {Hoeft}, M. and {Horellou}, C. and {Iacobelli}, M. and {Jamrozy}, M. and {Jeli{\'c}}, V. and {Kondapally}, R. and {Kukreti}, P. and {Kunert-Bajraszewska}, M. and {Magliocchetti}, M. and {Mahatma}, V. and {Ma{\l}ek}, K. and {Mandal}, S. and {Massaro}, F. and {Meyer-Zhao}, Z. and {Mingo}, B. and {Mostert}, R.~I.~J. and {Nair}, D.~G. and {Nakoneczny}, S.~J. and {Nikiel-Wroczy{\'n}ski}, B. and {Orr{\'u}}, E. and {Pajdosz-{\'S}mierciak}, U. and {Pasini}, T. and {Prandoni}, I. and {van Piggelen}, H.~E. and {Rajpurohit}, K. and {Retana-Montenegro}, E. and {Riseley}, C.~J. and {Rowlinson}, A. and {Saxena}, A. and {Schrijvers}, C. and {Sweijen}, F. and {Siewert}, T.~M. and {Timmerman}, R. and {Vaccari}, M. and {Vink}, J. and {West}, J.~L. and {Wo{\l}owska}, A. and {Zhang}, X. and {Zheng}, J.},
        title = "{The LOFAR Two-metre Sky Survey. V. Second data release}",
      journal = {\aap},
         year = 2022,
        month = mar,
       volume = {659},
          eid = {A1},
        pages = {A1},
          doi = {10.1051/0004-6361/202142484},
archivePrefix = {arXiv},
       eprint = {2202.11733},
 primaryClass = {astro-ph.GA},
       adsurl = {https://ui.adsabs.harvard.edu/abs/2022A&A...659A...1S}
}

@ARTICLE{2017A&A...598A.104S,
       author = {{Shimwell}, T.~W. and {R{\"o}ttgering}, H.~J.~A. and {Best}, P.~N. and {Williams}, W.~L. and {Dijkema}, T.~J. and {de Gasperin}, F. and {Hardcastle}, M.~J. and {Heald}, G.~H. and {Hoang}, D.~N. and {Horneffer}, A. and {Intema}, H. and {Mahony}, E.~K. and {Mandal}, S. and {Mechev}, A.~P. and {Morabito}, L. and {Oonk}, J.~B.~R. and {Rafferty}, D. and {Retana-Montenegro}, E. and {Sabater}, J. and {Tasse}, C. and {van Weeren}, R.~J. and {Br{\"u}ggen}, M. and {Brunetti}, G. and {Chy{\.z}y}, K.~T. and {Conway}, J.~E. and {Haverkorn}, M. and {Jackson}, N. and {Jarvis}, M.~J. and {McKean}, J.~P. and {Miley}, G.~K. and {Morganti}, R. and {White}, G.~J. and {Wise}, M.~W. and {van Bemmel}, I.~M. and {Beck}, R. and {Brienza}, M. and {Bonafede}, A. and {Calistro Rivera}, G. and {Cassano}, R. and {Clarke}, A.~O. and {Cseh}, D. and {Deller}, A. and {Drabent}, A. and {van Driel}, W. and {Engels}, D. and {Falcke}, H. and {Ferrari}, C. and {Fr{\"o}hlich}, S. and {Garrett}, M.~A. and {Harwood}, J.~J. and {Heesen}, V. and {Hoeft}, M. and {Horellou}, C. and {Israel}, F.~P. and {Kapi{\'n}ska}, A.~D. and {Kunert-Bajraszewska}, M. and {McKay}, D.~J. and {Mohan}, N.~R. and {Orr{\'u}}, E. and {Pizzo}, R.~F. and {Prandoni}, I. and {Schwarz}, D.~J. and {Shulevski}, A. and {Sipior}, M. and {Smith}, D.~J.~B. and {Sridhar}, S.~S. and {Steinmetz}, M. and {Stroe}, A. and {Varenius}, E. and {van der Werf}, P.~P. and {Zensus}, J.~A. and {Zwart}, J.~T.~L.},
        title = "{The LOFAR Two-metre Sky Survey. I. Survey description and preliminary data release}",
      journal = {\aap},
         year = 2017,
        month = feb,
       volume = {598},
          eid = {A104},
        pages = {A104},
          doi = {10.1051/0004-6361/201629313},
archivePrefix = {arXiv},
       eprint = {1611.02700},
 primaryClass = {astro-ph.IM},
       adsurl = {https://ui.adsabs.harvard.edu/abs/2017A&A...598A.104S}
}

@ARTICLE{2018MNRAS.480.2544R,
       author = {{Rowlands}, K. and {Heckman}, T. and {Wild}, V. and {Zakamska}, N.~L. and {Rodriguez-Gomez}, V. and {Barrera-Ballesteros}, J. and {Lotz}, J. and {Thilker}, D. and {Andrews}, B.~H. and {Boquien}, M. and {Brinkmann}, J. and {Brownstein}, J.~R. and {Hwang}, H. -C. and {Smethurst}, R.},
        title = "{SDSS-IV MaNGA: spatially resolved star formation histories and the connection to galaxy physical properties}",
      journal = {\mnras},
         year = 2018,
        month = oct,
       volume = {480},
       number = {2},
        pages = {2544-2561},
          doi = {10.1093/mnras/sty1916},
archivePrefix = {arXiv},
       eprint = {1807.06066},
 primaryClass = {astro-ph.GA},
       adsurl = {https://ui.adsabs.harvard.edu/abs/2018MNRAS.480.2544R}
}

@ARTICLE{2012MNRAS.421.1569B,
       author = {{Best}, P.~N. and {Heckman}, T.~M.},
        title = "{On the fundamental dichotomy in the local radio-AGN population: accretion, evolution and host galaxy properties}",
      journal = {\mnras},
         year = 2012,
        month = apr,
       volume = {421},
       number = {2},
        pages = {1569-1582},
          doi = {10.1111/j.1365-2966.2012.20414.x},
archivePrefix = {arXiv},
       eprint = {1201.2397},
 primaryClass = {astro-ph.CO},
       adsurl = {https://ui.adsabs.harvard.edu/abs/2012MNRAS.421.1569B}
}

@ARTICLE{2005MNRAS.362...25B,
       author = {{Best}, P.~N. and {Kauffmann}, G. and {Heckman}, T.~M. and {Brinchmann}, J. and {Charlot}, S. and {Ivezi{\'c}}, {\v{Z}}. and {White}, S.~D.~M.},
        title = "{The host galaxies of radio-loud active galactic nuclei: mass dependences, gas cooling and active galactic nuclei feedback}",
      journal = {\mnras},
         year = 2005,
        month = sep,
       volume = {362},
       number = {1},
        pages = {25-40},
          doi = {10.1111/j.1365-2966.2005.09192.x},
archivePrefix = {arXiv},
       eprint = {astro-ph/0506269},
 primaryClass = {astro-ph},
       adsurl = {https://ui.adsabs.harvard.edu/abs/2005MNRAS.362...25B}
}

@ARTICLE{2017MNRAS.465.3291W,
       author = {{Weinberger}, Rainer and {Springel}, Volker and {Hernquist}, Lars and {Pillepich}, Annalisa and {Marinacci}, Federico and {Pakmor}, R{\"u}diger and {Nelson}, Dylan and {Genel}, Shy and {Vogelsberger}, Mark and {Naiman}, Jill and {Torrey}, Paul},
        title = "{Simulating galaxy formation with black hole driven thermal and kinetic feedback}",
      journal = {\mnras},
         year = 2017,
        month = mar,
       volume = {465},
       number = {3},
        pages = {3291-3308},
          doi = {10.1093/mnras/stw2944},
archivePrefix = {arXiv},
       eprint = {1607.03486},
 primaryClass = {astro-ph.GA},
       adsurl = {https://ui.adsabs.harvard.edu/abs/2017MNRAS.465.3291W}
}

@ARTICLE{2005MNRAS.363L..91C,
       author = {{Churazov}, E. and {Sazonov}, S. and {Sunyaev}, R. and {Forman}, W. and {Jones}, C. and {B{\"o}hringer}, H.},
        title = "{Supermassive black holes in elliptical galaxies: switching from very bright to very dim}",
      journal = {\mnras},
         year = 2005,
        month = oct,
       volume = {363},
       number = {1},
        pages = {L91-L95},
          doi = {10.1111/j.1745-3933.2005.00093.x},
archivePrefix = {arXiv},
       eprint = {astro-ph/0507073},
 primaryClass = {astro-ph},
       adsurl = {https://ui.adsabs.harvard.edu/abs/2005MNRAS.363L..91C}
}

@ARTICLE{2015MNRAS.447.2857B,
       author = {{Bryant}, J.~J. and {Owers}, M.~S. and {Robotham}, A.~S.~G. and {Croom}, S.~M. and {Driver}, S.~P. and {Drinkwater}, M.~J. and {Lorente}, N.~P.~F. and {Cortese}, L. and {Scott}, N. and {Colless}, M. and {Schaefer}, A. and {Taylor}, E.~N. and {Konstantopoulos}, I.~S. and {Allen}, J.~T. and {Baldry}, I. and {Barnes}, L. and {Bauer}, A.~E. and {Bland-Hawthorn}, J. and {Bloom}, J.~V. and {Brooks}, A.~M. and {Brough}, S. and {Cecil}, G. and {Couch}, W. and {Croton}, D. and {Davies}, R. and {Ellis}, S. and {Fogarty}, L.~M.~R. and {Foster}, C. and {Glazebrook}, K. and {Goodwin}, M. and {Green}, A. and {Gunawardhana}, M.~L. and {Hampton}, E. and {Ho}, I. -T. and {Hopkins}, A.~M. and {Kewley}, L. and {Lawrence}, J.~S. and {Leon-Saval}, S.~G. and {Leslie}, S. and {McElroy}, R. and {Lewis}, G. and {Liske}, J. and {L{\'o}pez-S{\'a}nchez}, {\'A}. R. and {Mahajan}, S. and {Medling}, A.~M. and {Metcalfe}, N. and {Meyer}, M. and {Mould}, J. and {Obreschkow}, D. and {O'Toole}, S. and {Pracy}, M. and {Richards}, S.~N. and {Shanks}, T. and {Sharp}, R. and {Sweet}, S.~M. and {Thomas}, A.~D. and {Tonini}, C. and {Walcher}, C.~J.},
        title = "{The SAMI Galaxy Survey: instrument specification and target selection}",
      journal = {\mnras},
         year = 2015,
        month = mar,
       volume = {447},
       number = {3},
        pages = {2857-2879},
          doi = {10.1093/mnras/stu2635},
archivePrefix = {arXiv},
       eprint = {1407.7335},
 primaryClass = {astro-ph.GA},
       adsurl = {https://ui.adsabs.harvard.edu/abs/2015MNRAS.447.2857B}
}

@ARTICLE{2012A&A...538A...8S,
       author = {{S{\'a}nchez}, S.~F. and {Kennicutt}, R.~C. and {Gil de Paz}, A. and {van de Ven}, G. and {V{\'\i}lchez}, J.~M. and {Wisotzki}, L. and {Walcher}, C.~J. and {Mast}, D. and {Aguerri}, J.~A.~L. and {Albiol-P{\'e}rez}, S. and {Alonso-Herrero}, A. and {Alves}, J. and {Bakos}, J. and {Bart{\'a}kov{\'a}}, T. and {Bland-Hawthorn}, J. and {Boselli}, A. and {Bomans}, D.~J. and {Castillo-Morales}, A. and {Cortijo-Ferrero}, C. and {de Lorenzo-C{\'a}ceres}, A. and {Del Olmo}, A. and {Dettmar}, R. -J. and {D{\'\i}az}, A. and {Ellis}, S. and {Falc{\'o}n-Barroso}, J. and {Flores}, H. and {Gallazzi}, A. and {Garc{\'\i}a-Lorenzo}, B. and {Gonz{\'a}lez Delgado}, R. and {Gruel}, N. and {Haines}, T. and {Hao}, C. and {Husemann}, B. and {Igl{\'e}sias-P{\'a}ramo}, J. and {Jahnke}, K. and {Johnson}, B. and {Jungwiert}, B. and {Kalinova}, V. and {Kehrig}, C. and {Kupko}, D. and {L{\'o}pez-S{\'a}nchez}, {\'A}. R. and {Lyubenova}, M. and {Marino}, R.~A. and {M{\'a}rmol-Queralt{\'o}}, E. and {M{\'a}rquez}, I. and {Masegosa}, J. and {Meidt}, S. and {Mendez-Abreu}, J. and {Monreal-Ibero}, A. and {Montijo}, C. and {Mour{\~a}o}, A.~M. and {Palacios-Navarro}, G. and {Papaderos}, P. and {Pasquali}, A. and {Peletier}, R. and {P{\'e}rez}, E. and {P{\'e}rez}, I. and {Quirrenbach}, A. and {Rela{\~n}o}, M. and {Rosales-Ortega}, F.~F. and {Roth}, M.~M. and {Ruiz-Lara}, T. and {S{\'a}nchez-Bl{\'a}zquez}, P. and {Sengupta}, C. and {Singh}, R. and {Stanishev}, V. and {Trager}, S.~C. and {Vazdekis}, A. and {Viironen}, K. and {Wild}, V. and {Zibetti}, S. and {Ziegler}, B.},
        title = "{CALIFA, the Calar Alto Legacy Integral Field Area survey. I. Survey presentation}",
      journal = {\aap},
         year = 2012,
        month = feb,
       volume = {538},
          eid = {A8},
        pages = {A8},
          doi = {10.1051/0004-6361/201117353},
archivePrefix = {arXiv},
       eprint = {1111.0962},
 primaryClass = {astro-ph.CO},
       adsurl = {https://ui.adsabs.harvard.edu/abs/2012A&A...538A...8S}
}

@ARTICLE{2018MNRAS.477.3014B,
       author = {{Belfiore}, Francesco and {Maiolino}, Roberto and {Bundy}, Kevin and {Masters}, Karen and {Bershady}, Matthew and {Oyarz{\'u}n}, Grecco A. and {Lin}, Lihwai and {Cano-Diaz}, Mariana and {Wake}, David and {Spindler}, Ashley and {Thomas}, Daniel and {Brownstein}, Joel R. and {Drory}, Niv and {Yan}, Renbin},
        title = "{SDSS IV MaNGA - sSFR profiles and the slow quenching of discs in green valley galaxies}",
      journal = {\mnras},
         year = 2018,
        month = jul,
       volume = {477},
       number = {3},
        pages = {3014-3029},
          doi = {10.1093/mnras/sty768},
archivePrefix = {arXiv},
       eprint = {1710.05034},
 primaryClass = {astro-ph.GA},
       adsurl = {https://ui.adsabs.harvard.edu/abs/2018MNRAS.477.3014B}
}

@ARTICLE{1993ARA&A..31..473A,
       author = {{Antonucci}, Robert},
        title = "{Unified models for active galactic nuclei and quasars.}",
      journal = {\araa},
         year = 1993,
        month = jan,
       volume = {31},
        pages = {473-521},
          doi = {10.1146/annurev.aa.31.090193.002353},
       adsurl = {https://ui.adsabs.harvard.edu/abs/1993ARA&A..31..473A}
}

@ARTICLE{2020ApJ...901..159C,
       author = {{Comerford}, Julia M. and {Negus}, James and {M{\"u}ller-S{\'a}nchez}, Francisco and {Eracleous}, Michael and {Wylezalek}, Dominika and {Storchi-Bergmann}, Thaisa and {Greene}, Jenny E. and {Barrows}, R. Scott and {Nevin}, Rebecca and {Roy}, Namrata and {Stemo}, Aaron},
        title = "{A Catalog of 406 AGNs in MaNGA: A Connection between Radio-mode AGNs and Star Formation Quenching}",
      journal = {\apj},
         year = 2020,
        month = oct,
       volume = {901},
       number = {2},
          eid = {159},
        pages = {159},
          doi = {10.3847/1538-4357/abb2ae},
archivePrefix = {arXiv},
       eprint = {2008.11210},
 primaryClass = {astro-ph.GA},
       adsurl = {https://ui.adsabs.harvard.edu/abs/2020ApJ...901..159C}
}

@ARTICLE{2025MNRAS.539.1856H,
       author = {{Hardcastle}, M.~J. and {Pierce}, J.~C.~S. and {Duncan}, K.~J. and {G{\"u}rkan}, G. and {Gong}, Y. and {Horton}, M.~A. and {Mingo}, B. and {R{\"o}ttgering}, H.~J.~A. and {Smith}, D.~J.~B.},
        title = "{Radio AGN selection in LoTSS DR2}",
      journal = {\mnras},
         year = 2025,
        month = may,
       volume = {539},
       number = {2},
        pages = {1856-1878},
          doi = {10.1093/mnras/staf622},
archivePrefix = {arXiv},
       eprint = {2504.09303},
 primaryClass = {astro-ph.GA},
       adsurl = {https://ui.adsabs.harvard.edu/abs/2025MNRAS.539.1856H}
}

@ARTICLE{2018MNRAS.480.3201K,
       author = {{Kauffmann}, Guinevere},
        title = "{Properties of AGNs selected by their mid-IR colours: evidence for a physically distinct mode of black hole growth}",
      journal = {\mnras},
         year = 2018,
        month = nov,
       volume = {480},
       number = {3},
        pages = {3201-3214},
          doi = {10.1093/mnras/sty2029},
archivePrefix = {arXiv},
       eprint = {1808.03145},
 primaryClass = {astro-ph.GA},
       adsurl = {https://ui.adsabs.harvard.edu/abs/2018MNRAS.480.3201K}
}

@ARTICLE{2013MNRAS.433..622M,
       author = {{Mullaney}, J.~R. and {Alexander}, D.~M. and {Fine}, S. and {Goulding}, A.~D. and {Harrison}, C.~M. and {Hickox}, R.~C.},
        title = "{Narrow-line region gas kinematics of 24 264 optically selected AGN: the radio connection}",
      journal = {\mnras},
         year = 2013,
        month = jul,
       volume = {433},
       number = {1},
        pages = {622-638},
          doi = {10.1093/mnras/stt751},
archivePrefix = {arXiv},
       eprint = {1305.0263},
 primaryClass = {astro-ph.CO},
       adsurl = {https://ui.adsabs.harvard.edu/abs/2013MNRAS.433..622M}
}

@ARTICLE{2024Natur.632.1009W,
       author = {{Wang}, Tao and {Xu}, Ke and {Wu}, Yuxuan and {Shi}, Yong and {Elbaz}, David and {Ho}, Luis C. and {Zhang}, Zhi-Yu and {Gu}, Qiusheng and {Wang}, Yijun and {Shu}, Chenggang and {Yuan}, Feng and {Xia}, Xiaoyang and {Wang}, Kai},
        title = "{Black holes regulate cool gas accretion in massive galaxies}",
      journal = {\nat},
         year = 2024,
        month = aug,
       volume = {632},
       number = {8027},
        pages = {1009-1013},
          doi = {10.1038/s41586-024-07821-2},
archivePrefix = {arXiv},
       eprint = {2311.07653},
 primaryClass = {astro-ph.GA},
       adsurl = {https://ui.adsabs.harvard.edu/abs/2024Natur.632.1009W}
}

@ARTICLE{2012ARA&A..50..455F,
       author = {{Fabian}, A.~C.},
        title = "{Observational Evidence of Active Galactic Nuclei Feedback}",
      journal = {\araa},
         year = 2012,
        month = sep,
       volume = {50},
        pages = {455-489},
          doi = {10.1146/annurev-astro-081811-125521},
archivePrefix = {arXiv},
       eprint = {1204.4114},
 primaryClass = {astro-ph.CO},
       adsurl = {https://ui.adsabs.harvard.edu/abs/2012ARA&A..50..455F}
}

@ARTICLE{2025A&A...695A.185Z,
       author = {{Zanchettin}, M.~V. and {Ramos Almeida}, C. and {Audibert}, A. and {Acosta-Pulido}, J.~A. and {Cezar}, P.~H. and {Hicks}, E. and {Lapi}, A. and {Mullaney}, J.},
        title = "{Unveiling the warm molecular outflow component of type-2 quasars with SINFONI}",
      journal = {\aap},
         year = 2025,
        month = mar,
       volume = {695},
          eid = {A185},
        pages = {A185},
          doi = {10.1051/0004-6361/202453224},
archivePrefix = {arXiv},
       eprint = {2502.12800},
 primaryClass = {astro-ph.GA},
       adsurl = {https://ui.adsabs.harvard.edu/abs/2025A&A...695A.185Z}
}

@ARTICLE{2014MNRAS.441.3306H,
       author = {{Harrison}, C.~M. and {Alexander}, D.~M. and {Mullaney}, J.~R. and {Swinbank}, A.~M.},
        title = "{Kiloparsec-scale outflows are prevalent among luminous AGN: outflows and feedback in the context of the overall AGN population}",
      journal = {\mnras},
         year = 2014,
        month = jul,
       volume = {441},
       number = {4},
        pages = {3306-3347},
          doi = {10.1093/mnras/stu515},
archivePrefix = {arXiv},
       eprint = {1403.3086},
 primaryClass = {astro-ph.GA},
       adsurl = {https://ui.adsabs.harvard.edu/abs/2014MNRAS.441.3306H}
}

@ARTICLE{2010MNRAS.402.2211A,
       author = {{Alexander}, D.~M. and {Swinbank}, A.~M. and {Smail}, Ian and {McDermid}, R. and {Nesvadba}, N.~P.~H.},
        title = "{Searching for evidence of energetic feedback in distant galaxies: a galaxy wide outflow in a z \raisebox{-0.5ex}\textasciitilde 2 ultraluminous infrared galaxy}",
      journal = {\mnras},
         year = 2010,
        month = mar,
       volume = {402},
       number = {4},
        pages = {2211-2220},
          doi = {10.1111/j.1365-2966.2009.16046.x},
archivePrefix = {arXiv},
       eprint = {0911.0014},
 primaryClass = {astro-ph.CO},
       adsurl = {https://ui.adsabs.harvard.edu/abs/2010MNRAS.402.2211A}
}

@ARTICLE{2005MNRAS.361..776S,
       author = {{Springel}, Volker and {Di Matteo}, Tiziana and {Hernquist}, Lars},
        title = "{Modelling feedback from stars and black holes in galaxy mergers}",
      journal = {\mnras},
         year = 2005,
        month = aug,
       volume = {361},
       number = {3},
        pages = {776-794},
          doi = {10.1111/j.1365-2966.2005.09238.x},
archivePrefix = {arXiv},
       eprint = {astro-ph/0411108},
 primaryClass = {astro-ph},
       adsurl = {https://ui.adsabs.harvard.edu/abs/2005MNRAS.361..776S}
}

@ARTICLE{1990ApJS...74..833H,
       author = {{Heckman}, Timothy M. and {Armus}, Lee and {Miley}, George K.},
        title = "{On the Nature and Implications of Starburst-driven Galactic Superwinds}",
      journal = {\apjs},
         year = 1990,
        month = dec,
       volume = {74},
        pages = {833},
          doi = {10.1086/191522},
       adsurl = {https://ui.adsabs.harvard.edu/abs/1990ApJS...74..833H}
}

@ARTICLE{2005Natur.433..604D,
       author = {{Di Matteo}, Tiziana and {Springel}, Volker and {Hernquist}, Lars},
        title = "{Energy input from quasars regulates the growth and activity of black holes and their host galaxies}",
      journal = {\nat},
         year = 2005,
        month = feb,
       volume = {433},
       number = {7026},
        pages = {604-607},
          doi = {10.1038/nature03335},
archivePrefix = {arXiv},
       eprint = {astro-ph/0502199},
 primaryClass = {astro-ph},
       adsurl = {https://ui.adsabs.harvard.edu/abs/2005Natur.433..604D}
}

@ARTICLE{2006ApJS..163....1H,
       author = {{Hopkins}, Philip F. and {Hernquist}, Lars and {Cox}, Thomas J. and {Di Matteo}, Tiziana and {Robertson}, Brant and {Springel}, Volker},
        title = "{A Unified, Merger-driven Model of the Origin of Starbursts, Quasars, the Cosmic X-Ray Background, Supermassive Black Holes, and Galaxy Spheroids}",
      journal = {\apjs},
         year = 2006,
        month = mar,
       volume = {163},
       number = {1},
        pages = {1-49},
          doi = {10.1086/499298},
archivePrefix = {arXiv},
       eprint = {astro-ph/0506398},
 primaryClass = {astro-ph},
       adsurl = {https://ui.adsabs.harvard.edu/abs/2006ApJS..163....1H}
}

@ARTICLE{2017NatAs...1E.165H,
       author = {{Harrison}, C.~M.},
        title = "{Impact of supermassive black hole growth on star formation}",
      journal = {Nature Astronomy},
         year = 2017,
        month = jul,
       volume = {1},
          eid = {0165},
        pages = {0165},
          doi = {10.1038/s41550-017-0165},
archivePrefix = {arXiv},
       eprint = {1703.06889},
 primaryClass = {astro-ph.GA},
       adsurl = {https://ui.adsabs.harvard.edu/abs/2017NatAs...1E.165H}
}

@ARTICLE{1998AJ....115.2285M,
       author = {{Magorrian}, John and {Tremaine}, Scott and {Richstone}, Douglas and {Bender}, Ralf and {Bower}, Gary and {Dressler}, Alan and {Faber}, S.~M. and {Gebhardt}, Karl and {Green}, Richard and {Grillmair}, Carl and {Kormendy}, John and {Lauer}, Tod},
        title = "{The Demography of Massive Dark Objects in Galaxy Centers}",
      journal = {\aj},
         year = 1998,
        month = jun,
       volume = {115},
       number = {6},
        pages = {2285-2305},
          doi = {10.1086/300353},
archivePrefix = {arXiv},
       eprint = {astro-ph/9708072},
 primaryClass = {astro-ph},
       adsurl = {https://ui.adsabs.harvard.edu/abs/1998AJ....115.2285M}
}

@ARTICLE{2015ApJ...813...82R,
       author = {{Reines}, Amy E. and {Volonteri}, Marta},
        title = "{Relations between Central Black Hole Mass and Total Galaxy Stellar Mass in the Local Universe}",
      journal = {\apj},
         year = 2015,
        month = nov,
       volume = {813},
       number = {2},
          eid = {82},
        pages = {82},
          doi = {10.1088/0004-637X/813/2/82},
archivePrefix = {arXiv},
       eprint = {1508.06274},
 primaryClass = {astro-ph.GA},
       adsurl = {https://ui.adsabs.harvard.edu/abs/2015ApJ...813...82R}
}

@ARTICLE{2013MNRAS.436.2708M,
       author = {{McGee}, Sean L.},
        title = "{The strong environmental dependence of black hole scaling relations}",
      journal = {\mnras},
         year = 2013,
        month = dec,
       volume = {436},
       number = {3},
        pages = {2708-2721},
          doi = {10.1093/mnras/stt1769},
archivePrefix = {arXiv},
       eprint = {1302.6237},
 primaryClass = {astro-ph.CO},
       adsurl = {https://ui.adsabs.harvard.edu/abs/2013MNRAS.436.2708M}
}

@ARTICLE{2025MNRAS.536..777V,
       author = {{Vani}, Akash and {Ayromlou}, Mohammadreza and {Kauffmann}, Guinevere and {Springel}, Volker},
        title = "{Probing galaxy evolution from z = 0 to z ≃ 10 through galaxy scaling relations in three L-GALAXIES flavours}",
      journal = {\mnras},
         year = 2025,
        month = jan,
       volume = {536},
       number = {1},
        pages = {777-806},
          doi = {10.1093/mnras/stae2625},
archivePrefix = {arXiv},
       eprint = {2408.00824},
 primaryClass = {astro-ph.GA},
       adsurl = {https://ui.adsabs.harvard.edu/abs/2025MNRAS.536..777V}
}

@ARTICLE{2024MNRAS.528.4891G,
       author = {{Goubert}, Paul H. and {Bluck}, Asa F.~L. and {Piotrowska}, Joanna M. and {Maiolino}, Roberto},
        title = "{The role of environment and AGN feedback in quenching local galaxies: comparing cosmological hydrodynamical simulations to the SDSS}",
      journal = {\mnras},
         year = 2024,
        month = mar,
       volume = {528},
       number = {3},
        pages = {4891-4921},
          doi = {10.1093/mnras/stae269},
archivePrefix = {arXiv},
       eprint = {2401.12953},
 primaryClass = {astro-ph.GA},
       adsurl = {https://ui.adsabs.harvard.edu/abs/2024MNRAS.528.4891G}
}

@ARTICLE{2015A&A...578A.110A,
       author = {{Argudo-Fern{\'a}ndez}, M. and {Verley}, S. and {Bergond}, G. and {Duarte Puertas}, S. and {Ramos Carmona}, E. and {Sabater}, J. and {Fern{\'a}ndez Lorenzo}, M. and {Espada}, D. and {Sulentic}, J. and {Ruiz}, J.~E. and {Leon}, S.},
        title = "{Catalogues of isolated galaxies, isolated pairs, and isolated triplets in the local Universe}",
      journal = {\aap},
         year = 2015,
        month = jun,
       volume = {578},
          eid = {A110},
        pages = {A110},
          doi = {10.1051/0004-6361/201526016},
archivePrefix = {arXiv},
       eprint = {1504.00117},
 primaryClass = {astro-ph.GA},
       adsurl = {https://ui.adsabs.harvard.edu/abs/2015A&A...578A.110A}
}

@ARTICLE{2010ApJ...720.1066C,
       author = {{Cavagnolo}, K.~W. and {McNamara}, B.~R. and {Nulsen}, P.~E.~J. and {Carilli}, C.~L. and {Jones}, C. and {B{\^\i}rzan}, L.},
        title = "{A Relationship Between AGN Jet Power and Radio Power}",
      journal = {\apj},
         year = 2010,
        month = sep,
       volume = {720},
       number = {2},
        pages = {1066-1072},
          doi = {10.1088/0004-637X/720/2/1066},
archivePrefix = {arXiv},
       eprint = {1006.5699},
 primaryClass = {astro-ph.CO},
       adsurl = {https://ui.adsabs.harvard.edu/abs/2010ApJ...720.1066C}
}

@ARTICLE{2010ApJ...720..786L,
       author = {{LaMassa}, Stephanie M. and {Heckman}, Tim M. and {Ptak}, Andrew and {Martins}, Lucimara and {Wild}, Vivienne and {Sonnentrucker}, Paule},
        title = "{Indicators of Intrinsic Active Galactic Nucleus Luminosity: A Multi-wavelength Approach}",
      journal = {\apj},
         year = 2010,
        month = sep,
       volume = {720},
       number = {1},
        pages = {786-810},
          doi = {10.1088/0004-637X/720/1/786},
archivePrefix = {arXiv},
       eprint = {1007.0900},
 primaryClass = {astro-ph.CO},
       adsurl = {https://ui.adsabs.harvard.edu/abs/2010ApJ...720..786L}
}

@ARTICLE{2013MNRAS.432.1709C,
       author = {{Cappellari}, Michele and {Scott}, Nicholas and {Alatalo}, Katherine and {Blitz}, Leo and {Bois}, Maxime and {Bournaud}, Fr{\'e}d{\'e}ric and {Bureau}, M. and {Crocker}, Alison F. and {Davies}, Roger L. and {Davis}, Timothy A. and {de Zeeuw}, P.~T. and {Duc}, Pierre-Alain and {Emsellem}, Eric and {Khochfar}, Sadegh and {Krajnovi{\'c}}, Davor and {Kuntschner}, Harald and {McDermid}, Richard M. and {Morganti}, Raffaella and {Naab}, Thorsten and {Oosterloo}, Tom and {Sarzi}, Marc and {Serra}, Paolo and {Weijmans}, Anne-Marie and {Young}, Lisa M.},
        title = "{The ATLAS$^{3D}$ project - XV. Benchmark for early-type galaxies scaling relations from 260 dynamical models: mass-to-light ratio, dark matter, Fundamental Plane and Mass Plane}",
      journal = {\mnras},
         year = 2013,
        month = jul,
       volume = {432},
       number = {3},
        pages = {1709-1741},
          doi = {10.1093/mnras/stt562},
archivePrefix = {arXiv},
       eprint = {1208.3522},
 primaryClass = {astro-ph.CO},
       adsurl = {https://ui.adsabs.harvard.edu/abs/2013MNRAS.432.1709C}
}

@ARTICLE{2013ARA&A..51..511K,
       author = {{Kormendy}, John and {Ho}, Luis C.},
        title = "{Coevolution (Or Not) of Supermassive Black Holes and Host Galaxies}",
      journal = {\araa},
         year = 2013,
        month = aug,
       volume = {51},
       number = {1},
        pages = {511-653},
          doi = {10.1146/annurev-astro-082708-101811},
archivePrefix = {arXiv},
       eprint = {1304.7762},
 primaryClass = {astro-ph.CO},
       adsurl = {https://ui.adsabs.harvard.edu/abs/2013ARA&A..51..511K}
}

@ARTICLE{2003ApJ...589L..21M,
       author = {{Marconi}, Alessandro and {Hunt}, Leslie K.},
        title = "{The Relation between Black Hole Mass, Bulge Mass, and Near-Infrared Luminosity}",
      journal = {\apjl},
         year = 2003,
        month = may,
       volume = {589},
       number = {1},
        pages = {L21-L24},
          doi = {10.1086/375804},
archivePrefix = {arXiv},
       eprint = {astro-ph/0304274},
 primaryClass = {astro-ph},
       adsurl = {https://ui.adsabs.harvard.edu/abs/2003ApJ...589L..21M}
}

@ARTICLE{2013MNRAS.436.2576L,
       author = {{Liu}, Guilin and {Zakamska}, Nadia L. and {Greene}, Jenny E. and {Nesvadba}, Nicole P.~H. and {Liu}, Xin},
        title = "{Observations of feedback from radio-quiet quasars - II. Kinematics of ionized gas nebulae}",
      journal = {\mnras},
         year = 2013,
        month = dec,
       volume = {436},
       number = {3},
        pages = {2576-2597},
          doi = {10.1093/mnras/stt1755},
archivePrefix = {arXiv},
       eprint = {1305.6922},
 primaryClass = {astro-ph.CO},
       adsurl = {https://ui.adsabs.harvard.edu/abs/2013MNRAS.436.2576L}
}

@ARTICLE{2007MNRAS.380..877S,
       author = {{Sijacki}, Debora and {Springel}, Volker and {Di Matteo}, Tiziana and {Hernquist}, Lars},
        title = "{A unified model for AGN feedback in cosmological simulations of structure formation}",
      journal = {\mnras},
         year = 2007,
        month = sep,
       volume = {380},
       number = {3},
        pages = {877-900},
          doi = {10.1111/j.1365-2966.2007.12153.x},
archivePrefix = {arXiv},
       eprint = {0705.2238},
 primaryClass = {astro-ph},
       adsurl = {https://ui.adsabs.harvard.edu/abs/2007MNRAS.380..877S}
}

@ARTICLE{2014MNRAS.442..784Z,
       author = {{Zakamska}, Nadia L. and {Greene}, Jenny E.},
        title = "{Quasar feedback and the origin of radio emission in radio-quiet quasars}",
      journal = {\mnras},
         year = 2014,
        month = jul,
       volume = {442},
       number = {1},
        pages = {784-804},
          doi = {10.1093/mnras/stu842},
archivePrefix = {arXiv},
       eprint = {1402.6736},
 primaryClass = {astro-ph.GA},
       adsurl = {https://ui.adsabs.harvard.edu/abs/2014MNRAS.442..784Z}
}

@ARTICLE{1995ApJ...452..710N,
       author = {{Narayan}, Ramesh and {Yi}, Insu},
        title = "{Advection-dominated Accretion: Underfed Black Holes and Neutron Stars}",
      journal = {\apj},
         year = 1995,
        month = oct,
       volume = {452},
        pages = {710},
          doi = {10.1086/176343},
archivePrefix = {arXiv},
       eprint = {astro-ph/9411059},
 primaryClass = {astro-ph},
       adsurl = {https://ui.adsabs.harvard.edu/abs/1995ApJ...452..710N}
}

@ARTICLE{1973A&A....24..337S,
       author = {{Shakura}, N.~I. and {Sunyaev}, R.~A.},
        title = "{Black holes in binary systems. Observational appearance.}",
      journal = {\aap},
         year = 1973,
        month = jan,
       volume = {24},
        pages = {337-355},
       adsurl = {https://ui.adsabs.harvard.edu/abs/1973A&A....24..337S}
}

@ARTICLE{2024A&A...691A.124A,
       author = {{Alb{\'a}n}, M. and {Wylezalek}, D. and {Comerford}, J.~M. and {Greene}, J.~E. and {Riffel}, R.~A.},
        title = "{Mapping AGN winds: A connection between radio-mode AGNs and the AGN feedback cycle}",
      journal = {\aap},
         year = 2024,
        month = nov,
       volume = {691},
          eid = {A124},
        pages = {A124},
          doi = {10.1051/0004-6361/202451738},
archivePrefix = {arXiv},
       eprint = {2408.16831},
 primaryClass = {astro-ph.GA},
       adsurl = {https://ui.adsabs.harvard.edu/abs/2024A&A...691A.124A}
}

@ARTICLE{2025A&A...698A..99K,
       author = {{Kukreti}, Pranav and {Wylezalek}, Dominika and {Alb{\'a}n}, Marco and {Dall'Agnol de Oliveira}, Bruno},
        title = "{Feedback from low-to-moderate-luminosity radio-active galactic nuclei with MaNGA}",
      journal = {\aap},
         year = 2025,
        month = jun,
       volume = {698},
          eid = {A99},
        pages = {A99},
          doi = {10.1051/0004-6361/202453307},
archivePrefix = {arXiv},
       eprint = {2503.20889},
 primaryClass = {astro-ph.GA},
       adsurl = {https://ui.adsabs.harvard.edu/abs/2025A&A...698A..99K}
}

@ARTICLE{2020MNRAS.492.4680W,
       author = {{Wylezalek}, Dominika and {Flores}, Anthony M. and {Zakamska}, Nadia L. and {Greene}, Jenny E. and {Riffel}, Rogemar A.},
        title = "{Ionized gas outflow signatures in SDSS-IV MaNGA active galactic nuclei}",
      journal = {\mnras},
         year = 2020,
        month = mar,
       volume = {492},
       number = {4},
        pages = {4680-4696},
          doi = {10.1093/mnras/staa062},
archivePrefix = {arXiv},
       eprint = {1911.10212},
 primaryClass = {astro-ph.GA},
       adsurl = {https://ui.adsabs.harvard.edu/abs/2020MNRAS.492.4680W}
}

@ARTICLE{2011MNRAS.415.2359N,
       author = {{Nesvadba}, N.~P.~H. and {Polletta}, M. and {Lehnert}, M.~D. and {Bergeron}, J. and {De Breuck}, C. and {Lagache}, G. and {Omont}, A.},
        title = "{The dynamics of the ionized and molecular interstellar medium in powerful obscured quasars at z{\ensuremath{\geq}} 3.5}",
      journal = {\mnras},
         year = 2011,
        month = aug,
       volume = {415},
       number = {3},
        pages = {2359-2372},
          doi = {10.1111/j.1365-2966.2011.18862.x},
archivePrefix = {arXiv},
       eprint = {1104.0937},
 primaryClass = {astro-ph.CO},
       adsurl = {https://ui.adsabs.harvard.edu/abs/2011MNRAS.415.2359N}
}

@ARTICLE{2016ApJ...817..108W,
       author = {{Woo}, Jong-Hak and {Bae}, Hyun-Jin and {Son}, Donghoon and {Karouzos}, Marios},
        title = "{The Prevalence of Gas Outflows in Type 2 AGNs}",
      journal = {\apj},
         year = 2016,
        month = feb,
       volume = {817},
       number = {2},
          eid = {108},
        pages = {108},
          doi = {10.3847/0004-637X/817/2/108},
archivePrefix = {arXiv},
       eprint = {1511.05142},
 primaryClass = {astro-ph.GA},
       adsurl = {https://ui.adsabs.harvard.edu/abs/2016ApJ...817..108W}
}

@ARTICLE{2022A&A...659A.160B,
       author = {{Bluck}, Asa F.~L. and {Maiolino}, Roberto and {Brownson}, Simcha and {Conselice}, Christopher J. and {Ellison}, Sara L. and {Piotrowska}, Joanna M. and {Thorp}, Mallory D.},
        title = "{The quenching of galaxies, bulges, and disks since cosmic noon. A machine learning approach for identifying causality in astronomical data}",
      journal = {\aap},
         year = 2022,
        month = mar,
       volume = {659},
          eid = {A160},
        pages = {A160},
          doi = {10.1051/0004-6361/202142643},
archivePrefix = {arXiv},
       eprint = {2201.07814},
 primaryClass = {astro-ph.GA},
       adsurl = {https://ui.adsabs.harvard.edu/abs/2022A&A...659A.160B}
}

@ARTICLE{2016MNRAS.463.2799I,
       author = {{Ibarra-Medel}, H{\'e}ctor J. and {S{\'a}nchez}, Sebasti{\'a}n F. and {Avila-Reese}, Vladimir and {Hern{\'a}ndez-Toledo}, H{\'e}ctor M. and {Gonz{\'a}lez}, J. Jes{\'u}s and {Drory}, Niv and {Bundy}, Kevin and {Bizyaev}, Dmitry and {Cano-D{\'\i}az}, Mariana and {Malanushenko}, Elena and {Pan}, Kaike and {Roman-Lopes}, Alexandre and {Thomas}, Daniel},
        title = "{SDSS IV MaNGA: the global and local stellar mass assemby histories of galaxies}",
      journal = {\mnras},
         year = 2016,
        month = dec,
       volume = {463},
       number = {3},
        pages = {2799-2818},
          doi = {10.1093/mnras/stw2126},
archivePrefix = {arXiv},
       eprint = {1609.01304},
 primaryClass = {astro-ph.GA},
       adsurl = {https://ui.adsabs.harvard.edu/abs/2016MNRAS.463.2799I}
}

@ARTICLE{2002ApJS..142...35K,
       author = {{Kewley}, L.~J. and {Dopita}, M.~A.},
        title = "{Using Strong Lines to Estimate Abundances in Extragalactic H II Regions and Starburst Galaxies}",
      journal = {\apjs},
         year = 2002,
        month = sep,
       volume = {142},
       number = {1},
        pages = {35-52},
          doi = {10.1086/341326},
archivePrefix = {arXiv},
       eprint = {astro-ph/0206495},
 primaryClass = {astro-ph},
       adsurl = {https://ui.adsabs.harvard.edu/abs/2002ApJS..142...35K}
}

@ARTICLE{1999ApJ...527...54B,
       author = {{Balogh}, Michael L. and {Morris}, Simon L. and {Yee}, H.~K.~C. and {Carlberg}, R.~G. and {Ellingson}, Erica},
        title = "{Differential Galaxy Evolution in Cluster and Field Galaxies at z\raisebox{-0.5ex}\textasciitilde0.3}",
      journal = {\apj},
         year = 1999,
        month = dec,
       volume = {527},
       number = {1},
        pages = {54-79},
          doi = {10.1086/308056},
archivePrefix = {arXiv},
       eprint = {astro-ph/9906470},
 primaryClass = {astro-ph},
       adsurl = {https://ui.adsabs.harvard.edu/abs/1999ApJ...527...54B}
}

@ARTICLE{2025A&A...697A.196I,
       author = {{Igo}, Z. and {Merloni}, A.},
        title = "{The global energetics of radio AGN kinetic feedback in the local Universe}",
      journal = {\aap},
         year = 2025,
        month = may,
       volume = {697},
          eid = {A196},
        pages = {A196},
          doi = {10.1051/0004-6361/202452888},
archivePrefix = {arXiv},
       eprint = {2504.00090},
 primaryClass = {astro-ph.GA},
       adsurl = {https://ui.adsabs.harvard.edu/abs/2025A&A...697A.196I}
}

@ARTICLE{2024A&A...686A..43I,
       author = {{Igo}, Z. and {Merloni}, A. and {Hoang}, D. and {Buchner}, J. and {Liu}, T. and {Salvato}, M. and {Arcodia}, R. and {Bellstedt}, S. and {Br{\"u}ggen}, M. and {Croston}, J.~H. and et al.},
        title = "{The LOFAR - eFEDS survey: The incidence of radio and X-ray AGN and the disk-jet connection}",
      journal = {\aap},
         year = 2024,
        month = jun,
       volume = {686},
          eid = {A43},
        pages = {A43},
          doi = {10.1051/0004-6361/202349069},
archivePrefix = {arXiv},
       eprint = {2402.16943},
 primaryClass = {astro-ph.HE},
       adsurl = {https://ui.adsabs.harvard.edu/abs/2024A&A...686A..43I}
}

@ARTICLE{2019A&A...622A..17S,
       author = {{Sabater}, J. and {Best}, P.~N. and {Hardcastle}, M.~J. and {Shimwell}, T.~W. and {Tasse}, C. and {Williams}, W.~L. and {Br{\"u}ggen}, M. and {Cochrane}, R.~K. and {Croston}, J.~H. and {de Gasperin}, F. and {Duncan}, K.~J. and {G{\"u}rkan}, G. and {Mechev}, A.~P. and {Morabito}, L.~K. and {Prandoni}, I. and {R{\"o}ttgering}, H.~J.~A. and {Smith}, D.~J.~B. and {Harwood}, J.~J. and {Mingo}, B. and {Mooney}, S. and {Saxena}, A.},
        title = "{The LoTSS view of radio AGN in the local Universe. The most massive galaxies are always switched on}",
      journal = {\aap},
         year = 2019,
        month = feb,
       volume = {622},
          eid = {A17},
        pages = {A17},
          doi = {10.1051/0004-6361/201833883},
archivePrefix = {arXiv},
       eprint = {1811.05528},
 primaryClass = {astro-ph.GA},
       adsurl = {https://ui.adsabs.harvard.edu/abs/2019A&A...622A..17S}
}

@ARTICLE{2021ApJ...906...38Z,
       author = {{Zhuang}, Ming-Yang and {Ho}, Luis C. and {Shangguan}, Jinyi},
        title = "{Black Hole Accretion Correlates with Star Formation Rate and Star Formation Efficiency in Nearby Luminous Type 1 Active Galaxies}",
      journal = {\apj},
         year = 2021,
        month = jan,
       volume = {906},
       number = {1},
          eid = {38},
        pages = {38},
          doi = {10.3847/1538-4357/abc94d},
archivePrefix = {arXiv},
       eprint = {2007.11285},
 primaryClass = {astro-ph.GA},
       adsurl = {https://ui.adsabs.harvard.edu/abs/2021ApJ...906...38Z}
}

@ARTICLE{2018MNRAS.478.4238D,
       author = {{Dai}, Y. Sophia and {Wilkes}, Belinda J. and {Bergeron}, Jacqueline and {Kuraszkiewicz}, Joanna and {Omont}, Alain and {Atanas}, Adam and {Teplitz}, Harry I.},
        title = "{Is there a relationship between AGN and star formationin IR-bright AGNs?}",
      journal = {\mnras},
         year = 2018,
        month = aug,
       volume = {478},
       number = {3},
        pages = {4238-4254},
          doi = {10.1093/mnras/sty1341},
archivePrefix = {arXiv},
       eprint = {1511.06761},
 primaryClass = {astro-ph.GA},
       adsurl = {https://ui.adsabs.harvard.edu/abs/2018MNRAS.478.4238D}
}

@ARTICLE{2008ApJS..175..356H,
       author = {{Hopkins}, Philip F. and {Hernquist}, Lars and {Cox}, Thomas J. and {Kere{\v{s}}}, Du{\v{s}}an},
        title = "{A Cosmological Framework for the Co-Evolution of Quasars, Supermassive Black Holes, and Elliptical Galaxies. I. Galaxy Mergers and Quasar Activity}",
      journal = {\apjs},
         year = 2008,
        month = apr,
       volume = {175},
       number = {2},
        pages = {356-389},
          doi = {10.1086/524362},
archivePrefix = {arXiv},
       eprint = {0706.1243},
 primaryClass = {astro-ph},
       adsurl = {https://ui.adsabs.harvard.edu/abs/2008ApJS..175..356H}
}

@ARTICLE{2009ARA&A..47..159B,
       author = {{Blanton}, Michael R. and {Moustakas}, John},
        title = "{Physical Properties and Environments of Nearby Galaxies}",
      journal = {\araa},
         year = 2009,
        month = sep,
       volume = {47},
       number = {1},
        pages = {159-210},
          doi = {10.1146/annurev-astro-082708-101734},
archivePrefix = {arXiv},
       eprint = {0908.3017},
 primaryClass = {astro-ph.GA},
       adsurl = {https://ui.adsabs.harvard.edu/abs/2009ARA&A..47..159B}
}

@ARTICLE{2014MNRAS.440..889S,
       author = {{Schawinski}, Kevin and {Urry}, C. Megan and {Simmons}, Brooke D. and {Fortson}, Lucy and {Kaviraj}, Sugata and {Keel}, William C. and {Lintott}, Chris J. and {Masters}, Karen L. and {Nichol}, Robert C. and {Sarzi}, Marc and {Skibba}, Ramin and {Treister}, Ezequiel and {Willett}, Kyle W. and {Wong}, O. Ivy and {Yi}, Sukyoung K.},
        title = "{The green valley is a red herring: Galaxy Zoo reveals two evolutionary pathways towards quenching of star formation in early- and late-type galaxies}",
      journal = {\mnras},
         year = 2014,
        month = may,
       volume = {440},
       number = {1},
        pages = {889-907},
          doi = {10.1093/mnras/stu327},
archivePrefix = {arXiv},
       eprint = {1402.4814},
 primaryClass = {astro-ph.GA},
       adsurl = {https://ui.adsabs.harvard.edu/abs/2014MNRAS.440..889S}
}

@ARTICLE{2007ApJ...660L..43N,
       author = {{Noeske}, K.~G. and {Weiner}, B.~J. and {Faber}, S.~M. and {Papovich}, C. and {Koo}, D.~C. and {Somerville}, R.~S. and {Bundy}, K. and {Conselice}, C.~J. and {Newman}, J.~A. and {Schiminovich}, D. and {Le Floc'h}, E. and {Coil}, A.~L. and {Rieke}, G.~H. and {Lotz}, J.~M. and {Primack}, J.~R. and {Barmby}, P. and {Cooper}, M.~C. and {Davis}, M. and {Ellis}, R.~S. and {Fazio}, G.~G. and {Guhathakurta}, P. and {Huang}, J. and {Kassin}, S.~A. and {Martin}, D.~C. and {Phillips}, A.~C. and {Rich}, R.~M. and {Small}, T.~A. and {Willmer}, C.~N.~A. and {Wilson}, G.},
        title = "{Star Formation in AEGIS Field Galaxies since z=1.1: The Dominance of Gradually Declining Star Formation, and the Main Sequence of Star-forming Galaxies}",
      journal = {\apjl},
         year = 2007,
        month = may,
       volume = {660},
       number = {1},
        pages = {L43-L46},
          doi = {10.1086/517926},
archivePrefix = {arXiv},
       eprint = {astro-ph/0701924},
 primaryClass = {astro-ph},
       adsurl = {https://ui.adsabs.harvard.edu/abs/2007ApJ...660L..43N}
}

@ARTICLE{2002AJ....124..675C,
       author = {{Condon}, J.~J. and {Cotton}, W.~D. and {Broderick}, J.~J.},
        title = "{Radio Sources and Star Formation in the Local Universe}",
      journal = {\aj},
         year = 2002,
        month = aug,
       volume = {124},
       number = {2},
        pages = {675-689},
          doi = {10.1086/341650},
       adsurl = {https://ui.adsabs.harvard.edu/abs/2002AJ....124..675C}
}

@ARTICLE{2009MNRAS.397..135K,
       author = {{Kauffmann}, Guinevere and {Heckman}, Timothy M.},
        title = "{Feast and Famine: regulation of black hole growth in low-redshift galaxies}",
      journal = {\mnras},
         year = 2009,
        month = jul,
       volume = {397},
       number = {1},
        pages = {135-147},
          doi = {10.1111/j.1365-2966.2009.14960.x},
archivePrefix = {arXiv},
       eprint = {0812.1224},
 primaryClass = {astro-ph},
       adsurl = {https://ui.adsabs.harvard.edu/abs/2009MNRAS.397..135K}
}

@ARTICLE{2013ApJ...764..184M,
       author = {{McConnell}, Nicholas J. and {Ma}, Chung-Pei},
        title = "{Revisiting the Scaling Relations of Black Hole Masses and Host Galaxy Properties}",
      journal = {\apj},
         year = 2013,
        month = feb,
       volume = {764},
       number = {2},
          eid = {184},
        pages = {184},
          doi = {10.1088/0004-637X/764/2/184},
archivePrefix = {arXiv},
       eprint = {1211.2816},
 primaryClass = {astro-ph.CO},
       adsurl = {https://ui.adsabs.harvard.edu/abs/2013ApJ...764..184M}
}

@ARTICLE{2022NewA...9701895L,
       author = {{Lacerda}, Eduardo A.~D. and {S{\'a}nchez}, S.~F. and {Mej{\'\i}a-Narv{\'a}ez}, A. and {Camps-Fari{\~n}a}, A. and {Espinosa-Ponce}, C. and {Barrera-Ballesteros}, J.~K. and {Ibarra-Medel}, H. and {Lugo-Aranda}, A.~Z.},
        title = "{pyFIT3D and pyPipe3D - The new version of the integral field spectroscopy data analysis pipeline}",
      journal = {\na},
         year = 2022,
        month = nov,
       volume = {97},
          eid = {101895},
        pages = {101895},
          doi = {10.1016/j.newast.2022.101895},
archivePrefix = {arXiv},
       eprint = {2202.08027},
 primaryClass = {astro-ph.GA},
       adsurl = {https://ui.adsabs.harvard.edu/abs/2022NewA...9701895L}
}

@ARTICLE{2022ApJS..262...36S,
       author = {{S{\'a}nchez}, S.~F. and {Barrera-Ballesteros}, J.~K. and {Lacerda}, E. and {Mej{\'\i}a-Narvaez}, A. and {Camps-Fari{\~n}a}, A. and {Bruzual}, Gustavo and {Espinosa-Ponce}, C. and {Rodr{\'\i}guez-Puebla}, A. and {Calette}, A.~R. and {Ibarra-Medel}, H. and {Avila-Reese}, V. and {Hernandez-Toledo}, H. and {Bershady}, M.~A. and {Cano-Diaz}, M. and {Munguia-Cordova}, A.~M.},
        title = "{SDSS-IV MaNGA: pyPipe3D Analysis Release for 10,000 Galaxies}",
      journal = {\apjs},
         year = 2022,
        month = oct,
       volume = {262},
       number = {2},
          eid = {36},
        pages = {36},
          doi = {10.3847/1538-4365/ac7b8f},
archivePrefix = {arXiv},
       eprint = {2206.07062},
 primaryClass = {astro-ph.GA},
       adsurl = {https://ui.adsabs.harvard.edu/abs/2022ApJS..262...36S}
}

@ARTICLE{1996AJ....111.1748F,
       author = {{Fukugita}, M. and {Ichikawa}, T. and {Gunn}, J.~E. and {Doi}, M. and {Shimasaku}, K. and {Schneider}, D.~P.},
        title = "{The Sloan Digital Sky Survey Photometric System}",
      journal = {\aj},
         year = 1996,
        month = apr,
       volume = {111},
        pages = {1748},
          doi = {10.1086/117915},
       adsurl = {https://ui.adsabs.harvard.edu/abs/1996AJ....111.1748F}
}

@ARTICLE{2014ARA&A..52..415M,
       author = {{Madau}, Piero and {Dickinson}, Mark},
        title = "{Cosmic Star-Formation History}",
      journal = {\araa},
         year = 2014,
        month = aug,
       volume = {52},
        pages = {415-486},
          doi = {10.1146/annurev-astro-081811-125615},
archivePrefix = {arXiv},
       eprint = {1403.0007},
 primaryClass = {astro-ph.CO},
       adsurl = {https://ui.adsabs.harvard.edu/abs/2014ARA&A..52..415M}
}

@ARTICLE{2003PASP..115..763C,
       author = {{Chabrier}, Gilles},
        title = "{Galactic Stellar and Substellar Initial Mass Function}",
      journal = {\pasp},
         year = 2003,
        month = jul,
       volume = {115},
       number = {809},
        pages = {763-795},
          doi = {10.1086/376392},
archivePrefix = {arXiv},
       eprint = {astro-ph/0304382},
 primaryClass = {astro-ph},
       adsurl = {https://ui.adsabs.harvard.edu/abs/2003PASP..115..763C}
}

@ARTICLE{2000ApJ...533..682C,
       author = {{Calzetti}, Daniela and {Armus}, Lee and {Bohlin}, Ralph C. and {Kinney}, Anne L. and {Koornneef}, Jan and {Storchi-Bergmann}, Thaisa},
        title = "{The Dust Content and Opacity of Actively Star-forming Galaxies}",
      journal = {\apj},
         year = 2000,
        month = apr,
       volume = {533},
       number = {2},
        pages = {682-695},
          doi = {10.1086/308692},
archivePrefix = {arXiv},
       eprint = {astro-ph/9911459},
 primaryClass = {astro-ph},
       adsurl = {https://ui.adsabs.harvard.edu/abs/2000ApJ...533..682C}
}

@ARTICLE{2012ARA&A..50..531K,
       author = {{Kennicutt}, Robert C. and {Evans}, Neal J.},
        title = "{Star Formation in the Milky Way and Nearby Galaxies}",
      journal = {\araa},
         year = 2012,
        month = sep,
       volume = {50},
        pages = {531-608},
          doi = {10.1146/annurev-astro-081811-125610},
archivePrefix = {arXiv},
       eprint = {1204.3552},
 primaryClass = {astro-ph.GA},
       adsurl = {https://ui.adsabs.harvard.edu/abs/2012ARA&A..50..531K}
}

@ARTICLE{2017AJ....154...86W,
       author = {{Wake}, David A. and {Bundy}, Kevin and {Diamond-Stanic}, Aleksandar M. and {Yan}, Renbin and {Blanton}, Michael R. and {Bershady}, Matthew A. and {S{\'a}nchez-Gallego}, Jos{\'e} R. and {Drory}, Niv and {Jones}, Amy and {Kauffmann}, Guinevere and {Law}, David R. and {Li}, Cheng and {MacDonald}, Nicholas and {Masters}, Karen and {Thomas}, Daniel and {Tinker}, Jeremy and {Weijmans}, Anne-Marie and {Brownstein}, Joel R.},
        title = "{The SDSS-IV MaNGA Sample: Design, Optimization, and Usage Considerations}",
      journal = {\aj},
         year = 2017,
        month = sep,
       volume = {154},
       number = {3},
          eid = {86},
        pages = {86},
          doi = {10.3847/1538-3881/aa7ecc},
archivePrefix = {arXiv},
       eprint = {1707.02989},
 primaryClass = {astro-ph.GA},
       adsurl = {https://ui.adsabs.harvard.edu/abs/2017AJ....154...86W}
}

@ARTICLE{2015MNRAS.449.1422M,
       author = {{Mateos}, S. and {Carrera}, F.~J. and {Alonso-Herrero}, A. and {Rovilos}, E. and {Hern{\'a}n-Caballero}, A. and {Barcons}, X. and {Blain}, A. and {Caccianiga}, A. and {Della Ceca}, R. and {Severgnini}, P.},
        title = "{Revisiting the relationship between 6 {\ensuremath{\mu}}m and 2-10 keV continuum luminosities of AGN}",
      journal = {\mnras},
         year = 2015,
        month = may,
       volume = {449},
       number = {2},
        pages = {1422-1440},
          doi = {10.1093/mnras/stv299},
archivePrefix = {arXiv},
       eprint = {1501.04335},
 primaryClass = {astro-ph.GA},
       adsurl = {https://ui.adsabs.harvard.edu/abs/2015MNRAS.449.1422M}
}

@ARTICLE{2010ApJ...712..833C,
       author = {{Conroy}, Charlie and {Gunn}, James E.},
        title = "{The Propagation of Uncertainties in Stellar Population Synthesis Modeling. III. Model Calibration, Comparison, and Evaluation}",
      journal = {\apj},
         year = 2010,
        month = apr,
       volume = {712},
       number = {2},
        pages = {833-857},
          doi = {10.1088/0004-637X/712/2/833},
archivePrefix = {arXiv},
       eprint = {0911.3151},
 primaryClass = {astro-ph.CO},
       adsurl = {https://ui.adsabs.harvard.edu/abs/2010ApJ...712..833C}
}

@ARTICLE{2008ApJ...685..160N,
       author = {{Nenkova}, Maia and {Sirocky}, Matthew M. and {Nikutta}, Robert and {Ivezi{\'c}}, {\v{Z}}eljko and {Elitzur}, Moshe},
        title = "{AGN Dusty Tori. II. Observational Implications of Clumpiness}",
      journal = {\apj},
         year = 2008,
        month = sep,
       volume = {685},
       number = {1},
        pages = {160-180},
          doi = {10.1086/590483},
archivePrefix = {arXiv},
       eprint = {0806.0512},
 primaryClass = {astro-ph},
       adsurl = {https://ui.adsabs.harvard.edu/abs/2008ApJ...685..160N}
}

@ARTICLE{2007AJ....133..734B,
       author = {{Blanton}, Michael R. and {Roweis}, Sam},
        title = "{K-Corrections and Filter Transformations in the Ultraviolet, Optical, and Near-Infrared}",
      journal = {\aj},
         year = 2007,
        month = feb,
       volume = {133},
       number = {2},
        pages = {734-754},
          doi = {10.1086/510127},
archivePrefix = {arXiv},
       eprint = {astro-ph/0606170},
 primaryClass = {astro-ph},
       adsurl = {https://ui.adsabs.harvard.edu/abs/2007AJ....133..734B}
}

@ARTICLE{2004ApJ...613..109H,
       author = {{Heckman}, Timothy M. and {Kauffmann}, Guinevere and {Brinchmann}, Jarle and {Charlot}, St{\'e}phane and {Tremonti}, Christy and {White}, Simon D.~M.},
        title = "{Present-Day Growth of Black Holes and Bulges: The Sloan Digital Sky Survey Perspective}",
      journal = {\apj},
         year = 2004,
        month = sep,
       volume = {613},
       number = {1},
        pages = {109-118},
          doi = {10.1086/422872},
archivePrefix = {arXiv},
       eprint = {astro-ph/0406218},
 primaryClass = {astro-ph},
       adsurl = {https://ui.adsabs.harvard.edu/abs/2004ApJ...613..109H}
}

@ARTICLE{2017MNRAS.466..798C,
       author = {{Cappellari}, Michele},
        title = "{Improving the full spectrum fitting method: accurate convolution with Gauss-Hermite functions}",
      journal = {\mnras},
         year = 2017,
        month = apr,
       volume = {466},
       number = {1},
        pages = {798-811},
          doi = {10.1093/mnras/stw3020},
archivePrefix = {arXiv},
       eprint = {1607.08538},
 primaryClass = {astro-ph.GA},
       adsurl = {https://ui.adsabs.harvard.edu/abs/2017MNRAS.466..798C}
}

@ARTICLE{2022A&A...664A..83H,
       author = {{Heesen}, V. and {Staffehl}, M. and {Basu}, A. and {Beck}, R. and {Stein}, M. and {Tabatabaei}, F.~S. and {Hardcastle}, M.~J. and {Chy{\.z}y}, K.~T. and {Shimwell}, T.~W. and {Adebahr}, B. and {Beswick}, R. and {Bomans}, D.~J. and {Botteon}, A. and {Brinks}, E. and {Br{\"u}ggen}, M. and {Dettmar}, R. -J. and {Drabent}, A. and {de Gasperin}, F. and {G{\"u}rkan}, G. and {Heald}, G.~H. and {Horellou}, C. and {Nikiel-Wroczynski}, B. and {Paladino}, R. and {Piotrowska}, J. and {R{\"o}ttgering}, H.~J.~A. and {Smith}, D.~J.~B. and {Tasse}, C.},
        title = "{Nearby galaxies in the LOFAR Two-metre Sky Survey. I. Insights into the non-linearity of the radio-SFR relation}",
      journal = {\aap},
         year = 2022,
        month = aug,
       volume = {664},
          eid = {A83},
        pages = {A83},
          doi = {10.1051/0004-6361/202142878},
archivePrefix = {arXiv},
       eprint = {2204.00635},
 primaryClass = {astro-ph.GA},
       adsurl = {https://ui.adsabs.harvard.edu/abs/2022A&A...664A..83H}
}

@ARTICLE{2024MNRAS.531..977D,
       author = {{Das}, Soumyadeep and {Smith}, Daniel J.~B. and {Haskell}, Paul and {Hardcastle}, Martin J. and {Best}, Philip N. and {Duncan}, Kenneth J. and {Arnaudova}, Marina I. and {Shenoy}, Shravya and {Kondapally}, Rohit and {Cochrane}, Rachel K. and {Drake}, Alyssa B. and {G{\"u}rkan}, G{\"u}lay and {Ma{\l}ek}, Katarzyna and {Morabito}, Leah K. and {Prandoni}, Isabella},
        title = "{The LOFAR Two-metre Sky Survey: The nature of the faint source population and SFR-radio luminosity relation using PROSPECTOR}",
      journal = {\mnras},
         year = 2024,
        month = jun,
       volume = {531},
       number = {1},
        pages = {977-996},
          doi = {10.1093/mnras/stae1204},
archivePrefix = {arXiv},
       eprint = {2405.01624},
 primaryClass = {astro-ph.GA},
       adsurl = {https://ui.adsabs.harvard.edu/abs/2024MNRAS.531..977D}
}

@ARTICLE{2023MNRAS.523.1729B,
       author = {{Best}, P.~N. and {Kondapally}, R. and {Williams}, W.~L. and {Cochrane}, R.~K. and {Duncan}, K.~J. and {Hale}, C.~L. and {Haskell}, P. and {Ma{\l}ek}, K. and {McCheyne}, I. and {Smith}, D.~J.~B. and {Wang}, L. and {Botteon}, A. and {Bonato}, M. and {Bondi}, M. and {Calistro Rivera}, G. and {Gao}, F. and {G{\"u}rkan}, G. and {Hardcastle}, M.~J. and {Jarvis}, M.~J. and {Mingo}, B. and {Miraghaei}, H. and {Morabito}, L.~K. and {Nisbet}, D. and {Prandoni}, I. and {R{\"o}ttgering}, H.~J.~A. and {Sabater}, J. and {Shimwell}, T. and {Tasse}, C. and {van Weeren}, R.},
        title = "{The LOFAR Two-metre Sky Survey: Deep Fields data release 1. V. Survey description, source classifications, and host galaxy properties}",
      journal = {\mnras},
         year = 2023,
        month = aug,
       volume = {523},
       number = {2},
        pages = {1729-1755},
          doi = {10.1093/mnras/stad1308},
archivePrefix = {arXiv},
       eprint = {2305.05782},
 primaryClass = {astro-ph.GA},
       adsurl = {https://ui.adsabs.harvard.edu/abs/2023MNRAS.523.1729B}
}

@ARTICLE{1986ARA&A..24..171O,
       author = {{Osterbrock}, Donald E. and {Mathews}, William G.},
        title = "{Emission-line regions of active galaxies and QSOs.}",
      journal = {\araa},
         year = 1986,
        month = jan,
       volume = {24},
        pages = {171-203},
          doi = {10.1146/annurev.aa.24.090186.001131},
       adsurl = {https://ui.adsabs.harvard.edu/abs/1986ARA&A..24..171O}
}

@ARTICLE{2024ApJ...977..102P,
       author = {{Pai}, Aashay and {Blanton}, Michael R. and {Moustakas}, John},
        title = "{Mid-infrared Variability in Nearby Galaxies from the MaNGA Sample}",
      journal = {\apj},
         year = 2024,
        month = dec,
       volume = {977},
       number = {1},
          eid = {102},
        pages = {102},
          doi = {10.3847/1538-4357/ad89b8},
archivePrefix = {arXiv},
       eprint = {2407.05508},
 primaryClass = {astro-ph.GA},
       adsurl = {https://ui.adsabs.harvard.edu/abs/2024ApJ...977..102P}
}

@ARTICLE{2023MNRAS.524.5827F,
       author = {{Fu}, Youquan and {Cappellari}, Michele and {Mao}, Shude and {Lu}, Shengdong and {Zhu}, Kai and {Li}, Ran},
        title = "{A complete catalogue of broad-line AGNs and double-peaked emission lines from MaNGA integral-field spectroscopy of 10K galaxies: stellar population of AGNs, supermassive black holes, and dual AGNs}",
      journal = {\mnras},
         year = 2023,
        month = oct,
       volume = {524},
       number = {4},
        pages = {5827-5843},
          doi = {10.1093/mnras/stad2214},
archivePrefix = {arXiv},
       eprint = {2305.02676},
 primaryClass = {astro-ph.GA},
       adsurl = {https://ui.adsabs.harvard.edu/abs/2023MNRAS.524.5827F}
}

@ARTICLE{1987ApJS...63..295V,
       author = {{Veilleux}, Sylvain and {Osterbrock}, Donald E.},
        title = "{Spectral Classification of Emission-Line Galaxies}",
      journal = {\apjs},
         year = 1987,
        month = feb,
       volume = {63},
        pages = {295},
          doi = {10.1086/191166},
       adsurl = {https://ui.adsabs.harvard.edu/abs/1987ApJS...63..295V}
}

@ARTICLE{1981PASP...93....5B,
       author = {{Baldwin}, J.~A. and {Phillips}, M.~M. and {Terlevich}, R.},
        title = "{Classification parameters for the emission-line spectra of extragalactic objects.}",
      journal = {\pasp},
         year = 1981,
        month = feb,
       volume = {93},
        pages = {5-19},
          doi = {10.1086/130766},
       adsurl = {https://ui.adsabs.harvard.edu/abs/1981PASP...93....5B}
}

@ARTICLE{2020MNRAS.499.5749J,
       author = {{Ji}, Xihan and {Yan}, Renbin},
        title = "{Constraining photoionization models with a reprojected optical diagnostic diagram}",
      journal = {\mnras},
         year = 2020,
        month = dec,
       volume = {499},
       number = {4},
        pages = {5749-5764},
          doi = {10.1093/mnras/staa3259},
archivePrefix = {arXiv},
       eprint = {2007.09159},
 primaryClass = {astro-ph.GA},
       adsurl = {https://ui.adsabs.harvard.edu/abs/2020MNRAS.499.5749J}
}

@ARTICLE{2010MNRAS.403.1036C,
       author = {{Cid Fernandes}, R. and {Stasi{\'n}ska}, G. and {Schlickmann}, M.~S. and {Mateus}, A. and {Vale Asari}, N. and {Schoenell}, W. and {Sodr{\'e}}, L.},
        title = "{Alternative diagnostic diagrams and the `forgotten' population of weak line galaxies in the SDSS}",
      journal = {\mnras},
         year = 2010,
        month = apr,
       volume = {403},
       number = {2},
        pages = {1036-1053},
          doi = {10.1111/j.1365-2966.2009.16185.x},
archivePrefix = {arXiv},
       eprint = {0912.1643},
 primaryClass = {astro-ph.CO},
       adsurl = {https://ui.adsabs.harvard.edu/abs/2010MNRAS.403.1036C}
}

@ARTICLE{2015ARA&A..53..365N,
       author = {{Netzer}, Hagai},
        title = "{Revisiting the Unified Model of Active Galactic Nuclei}",
      journal = {\araa},
         year = 2015,
        month = aug,
       volume = {53},
        pages = {365-408},
          doi = {10.1146/annurev-astro-082214-122302},
archivePrefix = {arXiv},
       eprint = {1505.00811},
 primaryClass = {astro-ph.GA},
       adsurl = {https://ui.adsabs.harvard.edu/abs/2015ARA&A..53..365N}
}

@ARTICLE{2014ARA&A..52..589H,
       author = {{Heckman}, Timothy M. and {Best}, Philip N.},
        title = "{The Coevolution of Galaxies and Supermassive Black Holes: Insights from Surveys of the Contemporary Universe}",
      journal = {\araa},
         year = 2014,
        month = aug,
       volume = {52},
        pages = {589-660},
          doi = {10.1146/annurev-astro-081913-035722},
archivePrefix = {arXiv},
       eprint = {1403.4620},
 primaryClass = {astro-ph.GA},
       adsurl = {https://ui.adsabs.harvard.edu/abs/2014ARA&A..52..589H}
}

@ARTICLE{2015AJ....149...77D,
       author = {{Drory}, N. and {MacDonald}, N. and {Bershady}, M.~A. and {Bundy}, K. and {Gunn}, J. and {Law}, D.~R. and {Smith}, M. and {Stoll}, R. and {Tremonti}, C.~A. and {Wake}, D.~A. and {Yan}, R. and {Weijmans}, A.~M. and {Byler}, N. and {Cherinka}, B. and {Cope}, F. and {Eigenbrot}, A. and {Harding}, P. and {Holder}, D. and {Huehnerhoff}, J. and {Jaehnig}, K. and {Jansen}, T.~C. and {Klaene}, M. and {Paat}, A.~M. and {Percival}, J. and {Sayres}, C.},
        title = "{The MaNGA Integral Field Unit Fiber Feed System for the Sloan 2.5 m Telescope}",
      journal = {\aj},
         year = 2015,
        month = feb,
       volume = {149},
       number = {2},
          eid = {77},
        pages = {77},
          doi = {10.1088/0004-6256/149/2/77},
archivePrefix = {arXiv},
       eprint = {1412.1535},
 primaryClass = {astro-ph.IM},
       adsurl = {https://ui.adsabs.harvard.edu/abs/2015AJ....149...77D}
}

@ARTICLE{2019AJ....158..160B,
       author = {{Belfiore}, Francesco and {Westfall}, Kyle B. and {Schaefer}, Adam and {Cappellari}, Michele and {Ji}, Xihan and {Bershady}, Matthew A. and {Tremonti}, Christy and {Law}, David R. and {Yan}, Renbin and {Bundy}, Kevin and {Shetty}, Shravan and {Drory}, Niv and {Thomas}, Daniel and {Emsellem}, Eric and {S{\'a}nchez}, Sebasti{\'a}n F.},
        title = "{The Data Analysis Pipeline for the SDSS-IV MaNGA IFU Galaxy Survey: Emission-line Modeling}",
      journal = {\aj},
         year = 2019,
        month = oct,
       volume = {158},
       number = {4},
          eid = {160},
        pages = {160},
          doi = {10.3847/1538-3881/ab3e4e},
archivePrefix = {arXiv},
       eprint = {1901.00866},
 primaryClass = {astro-ph.GA},
       adsurl = {https://ui.adsabs.harvard.edu/abs/2019AJ....158..160B}
}

@ARTICLE{2024MNRAS.527.6540H,
       author = {{Hale}, C.~L. and {Schwarz}, D.~J. and {Best}, P.~N. and {Nakoneczny}, S.~J. and {Alonso}, D. and {Bacon}, D. and {B{\"o}hme}, L. and {Bhardwaj}, N. and {Bilicki}, M. and {Camera}, S. and {Heneka}, C.~S. and {Pashapour-Ahmadabadi}, M. and {Tiwari}, P. and {Zheng}, J. and {Duncan}, K.~J. and {Jarvis}, M.~J. and {Kondapally}, R. and {Magliocchetti}, M. and {Rottgering}, H.~J.~A. and {Shimwell}, T.~W.},
        title = "{Cosmology from LOFAR Two-metre Sky Survey Data Release 2: angular clustering of radio sources}",
      journal = {\mnras},
         year = 2024,
        month = jan,
       volume = {527},
       number = {3},
        pages = {6540-6568},
          doi = {10.1093/mnras/stad3088},
archivePrefix = {arXiv},
       eprint = {2310.07627},
 primaryClass = {astro-ph.CO},
       adsurl = {https://ui.adsabs.harvard.edu/abs/2024MNRAS.527.6540H}
}

@MISC{2015ascl.soft02007M,
       author = {{Mohan}, Niruj and {Rafferty}, David},
        title = "{PyBDSF: Python Blob Detection and Source Finder}",
 howpublished = {Astrophysics Source Code Library, record ascl:1502.007},
         year = 2015,
        month = feb,
          eid = {ascl:1502.007},
        pages = {ascl:1502.007},
archivePrefix = {ascl},
       eprint = {1502.007},
       adsurl = {https://ui.adsabs.harvard.edu/abs/2015ascl.soft02007M}
}

@ARTICLE{2023A&A...678A.151H,
       author = {{Hardcastle}, M.~J. and {Horton}, M.~A. and {Williams}, W.~L. and {Duncan}, K.~J. and {Alegre}, L. and {Barkus}, B. and {Croston}, J.~H. and {Dickinson}, H. and {Osinga}, E. and {R{\"o}ttgering}, H.~J.~A. and {Sabater}, J. and {Shimwell}, T.~W. and {Smith}, D.~J.~B. and {Best}, P.~N. and {Botteon}, A. and {Br{\"u}ggen}, M. and {Drabent}, A. and {de Gasperin}, F. and {G{\"u}rkan}, G. and {Hajduk}, M. and {Hale}, C.~L. and {Hoeft}, M. and {Jamrozy}, M. and {Kunert-Bajraszewska}, M. and {Kondapally}, R. and {Magliocchetti}, M. and {Mahatma}, V.~H. and {Mostert}, R.~I.~J. and {O'Sullivan}, S.~P. and {Pajdosz-{\'S}mierciak}, U. and {Petley}, J. and {Pierce}, J.~C.~S. and {Prandoni}, I. and {Schwarz}, D.~J. and {Shulewski}, A. and {Siewert}, T.~M. and {Stott}, J.~P. and {Tang}, H. and {Vaccari}, M. and {Zheng}, X. and {Bailey}, T. and {Desbled}, S. and {Goyal}, A. and {Gonano}, V. and {Hanset}, M. and {Kurtz}, W. and {Lim}, S.~M. and {Mielle}, L. and {Molloy}, C.~S. and {Roth}, R. and {Terentev}, I.~A. and {Torres}, M.},
        title = "{The LOFAR Two-Metre Sky Survey. VI. Optical identifications for the second data release}",
      journal = {\aap},
         year = 2023,
        month = oct,
       volume = {678},
          eid = {A151},
        pages = {A151},
          doi = {10.1051/0004-6361/202347333},
archivePrefix = {arXiv},
       eprint = {2309.00102},
 primaryClass = {astro-ph.GA},
       adsurl = {https://ui.adsabs.harvard.edu/abs/2023A&A...678A.151H}
}

@ARTICLE{2015ApJS..219....8C,
       author = {{Chang}, Yu-Yen and {van der Wel}, Arjen and {da Cunha}, Elisabete and {Rix}, Hans-Walter},
        title = "{Stellar Masses and Star Formation Rates for 1M Galaxies from SDSS+WISE}",
      journal = {\apjs},
         year = 2015,
        month = jul,
       volume = {219},
       number = {1},
          eid = {8},
        pages = {8},
          doi = {10.1088/0067-0049/219/1/8},
archivePrefix = {arXiv},
       eprint = {1506.00648},
 primaryClass = {astro-ph.GA},
       adsurl = {https://ui.adsabs.harvard.edu/abs/2015ApJS..219....8C}
}

@ARTICLE{2014AJ....147..108L,
       author = {{Lang}, Dustin},
        title = "{unWISE: Unblurred Coadds of the WISE Imaging}",
      journal = {\aj},
         year = 2014,
        month = may,
       volume = {147},
       number = {5},
          eid = {108},
        pages = {108},
          doi = {10.1088/0004-6256/147/5/108},
archivePrefix = {arXiv},
       eprint = {1405.0308},
 primaryClass = {astro-ph.IM},
       adsurl = {https://ui.adsabs.harvard.edu/abs/2014AJ....147..108L}
}

@ARTICLE{2014ApJ...792...30M,
       author = {{Mainzer}, A. and {Bauer}, J. and {Cutri}, R.~M. and {Grav}, T. and {Masiero}, J. and {Beck}, R. and {Clarkson}, P. and {Conrow}, T. and {Dailey}, J. and {Eisenhardt}, P. and {Fabinsky}, B. and {Fajardo-Acosta}, S. and {Fowler}, J. and {Gelino}, C. and {Grillmair}, C. and {Heinrichsen}, I. and {Kendall}, M. and {Kirkpatrick}, J. Davy and {Liu}, F. and {Masci}, F. and {McCallon}, H. and {Nugent}, C.~R. and {Papin}, M. and {Rice}, E. and {Royer}, D. and {Ryan}, T. and {Sevilla}, P. and {Sonnett}, S. and {Stevenson}, R. and {Thompson}, D.~B. and {Wheelock}, S. and {Wiemer}, D. and {Wittman}, M. and {Wright}, E. and {Yan}, L.},
        title = "{Initial Performance of the NEOWISE Reactivation Mission}",
      journal = {\apj},
         year = 2014,
        month = sep,
       volume = {792},
       number = {1},
          eid = {30},
        pages = {30},
          doi = {10.1088/0004-637X/792/1/30},
archivePrefix = {arXiv},
       eprint = {1406.6025},
 primaryClass = {astro-ph.EP},
       adsurl = {https://ui.adsabs.harvard.edu/abs/2014ApJ...792...30M}
}

@ARTICLE{2016ApJ...818...88L,
       author = {{LaMassa}, Stephanie M. and {Civano}, Francesca and {Brusa}, Marcella and {Stern}, Daniel and {Glikman}, Eilat and {Gallagher}, Sarah and {Urry}, C. Meg and {Cales}, Sabrina and {Cappelluti}, Nico and {Cardamone}, Carolin and {Comastri}, Andrea and {Farrah}, Duncan and {Greene}, Jenny E. and {Komossa}, S. and {Merloni}, Andrea and {Mroczkowski}, Tony and {Natarajan}, Priyamvada and {Richards}, Gordon and {Salvato}, Mara and {Schawinski}, Kevin and {Treister}, Ezequiel},
        title = "{On R-W1 as A Diagnostic to Discover Obscured Active Galactic Nuclei in Wide-area X-Ray Surveys}",
      journal = {\apj},
         year = 2016,
        month = feb,
       volume = {818},
       number = {1},
          eid = {88},
        pages = {88},
          doi = {10.3847/0004-637X/818/1/88},
archivePrefix = {arXiv},
       eprint = {1511.02883},
 primaryClass = {astro-ph.GA},
       adsurl = {https://ui.adsabs.harvard.edu/abs/2016ApJ...818...88L}
}

@MISC{2013wise.rept....1C,
       author = {{Cutri}, R.~M. and {Wright}, E.~L. and {Conrow}, T. and {Fowler}, J.~W. and {Eisenhardt}, P.~R.~M. and {Grillmair}, C. and {Kirkpatrick}, J.~D. and {Masci}, F. and {McCallon}, H.~L. and {Wheelock}, S.~L. and {Fajardo-Acosta}, S. and {Yan}, L. and {Benford}, D. and {Harbut}, M. and {Jarrett}, T. and {Lake}, S. and {Leisawitz}, D. and {Ressler}, M.~E. and {Stanford}, S.~A. and {Tsai}, C.~W. and {Liu}, F. and {Helou}, G. and {Mainzer}, A. and {Gettings}, D. and {Gonzalez}, A. and {Hoffman}, D. and {Marsh}, K.~A. and {Padgett}, D. and {Skrutskie}, M.~F. and {Beck}, R.~P. and {Papin}, M. and {Wittman}, M.},
        title = "{Explanatory Supplement to the AllWISE Data Release Products}",
 howpublished = {Explanatory Supplement to the AllWISE Data Release Products, by R. M. Cutri et al.},
         year = 2013,
        month = nov,
        pages = {1},
       adsurl = {https://ui.adsabs.harvard.edu/abs/2013wise.rept....1C}
}

@ARTICLE{2010AJ....140.1868W,
       author = {{Wright}, Edward L. and {Eisenhardt}, Peter R.~M. and {Mainzer}, Amy K. and {Ressler}, Michael E. and {Cutri}, Roc M. and {Jarrett}, Thomas and {Kirkpatrick}, J. Davy and {Padgett}, Deborah and {McMillan}, Robert S. and {Skrutskie}, Michael and {Stanford}, S.~A. and {Cohen}, Martin and {Walker}, Russell G. and {Mather}, John C. and {Leisawitz}, David and {Gautier}, III, Thomas N. and {McLean}, Ian and {Benford}, Dominic and {Lonsdale}, Carol J. and {Blain}, Andrew and {Mendez}, Bryan and {Irace}, William R. and {Duval}, Valerie and {Liu}, Fengchuan and {Royer}, Don and {Heinrichsen}, Ingolf and {Howard}, Joan and {Shannon}, Mark and {Kendall}, Martha and {Walsh}, Amy L. and {Larsen}, Mark and {Cardon}, Joel G. and {Schick}, Scott and {Schwalm}, Mark and {Abid}, Mohamed and {Fabinsky}, Beth and {Naes}, Larry and {Tsai}, Chao-Wei},
        title = "{The Wide-field Infrared Survey Explorer (WISE): Mission Description and Initial On-orbit Performance}",
      journal = {\aj},
         year = 2010,
        month = dec,
       volume = {140},
       number = {6},
        pages = {1868-1881},
          doi = {10.1088/0004-6256/140/6/1868},
archivePrefix = {arXiv},
       eprint = {1008.0031},
 primaryClass = {astro-ph.IM},
       adsurl = {https://ui.adsabs.harvard.edu/abs/2010AJ....140.1868W}
}

@ARTICLE{2019AJ....157..168D,
       author = {{Dey}, Arjun and {Schlegel}, David J. and {Lang}, Dustin and {Blum}, Robert and {Burleigh}, Kaylan and {Fan}, Xiaohui and {Findlay}, Joseph R. and {Finkbeiner}, Doug and {Herrera}, David and {Juneau}, St{\'e}phanie and {Landriau}, Martin and {Levi}, Michael and {McGreer}, Ian and {Meisner}, Aaron and {Myers}, Adam D. and {Moustakas}, John and {Nugent}, Peter and {Patej}, Anna and {Schlafly}, Edward F. and {Walker}, Alistair R. and {Valdes}, Francisco and {Weaver}, Benjamin A. and {Y{\`e}che}, Christophe and {Zou}, Hu and {Zhou}, Xu and {Abareshi}, Behzad and {Abbott}, T.~M.~C. and {Abolfathi}, Bela and {Aguilera}, C. and {Alam}, Shadab and {Allen}, Lori and {Alvarez}, A. and {Annis}, James and {Ansarinejad}, Behzad and {Aubert}, Marie and {Beechert}, Jacqueline and {Bell}, Eric F. and {BenZvi}, Segev Y. and {Beutler}, Florian and {Bielby}, Richard M. and {Bolton}, Adam S. and {Brice{\~n}o}, C{\'e}sar and {Buckley-Geer}, Elizabeth J. and {Butler}, Karen and {Calamida}, Annalisa and {Carlberg}, Raymond G. and {Carter}, Paul and {Casas}, Ricard and {Castander}, Francisco J. and {Choi}, Yumi and {Comparat}, Johan and {Cukanovaite}, Elena and {Delubac}, Timoth{\'e}e and {DeVries}, Kaitlin and {Dey}, Sharmila and {Dhungana}, Govinda and {Dickinson}, Mark and {Ding}, Zhejie and {Donaldson}, John B. and {Duan}, Yutong and {Duckworth}, Christopher J. and {Eftekharzadeh}, Sarah and {Eisenstein}, Daniel J. and {Etourneau}, Thomas and {Fagrelius}, Parker A. and {Farihi}, Jay and {Fitzpatrick}, Mike and {Font-Ribera}, Andreu and {Fulmer}, Leah and {G{\"a}nsicke}, Boris T. and {Gaztanaga}, Enrique and {George}, Koshy and {Gerdes}, David W. and {Gontcho}, Satya Gontcho A. and {Gorgoni}, Claudio and {Green}, Gregory and {Guy}, Julien and {Harmer}, Diane and {Hernandez}, M. and {Honscheid}, Klaus and {Huang}, Lijuan Wendy and {James}, David J. and {Jannuzi}, Buell T. and {Jiang}, Linhua and {Joyce}, Richard and {Karcher}, Armin and {Karkar}, Sonia and {Kehoe}, Robert and {Kneib}, Jean-Paul and {Kueter-Young}, Andrea and {Lan}, Ting-Wen and {Lauer}, Tod R. and {Le Guillou}, Laurent and {Le Van Suu}, Auguste and {Lee}, Jae Hyeon and {Lesser}, Michael and {Perreault Levasseur}, Laurence and {Li}, Ting S. and {Mann}, Justin L. and {Marshall}, Robert and {Mart{\'\i}nez-V{\'a}zquez}, C.~E. and {Martini}, Paul and {du Mas des Bourboux}, H{\'e}lion and {McManus}, Sean and {Meier}, Tobias Gabriel and {M{\'e}nard}, Brice and {Metcalfe}, Nigel and {Mu{\~n}oz-Guti{\'e}rrez}, Andrea and {Najita}, Joan and {Napier}, Kevin and {Narayan}, Gautham and {Newman}, Jeffrey A. and {Nie}, Jundan and {Nord}, Brian and {Norman}, Dara J. and {Olsen}, Knut A.~G. and {Paat}, Anthony and {Palanque-Delabrouille}, Nathalie and {Peng}, Xiyan and {Poppett}, Claire L. and {Poremba}, Megan R. and {Prakash}, Abhishek and {Rabinowitz}, David and {Raichoor}, Anand and {Rezaie}, Mehdi and {Robertson}, A.~N. and {Roe}, Natalie A. and {Ross}, Ashley J. and {Ross}, Nicholas P. and {Rudnick}, Gregory and {Safonova}, Sasha and {Saha}, Abhijit and {S{\'a}nchez}, F. Javier and {Savary}, Elodie and {Schweiker}, Heidi and {Scott}, Adam and {Seo}, Hee-Jong and {Shan}, Huanyuan and {Silva}, David R. and {Slepian}, Zachary and {Soto}, Christian and {Sprayberry}, David and {Staten}, Ryan and {Stillman}, Coley M. and {Stupak}, Robert J. and {Summers}, David L. and {Sien Tie}, Suk and {Tirado}, H. and {Vargas-Maga{\~n}a}, Mariana and {Vivas}, A. Katherina and {Wechsler}, Risa H. and {Williams}, Doug and {Yang}, Jinyi and {Yang}, Qian and {Yapici}, Tolga and {Zaritsky}, Dennis and {Zenteno}, A. and {Zhang}, Kai and {Zhang}, Tianmeng and {Zhou}, Rongpu and {Zhou}, Zhimin},
        title = "{Overview of the DESI Legacy Imaging Surveys}",
      journal = {\aj},
         year = 2019,
        month = may,
       volume = {157},
       number = {5},
          eid = {168},
        pages = {168},
          doi = {10.3847/1538-3881/ab089d},
archivePrefix = {arXiv},
       eprint = {1804.08657},
 primaryClass = {astro-ph.IM},
       adsurl = {https://ui.adsabs.harvard.edu/abs/2019AJ....157..168D}
}




\bsp	
\label{lastpage}
\end{document}